\documentclass[cernpreprint,texlive=2011,txfonts,UKenglish]{latex/atlasdoc} 
\pdfoutput=1
\usepackage[subfigure,subcaption,block=none]{latex/atlaspackage}
\usepackage{latex/atlasphysics}
\usepackage{latex/atlasbiblatex}
\usepackage{multirow}

\usepackage{boobo_paper-defs}

\graphicspath{{figures/}{logos/}}

\AtlasTitle{Identification of boosted, hadronically decaying $W$ bosons and comparisons with ATLAS data taken at $\sqrt{s} = 8$~\TeV}

\author{The ATLAS Collaboration}

\AtlasRefCode{PERF-2015-03}

\PreprintIdNumber{CERN-PH-EP-2015-204}

\AtlasJournal{Eur. Phys. J. C }

\AtlasAbstract{This paper reports a detailed study of techniques for identifying boosted, hadronically decaying $W$ bosons using \SI{20.3}{\per\fb} of proton--proton collision data collected by the ATLAS detector at the LHC at a centre-of-mass energy $\sqrt{s} = 8\TeV$. A range of techniques for optimising the signal jet mass resolution are combined with various jet substructure variables. The results of these studies in Monte Carlo simulations show that a simple pairwise combination of groomed jet mass and one substructure variable can provide a 50\% efficiency for identifying $W$ bosons with transverse momenta larger than $200\GeV$ while maintaining multijet background efficiencies of 2--4\% for jets with the same transverse momentum. These signal and background efficiencies are confirmed in data for a selection of tagging techniques.}

\AtlasCoverSupportingNote{ATL-COM-PHYS-2014-1450}{https://cds.cern.ch/record/1967511}
\AtlasCoverTwikiURL{}

\AtlasCoverCommentsDeadline{Spokesperson's discretion}

\AtlasCoverAnalysisTeam{Christoph~Anders, Ayana~Arce, Lily~Asquith~(*), Petar~Bokan, Nicolas~Bousson, Reina~Camacho, Mario~Campanelli~(*), Chris~Delitzsch~(*), Joseph~Haley, Enrique~Kajomovitz, Shih-Chieh~Hsu~(*), Francesco~De~Lorenzi, Sam~Meehan, Ben~Nachman, David~Neto, Jason~Veatch, Lidija~Zivkovic}

\AtlasCoverEdBoardMember{Marjorie~Shapiro~(*), Jean-Francois~Arguin, David~Miller, Hideki~Okawa}

\AtlasCoverEgroupEditors{atlas-perf-2015-03-editors@cern.ch}

\AtlasCoverEgroupEdBoard{atlas-perf-2015-03-editorial-board@cern.ch}

\hypersetup{pdftitle={ATLAS draft},pdfauthor={The ATLAS Collaboration}}

\begin{document}
\maketitle
\tableofcontents

\clearpage

\newcommand{\AtlasCoordFootnote}{
ATLAS uses a right-handed coordinate system with its origin at the nominal interaction point (IP)
in the centre of the detector and the $z$-axis along the beam pipe.
The $x$-axis points from the IP to the centre of the LHC ring,
and the $y$-axis points upwards.
Cylindrical coordinates $(r,\phi)$ are used in the transverse plane, 
$\phi$ being the azimuthal angle around the $z$-axis.
The pseudorapidity is defined in terms of the polar angle $\theta$ as $\eta = -\ln \tan(\theta/2)$.
Angular distance is measured in units of $\Delta R \equiv \sqrt{(\Delta\eta)^{2} + (\Delta\phi)^{2}}$.}

\section{Introduction}
\label{sec:intro}

The high collision energies at the Large Hadron Collider (LHC) can result in the production of particles with transverse\footnote{\AtlasCoordFootnote} momenta, \pt, much larger than their mass. Such particles are boosted: their decay products are highly collimated, and for fully hadronic decays they can be reconstructed as a single hadronic jet~\cite{Seymour:1993mx} (a useful rule of thumb is $2M/\pt \sim R$: twice the jet mass divided by the \pt\ is roughly equal to the maximum opening angle of the two decay products). Heavy new particles as predicted in many theories beyond the Standard Model can be a source of highly boosted particles. 

The work presented here is the result of a detailed study of a large number of techniques and substructure variables that have, over recent years, been proposed as effective methods for tagging hadronically decaying boosted particles. 
In 2012, the ATLAS experiment collected \SI{20.3}{\per\fb} of proton--proton collision data at a centre-of-mass energy of $\sqrt{s} = 8\TeV$, providing an opportunity to determine which of the many available techniques are most useful for identifying boosted, hadronically decaying $W$ bosons. In the studies presented here, jets that contain the $W$ boson decay products are referred to as $W$-jets.

A brief overview of the existing jet grooming and substructure techniques, along with references to more detailed information, are provided in \secref{primer}. The ATLAS detector is described in \secref{detector}, and details of Monte Carlo simulations (MC) in \secref{mc}. The event selection procedure and object definitions are given in \secref{objects}.

The body of the work detailing the $W$-jet tagging performance studies is divided into a broad study using MC (\secref{partOne}) and a detailed study of selected techniques in data (\secref{datamc}). 

In \secref{partOne} a two-stage optimisation procedure has been adopted: firstly more than 500 jet reconstruction and grooming algorithm configurations are investigated at a basic level, studying the groomed jet mass distributions only. Secondly, 27 configurations that are well-behaved and show potential for $W$-jet tagging are investigated using pairwise combinations of mass and one substructure variable. 

In \secref{datamc}, one of the four most promising jet grooming algorithms and three substructure variables are selected as a benchmark for more detailed studies of the $W$-jet tagging performance in data. Jet mass and energy calibrations are derived and uncertainties are evaluated for the mass and the three selected substructure variables. Signal and background efficiencies are measured in \ttbar events and multijet events, respectively. Efficiencies in different MC simulations and event topologies are compared, and various sources of systematic uncertainty and their effects on the measurements are discussed.

In \secref{conclusions} the conclusions of all the studies are presented.

\section{A brief introduction to jets, grooming, and substructure variables}
\label{sec:primer}

\subsection{Jet grooming algorithms}\label{sec:primer_grooming}

The jet grooming algorithms studied here fall into three main categories: trimming~\cite{Krohn2010}, pruning~\cite{Ellis2009a,pruning2009} and split-filtering~\cite{Butterworth:2008iy}. Within each category there are several tunable configuration parameters, in addition to the chosen initial jet reconstruction algorithm, Cambridge--Aachen~\cite{Cacciari:2008gn} (\camkt) or \antikt~\cite{Cacciari:2008gp}, and jet radius parameter $R$. The FastJet~\cite{Cacciari:2011ma} package is used for jet reconstruction and grooming.
Jet grooming algorithms generally have two uses; (\textit{i}): to remove contributions from pileup (additional \textit{pp} interactions in the same or adjacent bunch crossings within the detector readout window), and (\textit{ii}) to reveal hard substructure within jets resulting from massive particle decays by removing the soft component of the radiation.

The three major categories of jet grooming algorithms are described below:

\begin{itemize}

\item \textbf{Trimming}: Starting with constituents of jets 
initially reconstructed using the C/A or \antikt\ algorithm, smaller `subjets' are reconstructed using the \kt algorithm~\cite{Ellis1993,Ellis1993} with a radius parameter $R = R_{\rm sub}$, 
and removed if they carry less than a fraction \Fcut of the original, ungroomed, large-$R$ jet \pt.
For reference, the recommended trimming configuration from prior ATLAS studies~\cite{Aad:2013gja} is \antiktten, with \Fcut $\geq~5~\%$ and $R_{\rm sub}~=~0.3$.

\item \textbf{Pruning}: The constituents of jets initially reconstructed with the \camkt\ or \antikt\ algorithms are 
re-clustered with the \camkt\ algorithm with two parameters: \Rcut and \Zcut. The \kt algorithm was used for re-clustering in previous studies~\cite{Aad:2013gja}, but was not found to be as effective.
In each pairwise clustering, the secondary constituent is discarded if it is ($i$) wide-angled: $\Delta R_{12} >$ \Rcut$\times 2M/p_{\mathrm T}$, where $\Delta R_{12}$ is the angular separation of the two subjets; or ($ii$) soft: $f_{2} <$ \Zcut, where $M$ is the jet mass and $f_{2}$ is the \pt\ fraction of the softer constituent with respect to the \pt\ of the pair. A configuration of the pruning algorithm is favoured by the CMS experiment for $W$-jet tagging~\cite{Khachatryan:2014vla,CMSWZ}, using C/A jets with $R=0.8$ and pruning with \Zcut=10\% and \Rcut=$\frac{1}{2}$.

\item \textbf{Split-filtering}:  
This algorithm has two stages: the first (splitting) is based on the jet substructure, and the second (filtering) is a grooming stage to remove soft radiation.
For the first stage, \camkt jets are de-clustered through the clustering history of the jet. This declustering is an exact reversal of the \camkt clustering procedure, and can be thought of as splitting the jet into two pieces. 
The momentum balance, \MomentumBalance, is defined as:

\begin{equation}\label{eq:mombalance}
\MomentumBalance =  \frac{\min(p_{\mathrm T1}, p_{\mathrm T2})}{m_{12}} \Delta R_{12},
\end{equation}

where $p_{\mathrm T1}$ ($p_{\mathrm T2}$) is the piece with the highest (the lowest) \pt, and $m_{12}$ is the invariant mass of the two pieces.

The mass-drop fraction $\mu_{12}$ is the fraction of mass carried by the piece with the highest mass:

\begin{equation}\label{eq:massdrop}
\mu_{12} = \frac{\max(m_1,m_2)}{m_{12}}.
\end{equation}

If the requirements on the mass-drop \MassDrop $<$ \MaxMassDrop and momentum balance \MomentumBalance $>$ \MinMomentumBalance are met then the jet is accepted and can proceed to the filtering stage. Otherwise the de-clustering procedure continues with the highest mass piece: this is now split into two pieces and the \MassDrop and \MomentumBalance requirements are again checked. This process continues iteratively.
 In the filtering stage, the constituents of the surviving jet are reclustered with a subjet size of $R_{\rm sub} = {\rm{min}}(0.3, \Delta R_{12})$ where $\Delta R_{12}$ is taken from the splitting stage. Any remaining radiation outside the three hardest subjets is discarded. This algorithm differs somewhat from pruning and trimming in that it involves both grooming and jet selection. A version of this algorithm is favoured by ATLAS diboson resonance searches~\cite{Aad:2015owa,Aad:2014xka,Aad:2015ufa}.
\end{itemize}

\subsection{Substructure variables}\label{sec:primer_vars}

\begin{figure}
\begin{center}
\includegraphics[scale=0.25,angle=-90]{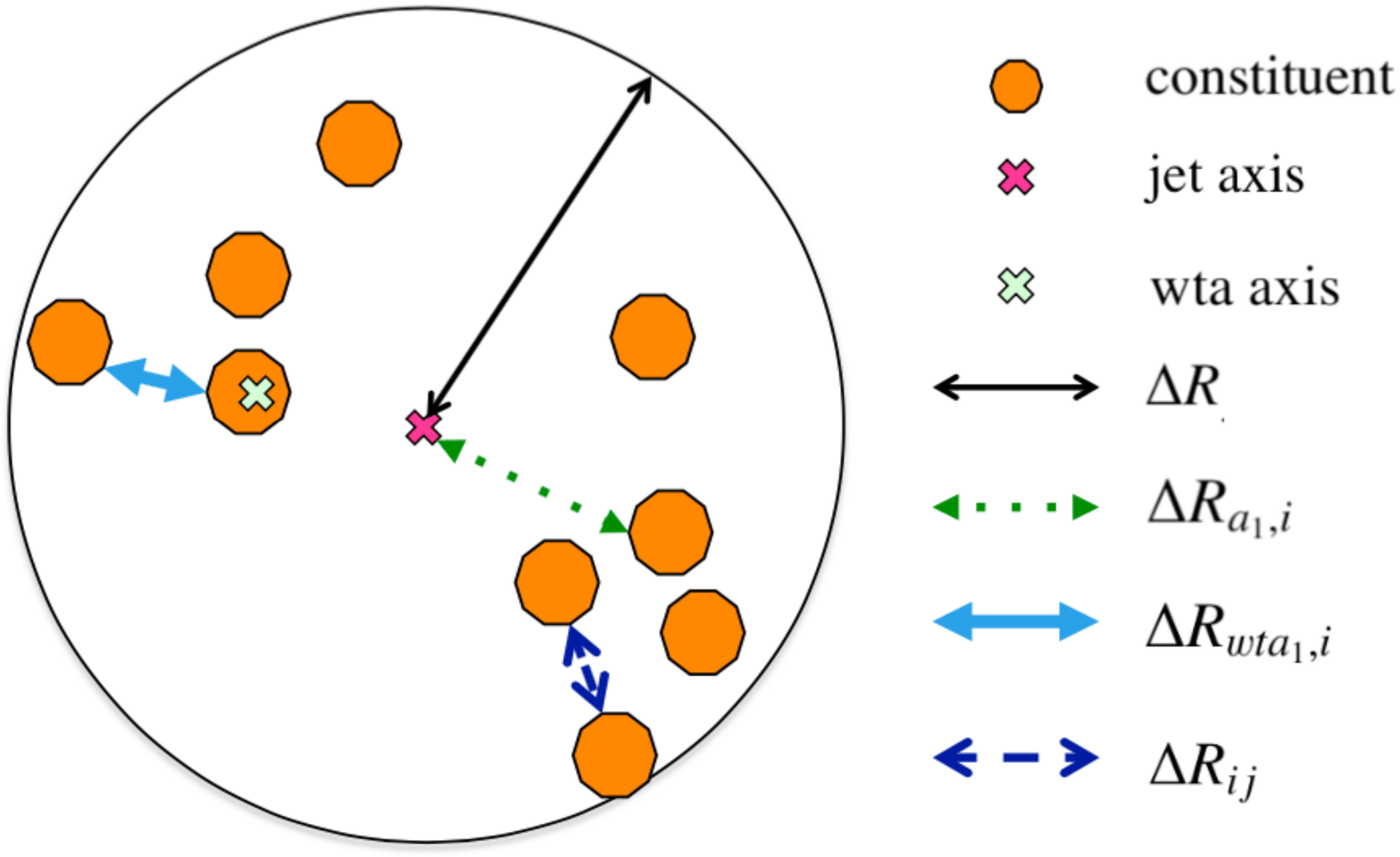}
\end{center}
\caption[Key to distance measures.]{Key to the various distance measures used in the calculation of substructure variables. The large black circle represents a jet in ($\eta$, $\phi$) space. The small, filled (orange) circles represent the constituents from which the jet is reconstructed. The various distance measures indicated are used by one or more of the algorithms described in the text. The abbreviation `wta' stands for `winner-takes-all'.}
\label{fig:key}
\end{figure}

Substructure variables are a set of jet properties that are designed to uncover hard substructure within jets. An important difference in the substructure variables comes from the choice of distance measure used in their calculation. The various distance measures available are illustrated in \figref{key}. The jet axis is usually defined as the thrust axis (along the jet momentum vector) and can also be defined as the `winner-takes-all' axis which is along the momentum vector of the constituent with the largest momentum.

The many jet substructure techniques can be roughly categorised as 
follows:

\begin{itemize}

\item \textbf{Jet shapes} use the relative positions and momenta of jet constituents with respect to each other, rather than defining subjets. The jet mass, $M$, energy correlation ratios \CTwoBeta~\cite{Larkoski:2013eya} and \DTwoBeta~\cite{Larkoski:2014pc,Larkoski:2015kga}, the mass-normalised angularity \Angularity~\cite{Almeida:2008yp}, and the planar flow, $P$~\cite{Almeida:2008yp}, all satisfy this description. The calculations of the jet mass and energy correlation ratios are described later in this section.

\item \textbf{Splitting scales} use the clustering history of the jet to define substructures (`natural subjets'). The splitting scales studied here are \SplitOneTwo~\cite{ButterworthSplit2002} and its mass-normalised form \ZOneTwo~\cite{Thaler2008}, and the momentum balance and mass-drop variables \MomentumBalance and \MassDrop, defined above in the description of the split-filtering algorithm. The soft-drop level \SoftDropLevel~\cite{Larkoski:2014sd} also belongs in this class of variables.

\item \textbf{Subjettiness} variables~\cite{Thaler:2010tr,Thaler:2011gf} 
force the constituents into substructure templates to see how well they fit 
(`synthetic subjets'), and are connected to how likely the corresponding 
jet is composed of n subjets. The calculations for two forms of 
2-subjettiness 
\TauTwo, \TauTwoWTA, and the corresponding ratios \TauTwoOne, \TauTwoOneWTA are given later in this section. The dipolarity~\cite{Hook2011}, D, uses a related method to define hard substructure.

\item \textbf{Centre-of-mass jet shapes} transform the constituents and then use them with respect to the jet axis. The variables considered are thrust, \ThrustMin, \ThrustMaj, sphericity, $S$, and aplanarity, $A$, which have been used in a previous ATLAS measurement~\cite{AtlasWtag7tev}.

\item \textbf{Quantum-jet variables} The quantum jets (`Q-jets') method~\cite{Ellis:2012sn} is unique in its class, using a non-deterministic approach to jet reconstruction. More information on the use of this method by ATLAS can be found in Ref.~\cite{QjetsPerf2012}.

\end{itemize}

The variables found in the following studies to be most interesting in terms of $W$-jet tagging are described here in more detail.

\textbf{Jet Mass}:\\
The mass of a jet is given by the difference between the squared sums of the energy $E_{i}$ and momenta $p_{i}$ of the constituents: 

\begin{equation}
M^2 = \left( \sum_{i} E_{i} \right) ^2 - \left( \sum_{i} p_{i} \right) ^2.
\label{eq:jetmasseqn}
\end{equation}

For a two-body decay, the jet mass can be approximated as:
\begin{equation}
M^2  \approx p_{\rm{T}1} p_{\rm{T}2} \Delta R^{2}_{12}.\label{eq:jetmasstwobody}
\end{equation}

\textbf{N-subjettiness}:\\
The ``N-subjettiness''~\cite{Thaler:2010tr,Thaler:2011gf} jet shape variables describe to what degree the substructure of a given jet $J$ is compatible with being composed of
$N$ or fewer subjets. The 0-, 1- and 2-subjettiness are defined as:

\begin{subequations}
\begin{align}
\tau_0 (\beta) = & \sum_{i \in J} p_{\mathrm{T}_{i}} \Delta R^{\beta}, \label{eq:tau0} \\[1ex]
\tau_1 (\beta) = & \frac{1}{\tau_{0} (\beta)} \sum_{i \in J} p_{\mathrm{T}_{i}} \Delta R_{a_{1},i}^{\beta}, \label{eq:tau1} \\[1ex]
\tau_2 (\beta) = & \frac{1}{\tau_{0} (\beta)} \sum_{i \in J} p_{\mathrm{T}_{i}} \min(\Delta R_{a_{1},i}^{\beta}, \Delta R_{a_{2},i}^{\beta}),\label{eq:tau2} 
\end{align}
\end{subequations}

where the distance $\Delta R$ refers to the distance between constituent $i$ and the jet axis, and the parameter $\beta$ can be used to give a weight to the angular separation of the jet constituents. In the studies presented here, the value of $\beta = 1$ is taken. The calculation of $\tau_N$ requires the definition of $N$ axes, such that the distance between each constituent and any of these axes is $R_{a_{N},i}$. In the above functions, the sum is performed over the constituents $i$ in the jet $J$, such that the normalisation factor $\tau_{0}$ (\eref{tau0}) is equivalent to the magnitude of the jet \pt multiplied by the $\beta-$exponentiated jet radius.  

Recent studies~\cite{Larkoski:2014wta} have shown that an effective alternative axis definition can increase the discrimination power of these variables. The `winner-takes-all' axis uses the direction of the hardest constituent in the exclusive \kt subjet instead of the subjet axis, such that the distance measure $\Delta R_{a_{1},i}$ changes in the calculation. 
The ratio of the N-subjettiness functions found with the standard subjet axes,
$\tau_{21}$, and with the `winner-takes-all' axes, \TauTwoOneWTA, can be used 
to generate the dimensionless variables that have been shown in 
particle-level MC to be particularly useful in identifying two-body 
structures within jets:

\begin{equation}
\TauTwoOne = \frac{\tau_{2}}{\tau_{1}}, \; \TauTwoOneWTA = \frac{\tau_{2}^{\rm{wta}}}{\tau_{1}^{\rm{wta}}}.\label{eq:tau21}
\end{equation}

\textbf{Energy correlation ratios}:\\
The 1-point, 2-point and 3-point energy correlation functions for a jet $J$ are given by:

\begin{subequations}
\begin{align}
E_{\mathrm CF0}  (\beta) = & 1, \label{eq:ecf0} \\
E_{\mathrm CF1}  (\beta) = & \sum_{i \in J} p_{\mathrm{T}_{i}}, \label{eq:ecf1} \\
E_{\mathrm CF2}  (\beta) = & \sum_{i < j \in J} p_{\mathrm{T}_{i}} p_{\mathrm{T}_{j}} \left(\Delta R_{ij} \right)^{\beta}, \label{eq:ecf2} \\  
E_{\mathrm CF3}  (\beta) = & \sum_{i < j < k \in J} p_{\mathrm{T}_{i}} p_{\mathrm{T}_{j}} p_{\mathrm{T}_{k}} \left(\Delta R_{ij} \Delta R_{ik} \Delta R_{jk}\right)^{\beta}, \label{eq:ecf3} 
\end{align}
\end{subequations}

where the parameter $\beta$ is used to give weight to the angular separation of the jet constituents.
In the above functions, the sum is over the constituents $i$ in the jet $J$, such that the 1-point correlation function \equref{ecf1} is approximately the jet \pT. Likewise, if one takes $\beta =2$, it is noted that the 2-point correlation functions are equivalent to the mass of a particle undergoing a two-body decay in collider coordinates. 

An abbreviated form of these definitions can be written as :

\begin{subequations}
\begin{align}
e_{2}^{(\beta)} = & \frac{ E_{\mathrm CF2} (\beta)}{E_{\mathrm CF1} (\beta) ^2}, \label{eq:ecf_rw2} \\[2ex] 
e_{3}^{(\beta)} = & \frac{ E_{\mathrm CF3} (\beta)}{E_{\mathrm CF1} (\beta) ^3}. \label{eq:ecf_rw3}
\end{align}
\end{subequations}

These ratios of the energy correlation functions can be used to generate the dimensionless variable \CTwoBeta~\cite{Larkoski:2013eya}, and its more recently modified version \DTwoBeta~\cite{Larkoski:2014pc,Larkoski:2015kga}, that have been shown in particle-level MC to be particularly useful in identifying two-body structures within jets:

\begin{subequations}
\begin{align}
C_{2}^{(\beta)} = & \frac{e_{3}^{(\beta)}}{(e_{2}^{(\beta)})^2}, \label{eq:c2} \\[2ex]
D_{2}^{(\beta)} = & \frac{e_{3}^{(\beta)}}{(e_{2}^{(\beta)})^3}. \label{eq:d2}   
\end{align}
\end{subequations}

Values of $\beta =$ 1 and 2 are studied here.

\section{The ATLAS detector}
\label{sec:detector}

The ATLAS detector~\cite{detPaper} at the LHC covers nearly the entire solid angle around the collision point.
It consists of an inner tracking detector surrounded by a thin superconducting solenoid, electromagnetic and hadronic calorimeters,
and a muon spectrometer incorporating three large superconducting toroid magnets.\par
The inner-detector system (ID) is immersed in a \SI{2}{\tesla} axial magnetic field 
and provides charged particle tracking in the range $|\eta| < 2.5$.
A high-granularity silicon pixel detector covers the vertex region and typically provides three measurements per track.
It is followed by a silicon microstrip tracker, which usually provides four two-dimensional measurement points per track.
These silicon detectors are complemented by a transition radiation tracker,
which enables radially extended track reconstruction up to $|\eta| = 2.0$. 
The transition radiation tracker also provides electron identification information 
based on the fraction of hits (typically 30 in total) above a higher energy-deposit threshold corresponding to transition radiation.

The calorimeter system covers the pseudorapidity range $|\eta| < 4.9$. Within the region $|\eta|< 3.2$, electromagnetic calorimetry is provided by barrel and 
endcap high-granularity lead/liquid-argon (LAr) electromagnetic calorimeters,
with an additional thin LAr presampler covering $|\eta| < 1.8$,
to correct for energy loss in material upstream of the calorimeters. For the jets measured here, the transverse granularity ranges from $0.003\times0.1$ to $0.1\times0.1$ in $\Delta \eta \times \Delta \phi$, depending on depth segment and pseudorapidity. Hadronic calorimetry is provided by a steel/scintillator-tile calorimeter,
segmented into three barrel structures within $|\eta| < 1.7$, and two copper/LAr hadronic endcap calorimeters. This system enables measurements of the shower energy deposition in three depth segments at a transverse granularity of typically $0.1\times0.1$. The solid angle coverage is extended with forward copper/LAr and tungsten/LAr calorimeter modules
optimised for electromagnetic and hadronic measurements respectively.

A muon spectrometer (MS) comprises separate trigger and
high-precision tracking chambers measuring the deflection of muons in a magnetic field generated by superconducting air-core toroids.
The precision chamber system covers the region $|\eta| < 2.7$ with three layers of monitored drift tubes, complemented by cathode strip chambers in the forward region, where the background is highest. The muon trigger system covers the range $|\eta| < 2.4$ with resistive-plate chambers in the barrel, and thin-gap chambers in the endcap regions.

A three-level trigger system is used to select interesting events~\cite{TriggerPerf2010}.
The Level-1 trigger is implemented in hardware and uses a subset of detector information
to reduce the event rate to a design value of at most \SI{75}{\kHz}.
This is followed by two software-based trigger levels which together reduce the event rate to about 400~Hz.

\section{Data and Monte Carlo simulations}
\label{sec:mc}

The data used for this analysis were collected during the $pp$ 
collision data-taking period in 2012, and correspond to an integrated luminosity of 20.3 fb$^{-1}$ with a mean number of $pp$ interactions per bunch crossing, $\avg{\mu}$, of about 20. The uncertainty on the integrated luminosity, 2.8\%, is derived following the same methodology as that detailed in Ref.~\cite{Aad:2013ucp} using beam-separation scans. Data quality and event selection requirements are given in \secref{objects}.

Events from Monte Carlo generator are passed through a \geant-based \cite{Agostinelli:2002hh} simulation of the ATLAS detector~\cite{Aad:2010ah}, and reconstructed using the same algorithm used as for data. All MC samples are produced with the addition of pileup, using hits from minimum-bias events that are produced with \Pythia (8.160)~\cite{pythia8} using the A2M set of tunable parameters (tune)~\cite{MC12AU2} and the MSTW2008LO~\cite{PDF-MSTW2008} PDF set. 
This simulated pileup does not exactly match the distribution of $\avg{\mu}$ measured in data. As such, event weights are derived as a function of $\avg{\mu}$ for the MC samples used in the data/MC comparisons, making the differences between the data and MC $\avg{\mu}$ distributions negligible.

\subsection{Monte Carlo samples for the $W$ signal}\label{sec:mc_signal}

Samples of the hypothetical process $W^\prime~\rightarrow~WZ~\rightarrow~qq\ell\ell$ are produced as a source of signal high-\pt\ $W$-jets, with the boost in \pT\ coming from the high mass of the parent $W^\prime$. These samples are produced using \Pythia (8.165) with the AU2~\cite{MC12AU2} tune and the MSTW20080LO~\cite{PDF-MSTW2008} PDF set. Nine separate signal samples are produced with $W^\prime$ masses ranging from 400 to 2000$\GeV$ in steps of 200$\GeV$. This ensures good coverage over a wide range of $W$-jet \pt. The nine samples are combined and the events are given weights such that when the event weights are applied, the \pT\ distribution of the combined signal $W$-jets sample matches that of the multijet background sample described in \secref{mc_background}. These are used as the signal samples in the preliminary optimisation studies presented in \secref{partOne}.

The $W$ boson tagging efficiency from top quark decays in data, detailed in \secref{datamc}, is measured using $t\bar{t}$ samples simulated with the \PowhegBox (version 1, r2330) NLO generator~\cite{Nason:2004rx} interfaced with \Pythia (6.427). A cross-check is performed with \Mcatnlo~\cite{mcatnlo} (4.03), with parton showers provided by \Herwig (6.520)~\cite{herwig}+\jimmy (4.31)~\cite{jimmy}. In both cases, the next-to-leading order CT10~\cite{PDF-CT10} PDF set is used, and the top quark mass is set to 172.5~\GeV. Single-top-quark events in the $s$-, $t$- and $Wt$-channels are simulated with \PowhegBox interfaced with \Pythia (6.426), with the Perugia 2011c~\cite{Perugia2010} tune. The $t$-channel is also generated with \PowhegBox in the four-flavour scheme. Background $W$+jet and $Z$+jet events are simulated using \Alpgen~\cite{alpgen} (2.14) in the four-flavour scheme ($b$-quarks are treated as massive) followed by \Pythia (6.426) for the parton shower. Up to five extra partons are considered in the matrix element. The CTEQ6L1~\cite{cteq6l1} PDF set and the Perugia 2011c tune are used. For diboson events, the \Sherpa~\cite{sherpa} (1.4.3) generator is used with up to three extra partons in the matrix element and the masses of the $b$- and $c$-quarks are taken into account.

The effects of differences between the $W^\prime~\rightarrow~WZ$ process used for $W$-jets in the preliminary optimisation studies and the \ttbar{} process used in the detailed comparisons with data are discussed in \secref{wprime_ttbar_comparison}.

\subsection{Monte Carlo samples for the multijet background}\label{sec:mc_background}

The background sample used in \secref{partOne} is made up of several high-\pt\ multijets event samples produced using \Pythia~\cite{pythia8} with the AU2~\cite{MC12AU2} tune and the CT10~\cite{PDF-CT10} PDF set. Eight samples in total are produced according to the leading jet's \pt, four of which are used in this analysis to cover the \pt range 200--2000$\GeV$. These samples are combined with event weights determined by their relative cross-sections to produce the smoothly falling \pt distribution predicted by \Pythia. The MC optimisation studies use the leading jets from these events. The jets in these background samples are initiated by light quarks and gluons, the interactions of which are described by Quantum Chromodynamics, QCD.

The $W$-tagging efficiency in multijet background events is studied on the 
same multijet samples as used for the optimisation studies, using \Pythia (8.165) with the AU2 tune and the CT10 PDF set, and also a \Herwigpp (2.6.3) sample with the EE3 tune~\cite{Gieseke:2012ft} and CTEQ6L1~\cite{cteq6l1} PDF set. It is these samples that are used for the comparisons with data in \secref{datamc}. 

The effects of differences between these samples due to using the leading jets (for the MC-based optimisation) or both leading and sub-leading jets (for the multijet background efficiency measurement in data) are discussed in \secref{wprime_ttbar_comparison}.

\section{Object reconstruction and event selection}\label{sec:objects}

In the studies presented here, calorimeter jets are reconstructed from three-dimensional topological clusters (topoclusters)~\cite{TopoClusters} which have been calibrated using the Local Cluster Weighting (LCW) scheme~\cite{EndcapTBelectronPion2002}. 
In MC simulated events, truth jets are built from generator-level particles that have a lifetime longer than 10~ps, excluding muons and neutrinos.
Jets are reconstructed using one of the iterative recombination jet reconstruction algorithms~\cite{Dokshitzer:1997in,Wobisch:1998wt} \camkt\ or \antikt. The \kt algorithm is also used by the jet trimming algorithm to reconstruct subjets.

In all following discussions, the term constituents means particles in the case of truth jets and LCW topoclusters in the case of calorimeter jets.

For the MC-based optimisation studies discussed in \secref{partOne}, events are characterised using the leading jet, reconstructed from generator-level particles with the \camkttwelve algorithm. 

Objects used to select \ttbar{} events in data and MC for the studies in \secref{datamc} include reconstructed leptons (electrons and muons), missing transverse momentum (\met{}), small-$R$ jets (reconstructed with the \antikt{} algorithm with radius parameter $R=0.4$), trimmed \antiktten jets and $b$-tagged jets, defined below.

\begin{itemize}

\item \textbf{Electrons}: Electron candidates are reconstructed from
  energy deposits in the EM calorimeter matched to reconstructed
  tracks in the ID. Candidates are required to be within $|\eta| <
  2.47$, excluding the barrel/endcap transition region, $1.37 < |\eta| < 1.52$, of
  the EM calorimeter, and must have a transverse energy $\et > 25\GeV$. They
  are required to satisfy tight identification
  criteria~\cite{Aad:2014fxa} and to fulfil isolation~\cite{Mann:2010ei}
  requirements; excluding its own track, the scalar sum of the $\pt{}$
  of charged tracks within a cone of size $\Delta R =
  \min(10\GeV/E_{\rm T},0.4)$ around the electron candidate must be
  less than 5\% of the \pt{} of the electron.
  
\item \textbf{Muons}: Muons are reconstructed by matching MS to ID
  tracks. Muons are required to be within $|\eta| < 2.5$ and have
  \pt{} $ > 25\GeV$. In order to reject non-prompt muons from hadron decays, the
  significance of their transverse impact parameter must be
  $|d_{0}|/\sigma_{d_0} < $ 3, the longitudinal impact parameter
  must be $|z_0| < $~2~mm, and the scalar sum of $\pt{}$ of the
  charged tracks within a cone of size $\Delta R =
  \min(10\GeV/\pt{},0.4)$ around the muon candidate, excluding its own
  track, must be less than 5\% of the \pt{} of muon.

\item \textbf{Trigger leptons}: Events are selected by requiring an 
un-prescaled single-lepton trigger for the electron and muon channels. Two single-electron triggers, with transverse energy thresholds of $E_{\rm T} > 24\GeV$ for isolated electrons and $E_{\rm T} > 60\GeV$ without isolation criteria, are used in combination with two single-muon triggers, with transverse momentum of $\pt > 24\GeV$ for isolated muons and $\pt > 36\GeV$ without isolation criteria. The selected muon (electron) must be matched to a trigger and is required to fulfil $ \pt{} > 25 (20)\GeV$ and $|\eta| < 2.5$. Events are rejected if any other electron or muon satisfying the identification criteria is found in the event.   

\item \textbf{Missing transverse momentum, \met{} and transverse mass, $m_\text{T}^{W}$}: 
The missing transverse momentum is calculated from the vector sum of the transverse energy of topological clusters in the calorimeter~\cite{MET7}. The clusters associated with the reconstructed electrons and small-$R$ jets are replaced by the calibrated energies of these objects. Muon \pt\ determined from the ID and the muon spectrometer are also included in the calculation. The 
\met is required to exceed 20$\GeV$. The sum of the \met{} and the transverse mass, $m_\text{T}^{W}=\sqrt{2\pt\met(1-\cos\Delta\phi)}$, reconstructed from the \met{} and the transverse momentum of the lepton, must be $\met{} + m_\text{T}^{W} > 60\GeV$.

\item \textbf{Small-$R$ Jets (\antiktfour)}: Using locally calibrated
  topological clusters as input, small-$R$ jets are formed using the
  \antikt\ algorithm with a radius parameter $R = 0.4$. Small-$R$
  jets are required to be within $|\eta| < 2.5$ and to have \pt{} $>
  25\GeV$.  To reject jets with significant pileup contributions, the
  jet vertex fraction~\cite{jetAreas2012}, defined as the scalar sum
  of the \pt{} of tracks associated with the jet that are assigned to
  the primary vertex divided by the scalar sum of the \pt{} of all
  tracks associated to the jet, is required to be greater than 0.5 for
  jets with \pt{} $< 50\GeV$. At least one small-$R$ jet must be found. In addition, at least one small-$R$ jet must lie within $\DeltaR = 1.5$  of the lepton.  The
leading small-$R$ jet within $\DeltaR = 1.5$ of the lepton is defined as the ``leptonic-top jet'' and denoted $j_{\ell t}$. Jets have to satisfy specific cleaning requirements~\cite{JetCleaning2011} to remove calorimeter signals coming from non-collision sources or calorimeter noise. Events containing any jets that fail these requirements are rejected. 

\item \textbf{$b$-jets (\antiktfour)}: The output of the MV1~\cite{AdvBtagMV1} algorithm is used to 
identify small-$R$ jets containing $b$-hadrons. Small-$R$ jets are tagged as $b$-jets if the MV1 weight is larger than the value corresponding to the 70\% $b$-tagging efficiency working point of the algorithm. At least one small-$R$ jet must be tagged as a $b$-jet. Loose $b$-jets are defined as having an MV1 weight larger than the value corresponding to the 80\% working point. All loose $b$-jets must be separated by $\DeltaR > 1.0$ from the $W$-jet candidate.

\item \textbf{Trimmed $R=1.0$ Jets}: 
Using locally calibrated topological clusters as inputs, \antiktten
jets are groomed using the trimming algorithm
with parameters \fcut = 5\% and \subjetr = 0.2. The pseudorapidity, energy and mass of these jets are calibrated using a simulation-based calibration scheme as mentioned in
\secref{findings_mc}. At least one trimmed \antiktten jet with $\pt > 200\GeV$ and $|\eta|<1.2$ is required. If more than one jet satisfies these criteria, the leading jet is used to reconstruct the \Wboson{} boson candidate, $J_{W}$.  This candidate, $J_{W}$, has to be well separated from the leptonic-top jet,
$\DeltaR(J_{W},j_{\ell t})>1.2$.  

\item \textbf{Overlapping jets and leptons}: An overlap removal procedure is applied to avoid double-counting of leptons and \antiktfour jets, along with an electron-in-jet subtraction procedure to recover prompt electrons that are used as constituents of a jet. If an electron lies $\Delta R < 0.4$ from the nearest jet, the electron four-momentum is subtracted from that of the jet. If the subtracted jet fails to meet the small-$R$ jet selection criteria outlined above, the jet is marked for removal. If the subtracted jet satisfies the jet selection criteria, the electron is removed and its four-momentum is added back into the jet. Next, muons are removed if $\Delta R({\rm{muon},jet}) < 0.04 + 10\GeV/p_{\rm{T},\rm{muon}}$ using jets that are not marked for removal after the electron subtraction process.

\end{itemize}

For the measurement of the multijet background efficiency, a different selection is used to ensure a multijet-enriched sample. The multijet sample is selected using a single, un-prescaled, $R = 1.0$ jet trigger that is 80\% efficient for jets with $\pt > 450\GeV$. No grooming is applied to jets at the trigger level. For events with a leading jet above the trigger threshold, both the leading and the sub-leading jets are used for this performance study, making it applicable for jets with \pt\ down to $200\GeV$. At least one \antiktten jet, trimmed with \fcut = 5\% and \subjetr = 0.2, is required to have $\pt > 200\GeV$ and $|\eta|<1.2$. Events containing fake jets from noise in the calorimeter or non-collision backgrounds, according to Refs.~\cite{Aad:2013zwa,Aad:2014bia}, are rejected.

For the \ttbar{} and multijet background selection, good data quality is required for events in data, meaning that all the detectors of ATLAS as well as the trigger and data acquisition system are required to be fully operational. Events are required to have at least one reconstructed primary vertex with at least five associated tracks, and this vertex must be consistent with the LHC beam spot.

\section{A comprehensive comparison of techniques in Monte Carlo simulations}
\label{sec:partOne}

The initial phase of this study evaluates the performance of a large number of grooming
and tagging algorithms in MC simulated events. 

To account for correlations between the $W$ boson \pt and the resulting jet substructure features, events are categorised by the \pt{} of the leading (highest \pt) jet reconstructed with the \camkt~\cite{Cacciari:2008gn} algorithm with radius parameter $R = 1.2$, using stable particles as inputs. These ranges in the ungroomed truth jet \pt, $\pt^\text{Truth}$, are: $[200, 350]\GeV, [350, 500]\GeV, [500, 1000]\GeV$. This large, ungroomed jet is considered a rough proxy for the $W$ boson, and this choice does not introduce a bias towards any particular grooming configuration for the $\pt^\text{Truth}$ ranges in question. Only events with a \camkttwelve truth jet within $|\eta|<1.2$ are considered, ensuring that jets are within the acceptance of the tracking detector, which is necessary for the derivation of the systematic uncertainties.

First, in \secref{groomers_perf}, more than $500$ jet reconstruction and grooming algorithm configurations are selected based on prior studies~\cite{CDF:2011prd,Aad:2012meb,JetMassAndSubstructure,Aad:2013gja,Cui2010,Khachatryan:2014vla}. The leading-groomed-jet mass distributions for $W$-jet signal and multijet background in MC are examined. An ordered list is built rating each configuration based on the background efficiency. The notation for the background efficiency at this grooming stage is $\epsilon_{\mathrm{QCD}}^{\mathrm{G}}$, and this is measured within a mass window that provides a signal efficiency of $68\%$, denoted $\epsilon_{W}^{\mathrm{G}} = 68\%$. The best performers for each category described in \secref{primer_grooming} (trimming, pruning, split-filtering) are retained for the next stage: a total of 27 jet collections.

Observations about pileup-dependence are summarised in \secref{pileup}. Jet grooming reduces the pileup-dependence of the jet mass and helps distinguish $W$-jets from those 
initiated by light quarks and gluons by improving the mass resolution, but does not provide strong background rejection. Further information coming from the distribution of energy deposits within a jet can be used to improve the ratio of signal to background. 

In the second stage, 26 substructure variables are studied for all 27 selected jet collections. These studies are detailed in \secref{substructure_vars}. Substructure variables can be calculated using jet constituents before or after grooming; in these studies all variables are calculated from the groomed jet's constituents, such that the potential sensitivity to pileup conditions is reduced.

The aim of these studies is to find an effective combination of groomed jet mass and one substructure variable. The background efficiency $\epsilon_{\mathrm{QCD}}^{\mathrm{G}\&\mathrm{T}}$ (where G$\&$T indicates grooming plus tagging) versus the signal efficiency $\epsilon_{W}^{\mathrm{G}\&\mathrm{T}}$ is calculated for all variables in each configuration, and background efficiencies for `medium' ($50\%$) and `tight' ($25\%$) signal efficiency working points are determined. Four grooming algorithms and three tagging variables are identified as having a particularly low background efficiency at the medium signal efficiency working point, $\epsilon_{W}^{\mathrm{G}\&\mathrm{T}} = 50\%$. 

In \secref{findings_mc} the conclusions of these preliminary studies of combined groomed mass and substructure taggers are presented.

\subsection{Performance of grooming algorithms}\label{sec:groomers_perf}

A set of more than 500 jet reconstruction and grooming algorithm configurations (introduced in \secref{primer_grooming}) are explored within the parameter space summarised in \Tabref{config}. 

\begin{figure}[htb]
\begin{center}
\begin{subfigure}[]{\label{fig:jetmass_r2}
\includegraphics[width=.4\textwidth]{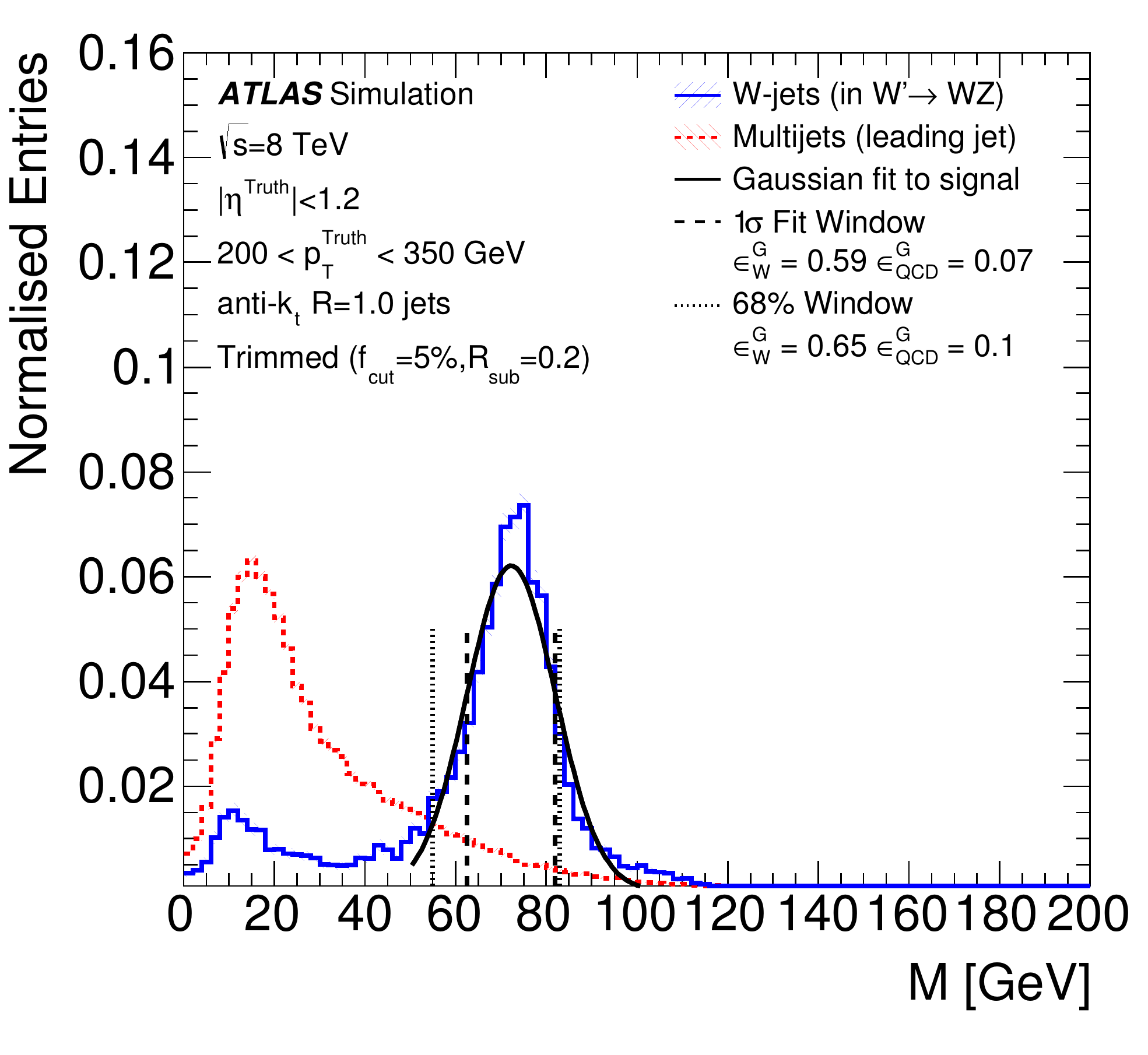}
}
\end{subfigure}
\begin{subfigure}[]{\label{fig:jetmass_r3}
\includegraphics[width=.4\textwidth]{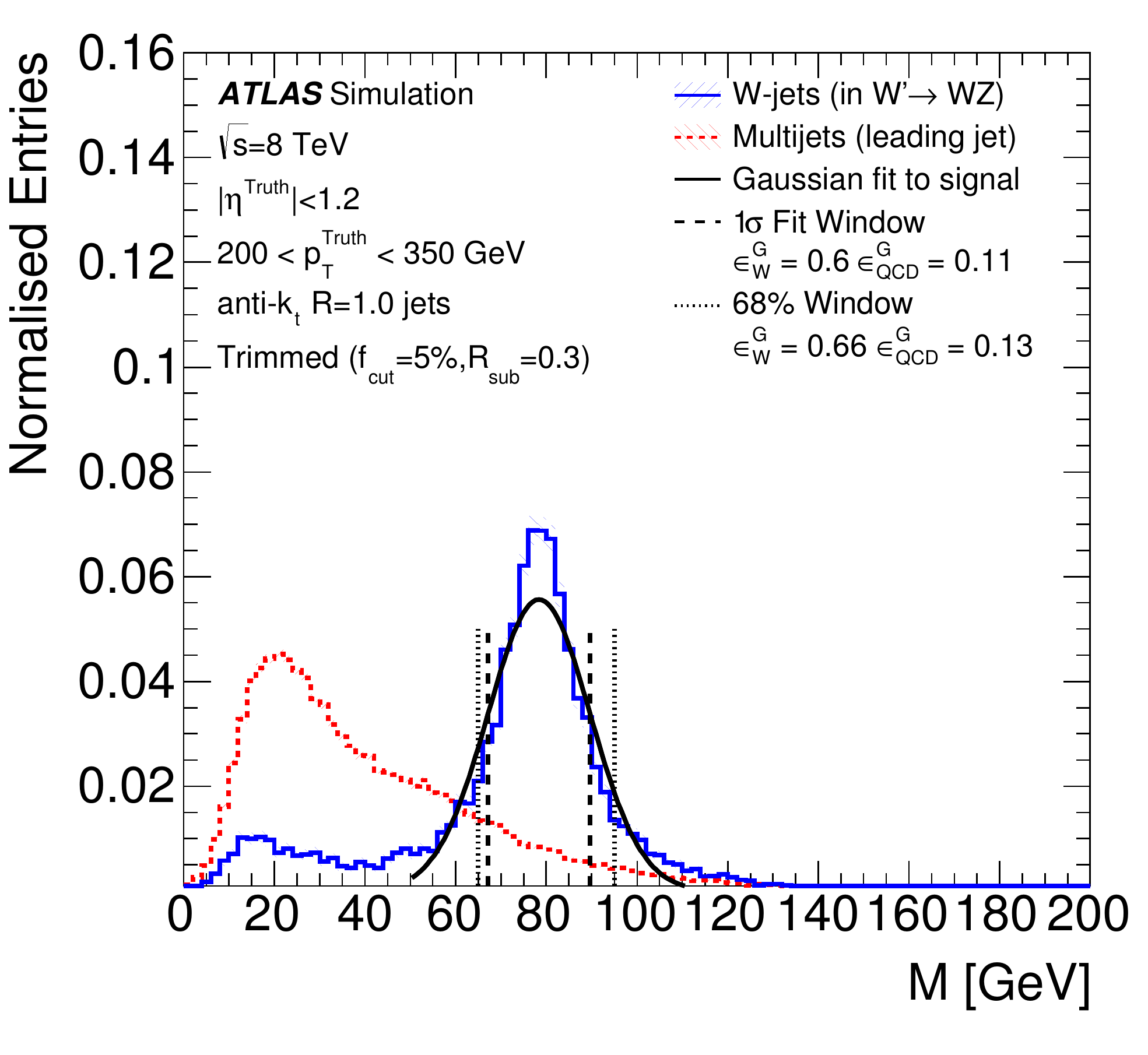}
}
\end{subfigure}
\begin{subfigure}[]{\label{fig:jetmass_prune}
\includegraphics[width=.4\textwidth]{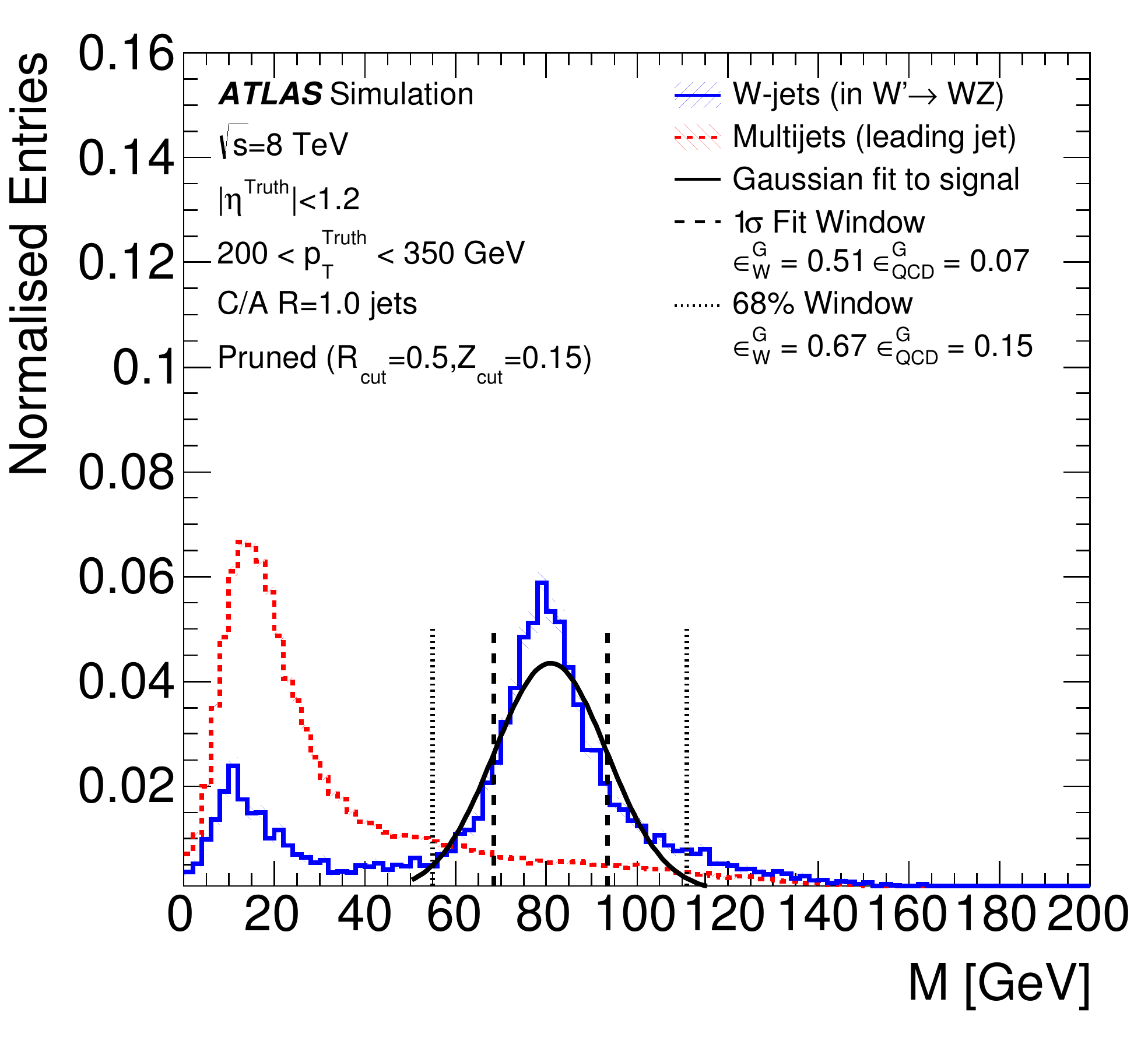}
}
\end{subfigure}
\begin{subfigure}[]{\label{fig:jetmass_bdrs}
\includegraphics[width=.4\textwidth]{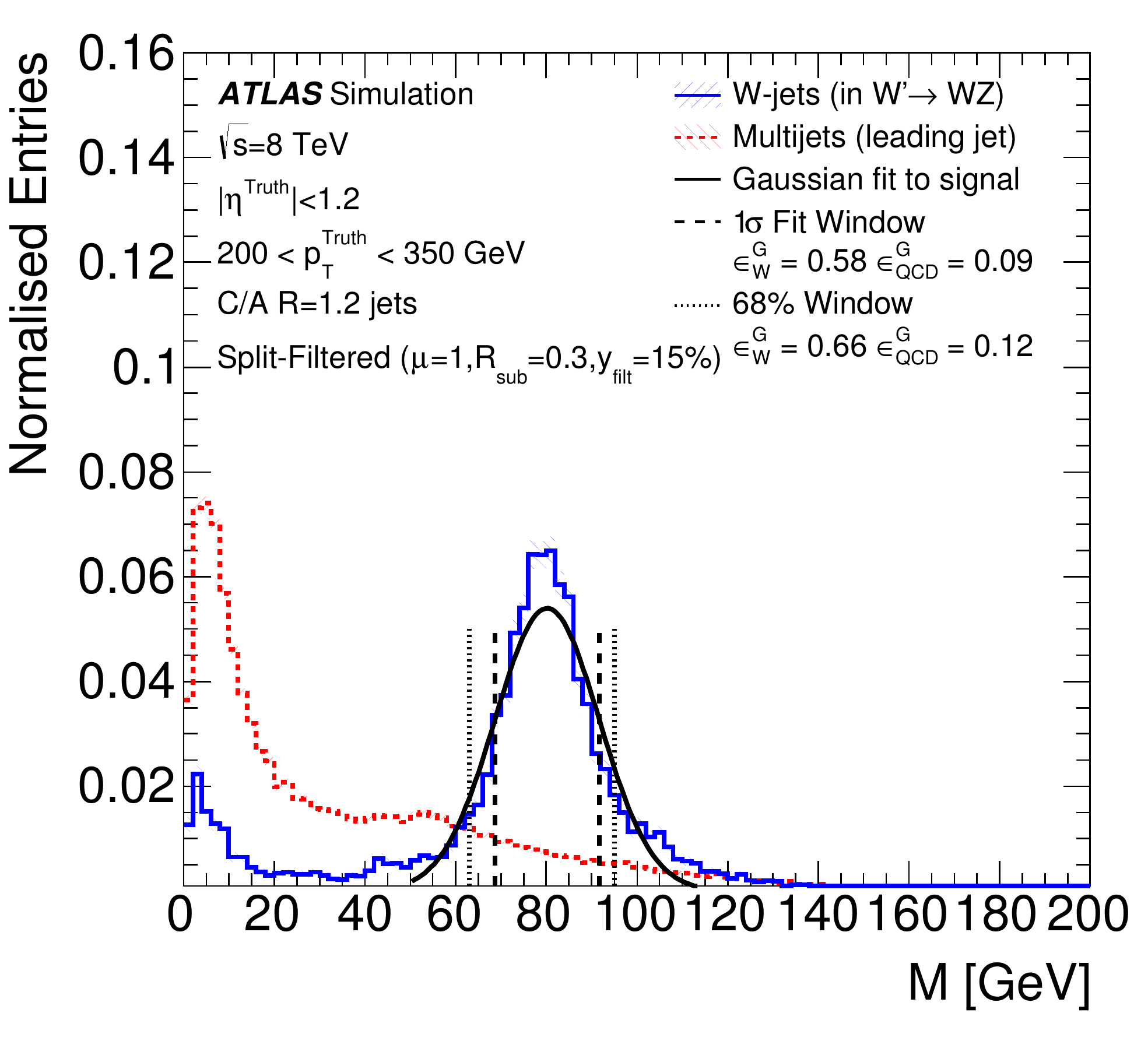}
}
\end{subfigure}
\end{center}
\caption[Uncalibrated mass distributions for 4 grooming algorithm.]{Uncalibrated mass distributions for various selected grooming configurations: (a) trimmed with \subjetr = 0.2, (b) trimmed with \subjetr = 0.3, (c) pruned, and (d) split-filtered. The transverse momentum range $\pt^\text{Truth} = [200, 350]\GeV$ is shown for $W$ signal (solid blue line) and multijet background (dashed red line). The (black) Gaussian fit uses an initial-condition mass set to 80.4$\GeV$. The dotted vertical lines indicate the 1$\sigma$ fit interval. The dashed lines contain 68\% of the signal and define the mass window. These are examples of grooming algorithms leading to satisfactory mass distributions. Uncertainty bands are statistical only.}
\label{fig:part1_summary1}
\end{figure}

The signal and background mass distributions for a selection of grooming configurations in the range $200 < \pt^\text{Truth} < 350\GeV$ are shown in \figref{part1_summary1}. 
A Gaussian fit to the $W$ boson mass peak (with the $W$ mass set as the initial condition)
is shown. Two alternative signal mass window definitions are considered: 
\begin{enumerate}
\item The 1$\sigma$ boundaries of the Gaussian fit.
\item The smallest interval that contains 68\% of the integral.
\end{enumerate}

Comparing the extent of these two mass windows allows an estimation of how closely the signal mass peak resembles a Gaussian distribution. The $W$-jet mass is required to be within the boundaries defined by this latter definition 
of the signal window; this leads, by definition, to a baseline signal efficiency of $\epsilon_{W}^{\mathrm{G}} = 68\%$ for all algorithms.

\begin{table}
\begin{center}
\resizebox{0.92\textwidth}{!}{\begin{minipage}{\textwidth}
\begin{tabular}{ c | c | c | c | c  }
\multicolumn{4}{l}{\textbf{Trimming configurations}} \\
\hline
\hline
 Input jet algorithms & $R$  &  $R_{\rm sub}$  & \multicolumn{2}{c}{\Fcut ($\%$)} \\
 \hline
  \multirow{ 2}{*}{C/A, \antikt}       & \multirow{ 2}{*}{0.6, 0.8, 1.0, 1.2}  & \multirow{ 2}{*}{0.1, 0.2, 0.3}   & \multicolumn{2}{c}{ \multirow{ 2}{*}{1, 2, 3, 4, 5, 7, 9, 11, 13, 15 } } \\  
  &&& \multicolumn{2}{c}{} \\    
\hline
\hline
\multicolumn{5}{l}{} \\
\multicolumn{4}{l}{\textbf{Pruning configurations}} \\
\hline
\hline
Input jet algorithm & $R$ & Reclust. alg.   & \Zcut ($\%$)          & \Rcut \\
 \hline
\multirow{ 2}{*}{C/A, \antikt }     & \multirow{ 2}{*}{0.8, 1.0, 1.2}      & \multirow{ 2}{*}{C/A }           & \multirow{ 2}{*}{10, 15, 20, 25, 30}    & \multirow{ 2}{*}{$\frac{1}{100}$, $\frac{1}{10}$, $\frac{1}{8}$, $\frac{1}{4}$, $\frac{1}{2}$, 1.0} \\
 &&&&\\
\hline
 \hline
\multicolumn{5}{l}{} \\
\multicolumn{4}{l}{\textbf{Split-filtering configurations}} \\
\hline
\hline
Input jet algorithm & $R$           & $R_{\rm sub}$                & \MaxMassDrop    & \ycut \\
 \hline
 \multirow{ 2}{*}{C/A}   &  \multirow{ 2}{*}{0.8, 1.0, 1.2}     &  \multirow{ 2}{*}{ 0.3, ${\rm min}(0.3,\Delta R/2)$} &  \multirow{ 2}{*}{67, 78, 89, 100}      &  \multirow{ 2}{*}{0.06, 0.07, ..., 0.20} \\
 &&&&\\
\hline
 \hline
\end{tabular}
\caption[Grooming types and configurations.]{Details of the different trimming, pruning and split-filtering configurations that were tried in order to define the best grooming algorithms. All combinations of the grooming parameters are explored in these studies.}
\label{tab:config}
\end{minipage}}
\end{center}
\end{table}

\begin{figure}[htb]
\begin{center}
\begin{subfigure}[]{ \label{fig:jetmass_bdrs_bad}
\includegraphics[width=.4\textwidth]{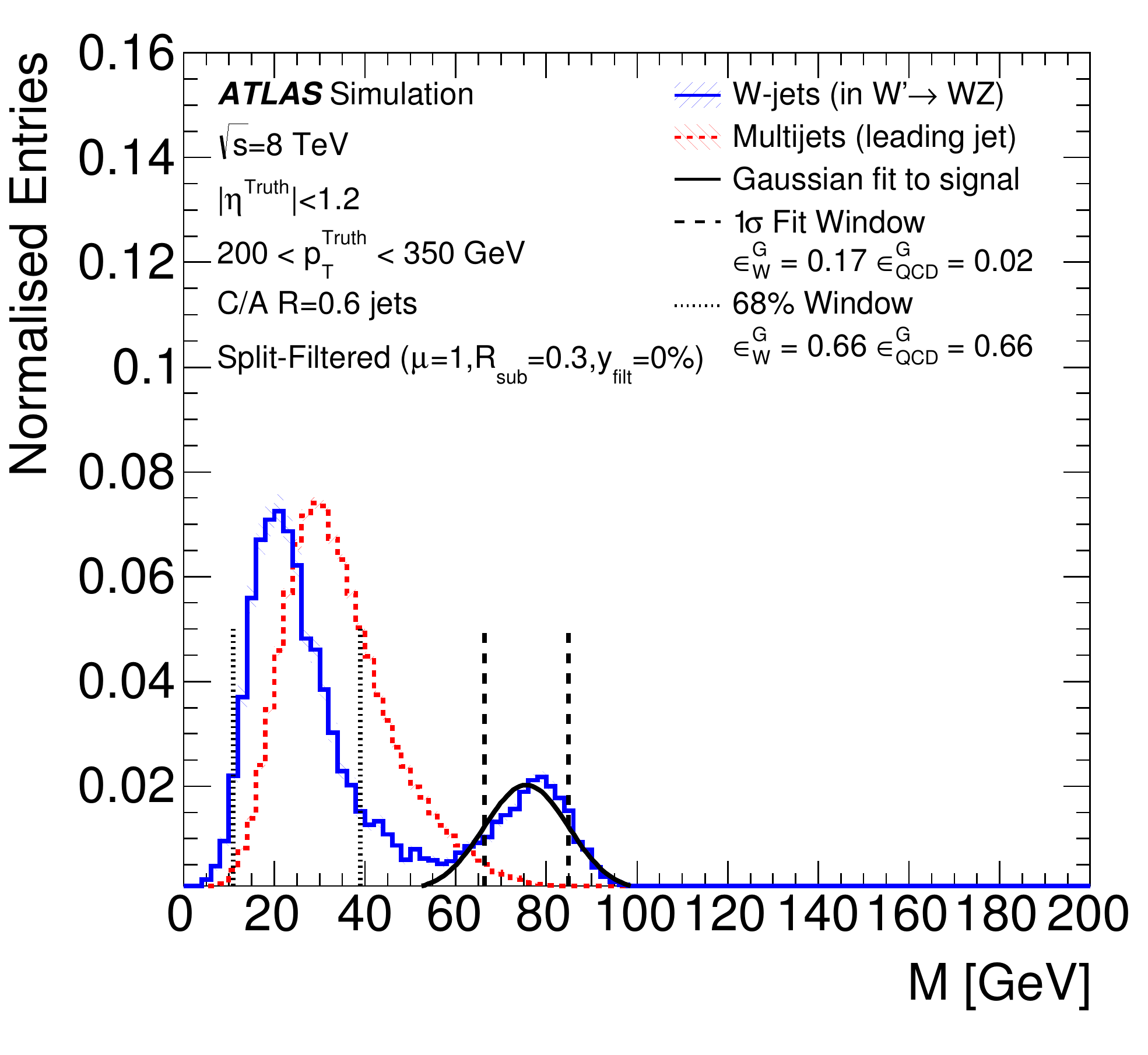}
}
\end{subfigure}
\begin{subfigure}[]{\label{fig:jetmass_prune_bad}
\includegraphics[width=.4\textwidth]{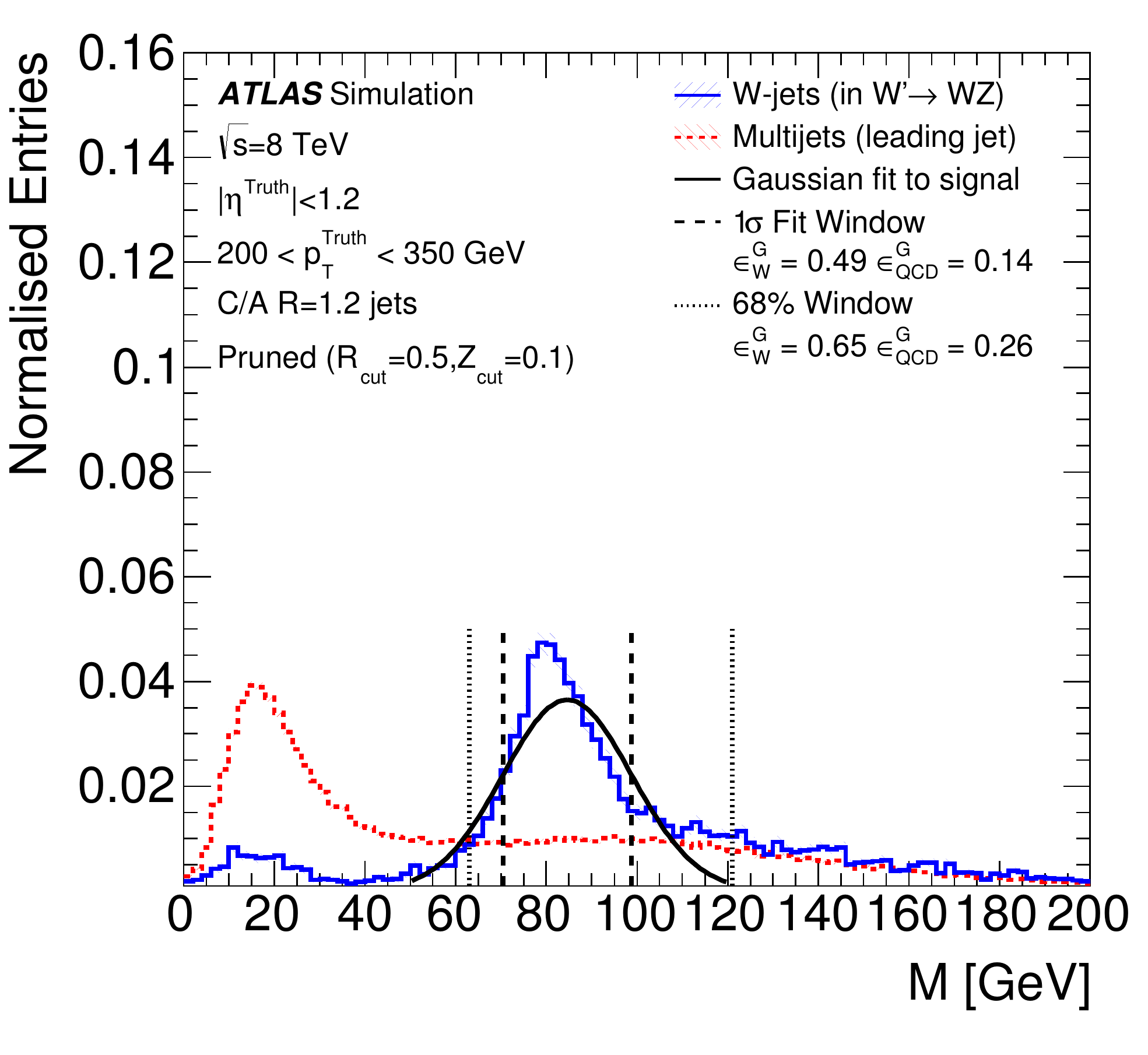}
}
\end{subfigure}
\end{center}
\caption[Uncalibrated mass distributions for 2 grooming algorithm.]{Uncalibrated mass distributions for two problematic grooming configurations in the transverse momentum range $\pt^\text{Truth} = [200, 350]\GeV$ for $W$ signal and multijet background. The Gaussian fit uses an initial-condition mass set to 80.4$\GeV$. The dotted vertical lines indicate the 1$\sigma$ fit interval. The dashed lines contain 68\% of the signal and define the mass window. These plots show examples of unwanted behaviours: in (a) most signal events are reconstructed with a small mass, indicating that the $W$ boson decay products are not fully contained in the jet; and in (b) the signal mass distribution is strongly asymmetric.}
\label{fig:part1_summary1_pathologies}
\end{figure}

The groomed jet mass distributions for leading jets are examined for all combinations of grooming configurations for $W$-jet signal and multijet background. The background efficiency, $\epsilon_{\mathrm{QCD}}^{\mathrm{G}}$ is defined as follows:
\begin{itemize}
\item The denominator is the total number of pre-selected events from the multijet background sample, where the pre-selection requires an ungroomed \camkttwelve truth jet with $\pt^\text{Truth} > 200\GeV$ and $|\eta^\text{Truth}| < 1.2$. 
\item The numerator is the number of pre-selected events where the groomed jet mass falls in the window that contains 68\% of the $W$-jet signal, $\epsilon_{W}^{\mathrm{G}} = 68\%$. 
\end{itemize}
The minimisation of  $\epsilon_{\mathrm{QCD}}^{\mathrm{G}}$ is the primary criterion for ordering the algorithms according to their performance. In addition, there are a number of possible pathologies revealed in the mass distributions: features that show obviously unsuitable configurations, or make it impossible to derive a jet mass calibration, or indicate the need for additional pileup removal techniques. These are:
\begin{itemize}
\item[($i$)] The $\epsilon_{W}^{\mathrm{G}} = 68\%$ window does not contain the $W$ boson mass~\cite{PDG2012}. An example of this is shown in \figref{jetmass_bdrs_bad}.
\item[($ii$)] The signal mass distribution is strongly non-Gaussian. An example of this is shown in \figref{jetmass_prune_bad}.
\item[($iii$)] The background mass distribution has an irregular shape (e.g. it has local maxima) in the region of the signal peak. An example of this is also shown in \figref{jetmass_prune_bad}. 
\item[($iv$)] The jet mass after grooming is strongly affected by pileup. Configurations where the average jet mass increases by $> 1\GeV $ times the number of primary vertices, NPV, are rejected. This issue is discussed in \secref{pileup}.
\end{itemize}
Algorithms that are susceptible to any of these pathologies are removed from the list of well-behaved algorithm configurations.

The $W$ boson tagging efficiency performance is studied independently for three different ranges in the \pt{} of the ungroomed truth jet reconstructed with the \camkttwelve algorithm: $[200, 350]$, $[350, 500]$, $[500, 1000]\GeV$. The results for the three grooming categories share some common features:
\begin{itemize}
\item The jets reconstructed with $R = 0.6$ and $R = 0.8$ are too small to contain all the decay products of a $W$-jet for \pt\ $<500\GeV$ and  \pt\ $<350\GeV$, respectively. The reconstructed jet mass is often much smaller than 80$\GeV$, indicating that some of the $W$ boson decay products are not clustered, and the 68\% signal mass window is wider, resulting in a higher background efficiency. Small radii jets can, however, have good performance at high \pt.
\item In the highest \pt\ bin, 500--1000$\GeV$, the various configurations result in a similar performance.
\end{itemize}

The unique features of each grooming category are presented below.

\begin{figure}
\begin{center}
\includegraphics[angle=0,width=0.95\textwidth]{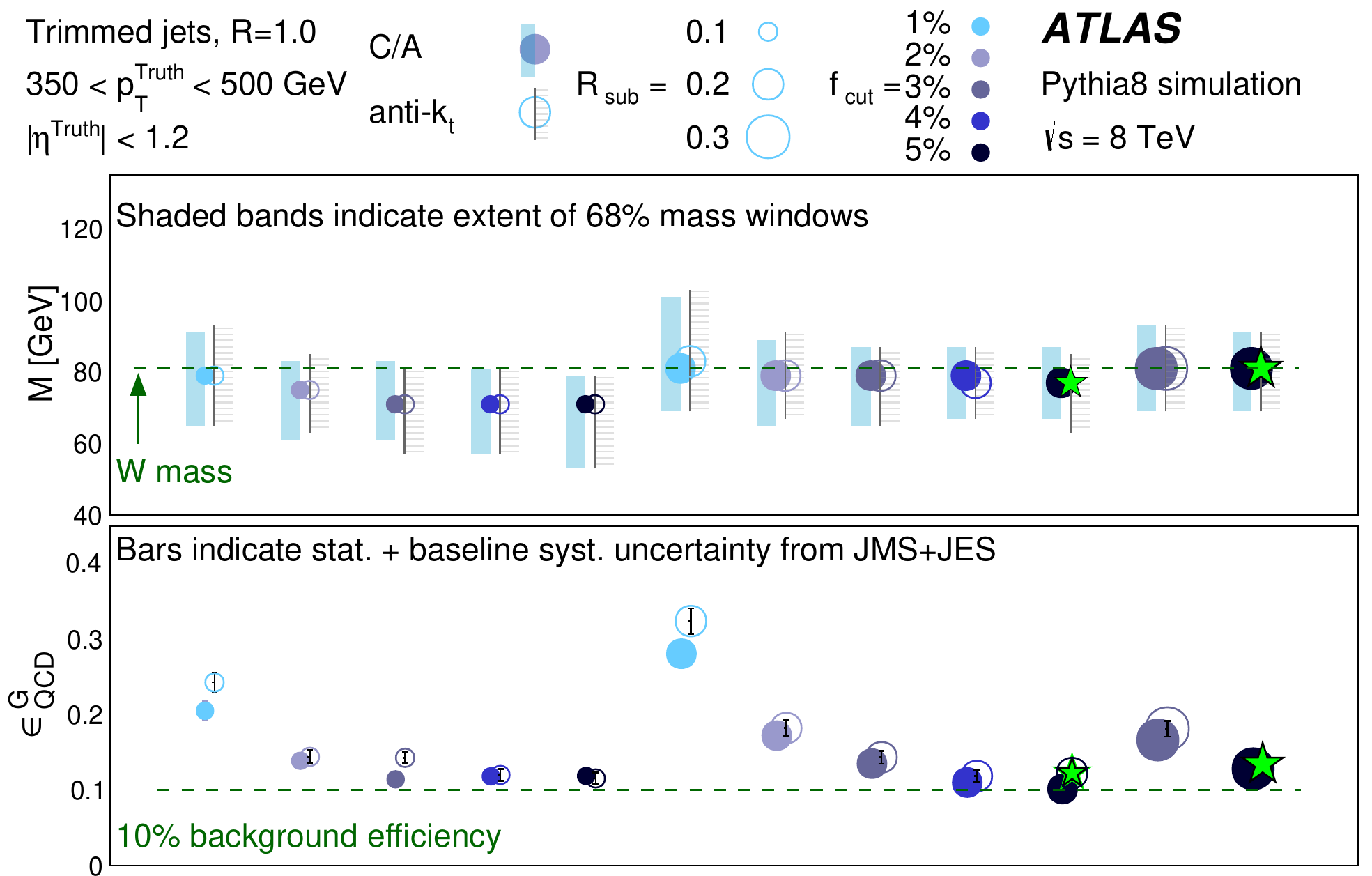}
\end{center}
\caption[Mass windows, MPVs and $\epsilon^{\mathrm{G}}_{\mathrm{QCD}}$ for trimming.]{ Mass windows and background efficiencies for various configurations of trimming ($R$=1.0 shown). The baseline systematic uncertainty on the background efficiency for the \pt bin in question (the range $350 < \pt\ < 500\GeV$ is shown here) is calculated by varying the jet mass scale (JMS) and jet energy scale (JES) by $\pm 1 \sigma$ for a representative jet collection. For trimming, this representative configuration is \subjetr$=0.2$ and \fcut$=5\%$. The stars indicate the favoured trimming configurations for $W$-tagging, as detailed in \secref{findings_mc}.} 
\label{fig:unreadable_trimmed}
\end{figure}

\textbf{Trimming}:\\
Various trimming configurations are studied, varying the algorithm and size of
the initial jet (\camkt with $R$~=~0.6--1.2, \antikt\ with $R$~=~0.8--1.2), and
the \subjetr and \fcut parameters summarised in \Tabref{config}. The background rejection and the boundaries of the 68\% signal mass windows obtained with a subset of trimming configurations for the range $350 < \pt < 500\GeV$ are shown in \figref{unreadable_trimmed} for \antiktten and \camktten jets. The systematic uncertainties resulting from the uncertainty on the jet mass and energy scale (described in detail in \secref{sysEff}) are provided to give the reader an idea of the relevance of the differences in performance between the grooming configurations.

The following characteristics are noted: 
\begin{itemize}
\item \camkt and \antikt\ jets have a similar performance under the same configurations.
\item The larger values of \fcut can lead to significantly lower background efficiency.
\item The dependence of the performance on \subjetr is less significant, but the background efficiency does decrease somewhat for smaller \subjetr values.
\end{itemize}

Based on the performance of these algorithms, the trimming implementations considered for further investigation are given in \Tabref{trimbest}.
Although promising, configurations with \subjetr = 0.1 are not pursued further in these studies, as this size is approaching the limiting granularity of the hadronic tile calorimeter, requiring further studies for a proper control of the systematic uncertainties.

\begin{table}
\begin{center}
\begin{tabular}{  c | c | c | c }
\hline
\hline
Initial algorithm & $R$ & \fcut & \subjetr \\
\hline
\antikt & 1.2 & 5\% & 0.2 \\
C/A & 1.2 & 5\% & 0.2 \\[1ex]
\antikt & 1.0 & 5\% & 0.2 \\
C/A & 1.0 & 5\% & 0.2 \\
\antikt & 1.0 & 5\% & 0.3 \\[1ex]
\antikt & 0.8 & 5\% & 0.2 \\
C/A & 0.8 & 5\% & 0.2 \\[1ex]
C/A & 0.6 & 5\% & 0.2 \\
\hline
\hline
\end{tabular}
\caption[Best trimmers.]{The best trimming configurations for $W$-tagging with each $R$ based on the first stage of the MC-based optimisation studies.}
\label{tab:trimbest}
\end{center}
\end{table}

\textbf{Pruning}:\\

\begin{figure}
\begin{center}
\includegraphics[angle=0,width=0.95\textwidth]{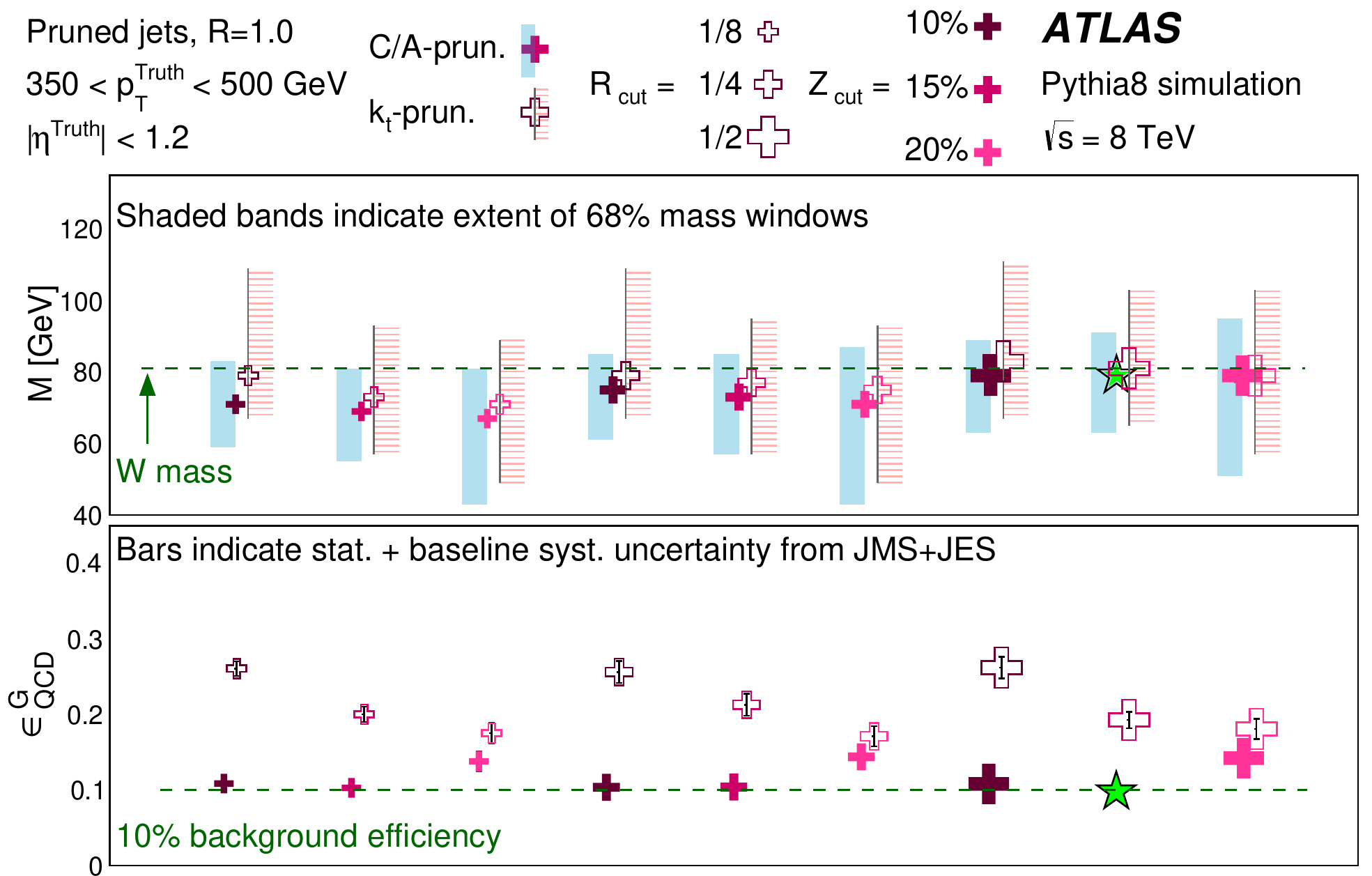}
\end{center}
\caption[Mass windows, MPVs and $\epsilon^{\mathrm{G}}_{\mathrm{QCD}}$ for lots of pruning configs.]{ Mass windows and background efficiencies for various configurations of pruning ($R$=1.0 shown). The baseline systematic uncertainty on the background efficiency for the \pt bin in question (the range $350 < \pt\ < 500\GeV$ is shown here) is calculated by varying the jet mass scale (JMS) and jet energy scale (JES) by $\pm 1 \sigma$ for a representative jet collection. For pruning, this representative configuration is \Rcut$=\frac{1}{2}$ and \Zcut$=15\%$. The star indicates the favoured pruning configuration for $W$-tagging, as detailed in \secref{findings_mc}.} 
\label{fig:unreadable_pruned}
\end{figure}

The performance of pruning is studied using both C/A and \antikt\ algorithms for the initial large-$R$ ($R$ = 0.6--1.2) jet finding, and C/A for the reclustering procedure. The background efficiencies and 68\% signal mass windows obtained with a subset of pruning configurations for the range $350 < \pt < 500\GeV$ are shown in \figref{unreadable_pruned}.

Several observations can be made:
\begin{itemize}
\item Using the \CamKt algorithm as the re-clustering algorithm for pruning is consistently better than using the \kt\ algorithm, for the same values of the \rcut and \Zcut parameters.
\item Pruning with smaller \rcut and/or higher \Zcut can be overly harsh, resulting in $W$-jet mass peaks at
 values lower than 80$\GeV$.
\item The background efficiency does not have strong dependence on \rcut or on \Zcut, but there is evidence for a \pt\-dependence of the optimal \Zcut, with \Zcut = 0.15 being preferable for the ranges $200 < \pt < 350\GeV$ and $350 < \pt < 500\GeV$, and \Zcut = 0.10 being preferred for \pt $> 500\GeV$.
\item For all pruning configurations, the performance is significantly worse in the lowest \pt\ bin.
\end{itemize}

Based on the performance of all the algorithms, the eight combinations retained for further
studies are given in \Tabref{prunebest}

\begin{table}
\begin{center}
\begin{tabular}{ c | c | c | c }
\hline
\hline
Initial algorithm & $R$ & \Zcut & \rcut \\
\hline
C/A & 1.2 & 10\% & 0.5 \\
C/A & 1.2 & 15\% & 0.5 \\[1ex]
C/A & 1.0 & 10\% & 0.5 \\
C/A & 1.0 & 15\% & 0.5 \\[1ex]
C/A & 0.8 & 10\% & 0.5 \\
C/A & 0.8 & 15\% & 0.5 \\[1ex]
C/A & 0.6 & 10\% & 0.5 \\
C/A & 0.6 & 15\% & 0.5 \\
\hline
 \hline
\end{tabular}
\caption[Best pruners.]{The best pruning configurations for $W$-tagging with each $R$ based on the first stage of the MC-based optimisation studies.}
\label{tab:prunebest}
\end{center}
\end{table}

\begin{figure}
\begin{center}
\includegraphics[angle=0,width=0.95\textwidth]{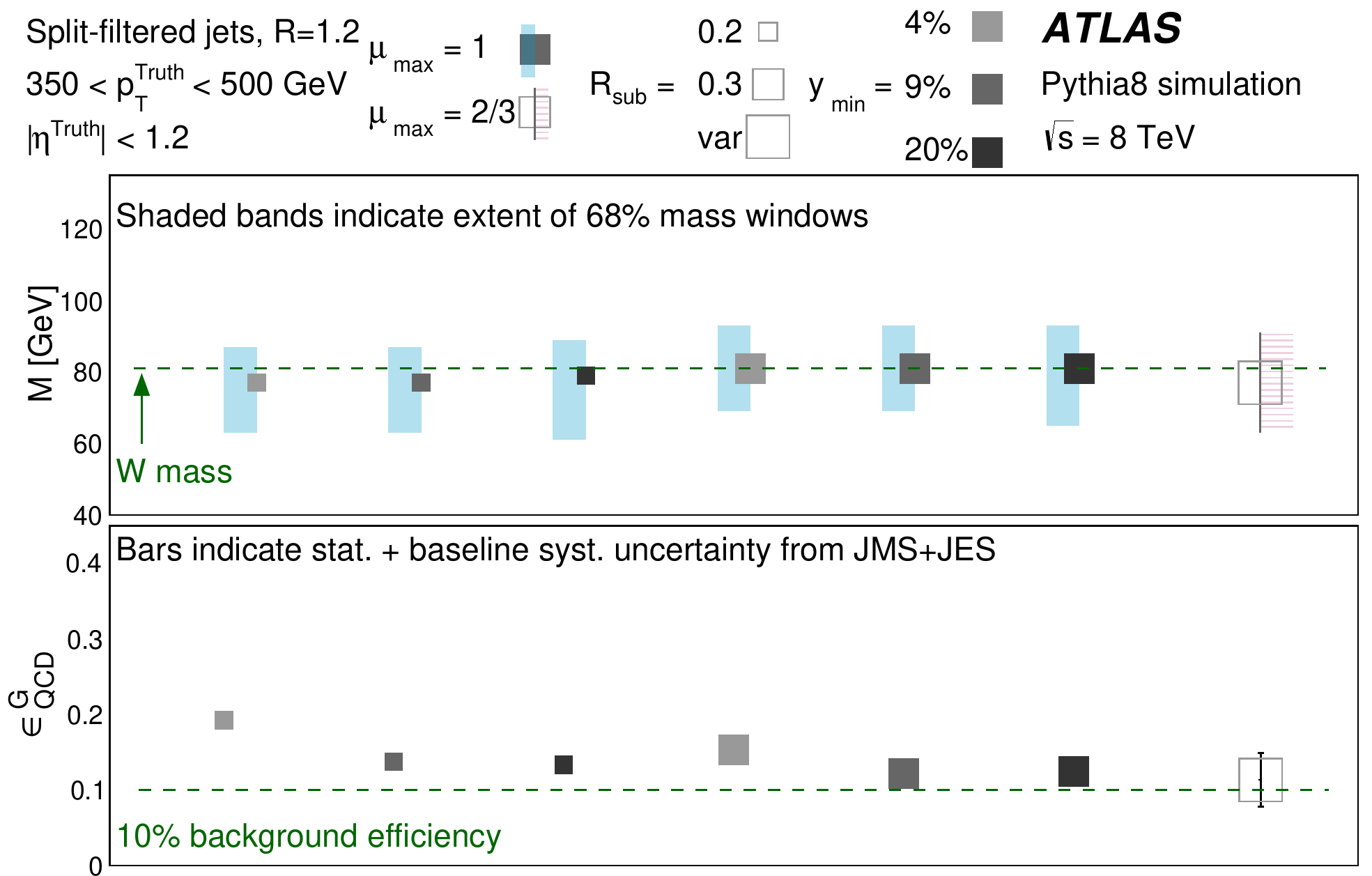}
\end{center}
\caption[Mass windows, MPVs and $\epsilon^{\mathrm{G}}_{\mathrm{QCD}}$ for more split-filtering configs.]{ Mass windows and background efficiencies for various additional configurations of split-filtering ($R$=1.2 shown). The baseline systematic uncertainty on the background efficiency for the \pt bin in question (the range $350 < \pt\ < 500\GeV$ is shown here) is calculated by varying the jet mass scale (JMS) and jet energy scale (JES) by $\pm 1 \sigma$ for a representative jet collection. For split-filtering, this representative configuration is \MaxMassDrop$=1$, \subjetr$=0.3$ and \ycut$=15\%$.}
\label{fig:unreadable_filtered_extra}
\end{figure}

\begin{figure}
\begin{center}
\includegraphics[angle=0,width=0.95\textwidth]{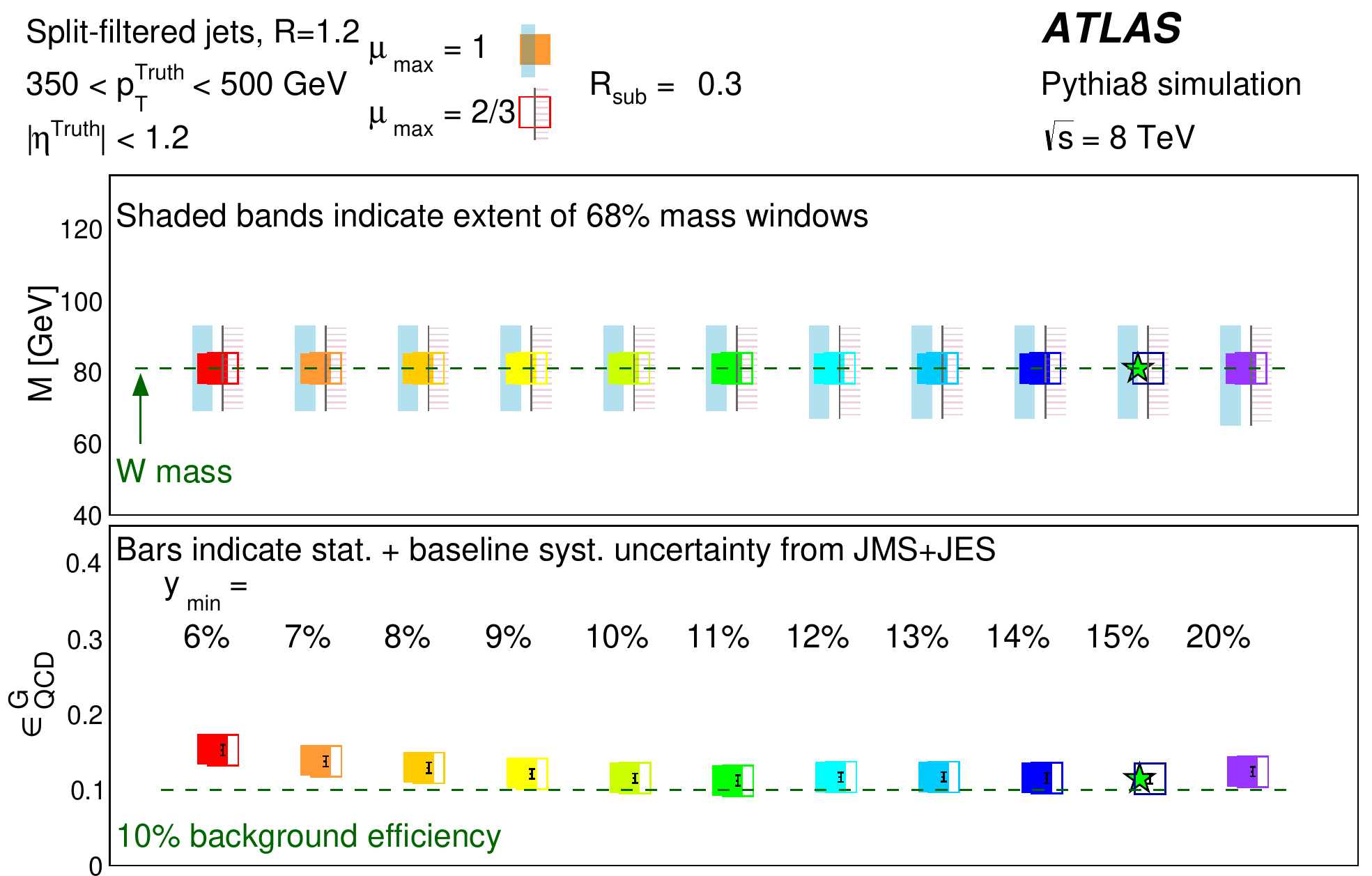}
\end{center}
\caption[Mass windows, MPVs and $\epsilon^{\mathrm{G}}_{\mathrm{QCD}}$ for lots of split-filtering configs.]{ Mass windows and background efficiencies for various configurations of split-filtering ($R$=1.2 shown). The baseline systematic uncertainty on the background efficiency for the \pt bin in question (the range $350 < \pt\ < 500\GeV$ is shown here) is calculated by varying the jet mass scale (JMS) and jet energy scale (JES) by $\pm 1 \sigma$ for a representative jet collection. For split-filtering, this representative configuration is \MaxMassDrop$=1$, \subjetr$=0.3$ and \ycut$=15\%$. The star indicates the favoured split-filtering configuration for $W$-tagging, as detailed in \secref{findings_mc}.} 
\label{fig:unreadable_filtered}
\end{figure}

\textbf{Split-filtering}:\\
Split-filtering is studied with \CamKt jets with $R =$ 1.2 and 1.0, and various
values of the parameters \MinMomentumBalance, \subjetr and \MaxMassDrop. The background efficiencies and 68\% signal mass windows obtained with a subset of split-filtering configurations for the range $350 < \pt < 500\GeV$ are shown in \figref{unreadable_filtered_extra} and \figref{unreadable_filtered}.

Observations from the results of these studies include the following:
\begin{itemize}
\item Larger \MinMomentumBalance values tend to result in lower background efficiencies.
\item The performance has a dependence on \MinMomentumBalance and the optimal requirement varies with jet \pt. For $\ycut \geq 0.09$, the background efficiency is relatively stable.
\item For a \MinMomentumBalance $>0.09$, there is not a strong dependence of the performance on \subjetr or \MaxMassDrop.
\end{itemize}

A total of 11 split-filtering jet collections are considered for further study, all with \MaxMassDrop $ = 100\%$ and
\subjetr = 0.3. These are given in \Tabref{filtbest}.

\begin{table}
\begin{center}
\begin{tabular}{ c | c | c | c | c }
\hline
\hline
Initial algorithm & $R$ & \MinMomentumBalance & \MaxMassDrop & \subjetr \\
\hline
C/A & 1.2 & 0\% & 100\% & 0.3 \\
C/A & 1.2 & 4\% & 100\% & 0.3 \\
C/A & 1.2 & 9\% & 100\% & 0.3 \\
C/A & 1.2 & 12\% & 100\% & 0.3 \\
C/A & 1.2 & 15\% & 100\% & 0.3 \\[1ex]
C/A & 0.8 & 0\% & 100\% & 0.3 \\
C/A & 0.8 & 4\% & 100\% & 0.3 \\
C/A & 0.8 & 9\% & 100\% & 0.3 \\[1ex]
C/A & 0.6 & 0\% & 100\% & 0.3 \\
C/A & 0.6 & 4\% & 100\% & 0.3 \\
C/A & 0.6 & 9\% & 100\% & 0.3 \\
\hline
 \hline
\end{tabular}
\caption[Best split-filterers.]{The best split-filtering configurations for $W$-tagging with each $R$ based on the first stage of the MC-based optimisation studies.}
\label{tab:filtbest}
\end{center}
\end{table}

\subsection{Pileup dependence}\label{sec:pileup}

\begin{figure}
\begin{center}
\includegraphics[width=0.6\textwidth]{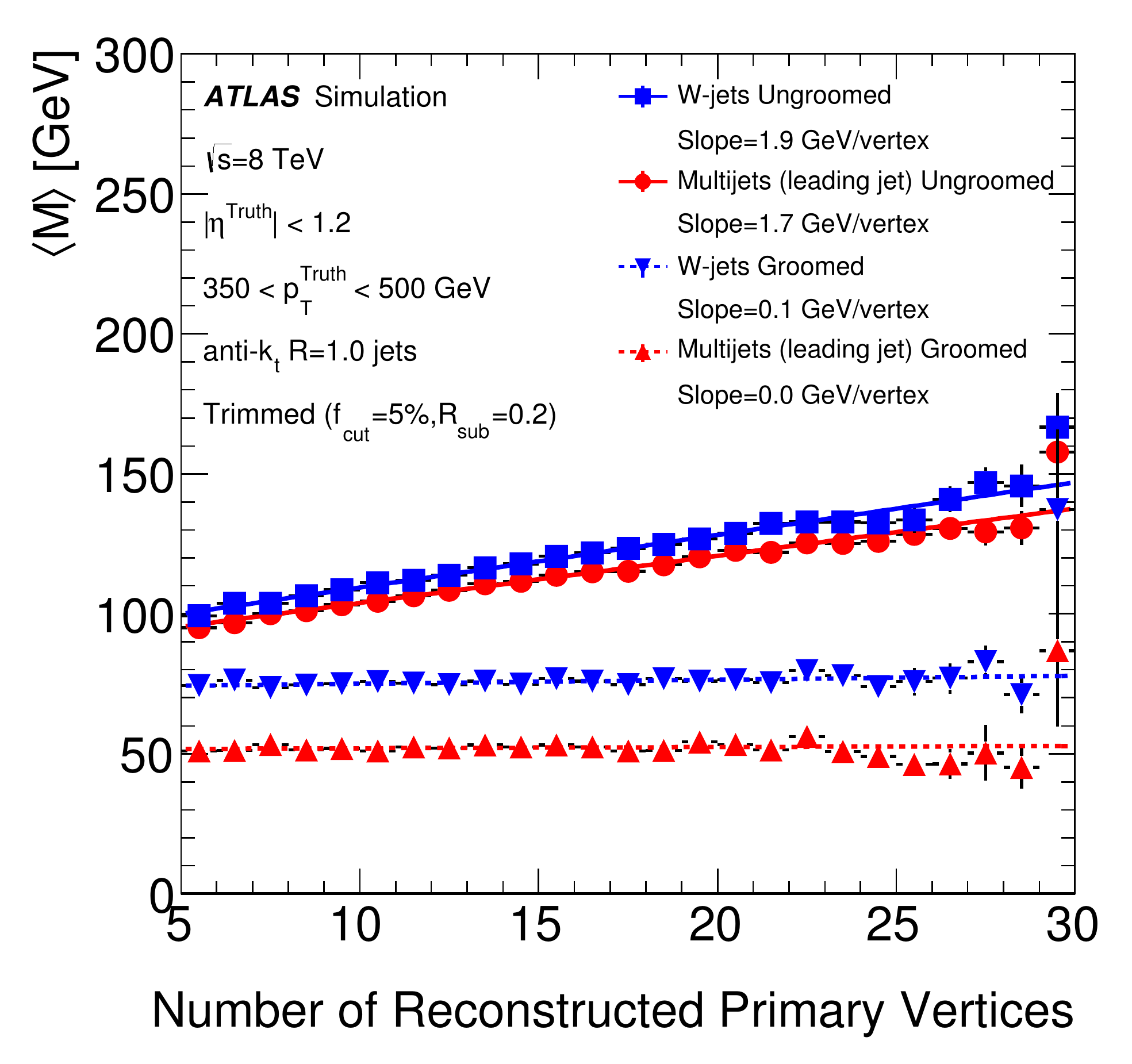}
\caption[Pileup dependence of jet mass after grooming.]{The average jet mass $\avg{M}$ as a function of the number of reconstructed primary vertices for $W$-jet signal and multijet background, before and after grooming using \antiktten trimmed with \fcut = 0.05 and \subjetr~=~0.2. The slopes of straight line fits are provided in each case: for ungroomed jets this is $\sim$ 2 \GeV\ per vertex, while for trimmed jets it is flat.}
\label{fig:part1_trim1b_top}
\end{center}
\end{figure}

The influence of pileup on the reconstructed groomed jets is examined during the first stage of algorithm optimisation, and configurations that show large susceptibility to pileup after grooming are discarded. There are a number of methods~\cite{Cacciari:2007fd,Soyez:2012hv,Aad:2012meb,groomedJetPileup2011,Cacciari:2014gra,Krohn:2013lba,Berta:2014eza,Bertolini:2014bba} available for reducing the effects of pileup, either on their own or combined with grooming; these techniques are not considered in this study. Most grooming configurations almost completely remove the effects of pileup from the mean jet mass as illustrated in \figref{part1_trim1b_top} in which the correlation between average jet mass $\avg{M}$ and number of primary vertices for a well-behaved trimming configuration is shown. The significant correlation between the average ungroomed jet mass and the number of reconstructed primary vertices is absent for trimmed jets in both signal and background.

\begin{figure}
\begin{center}
\includegraphics[width=0.9\textwidth]{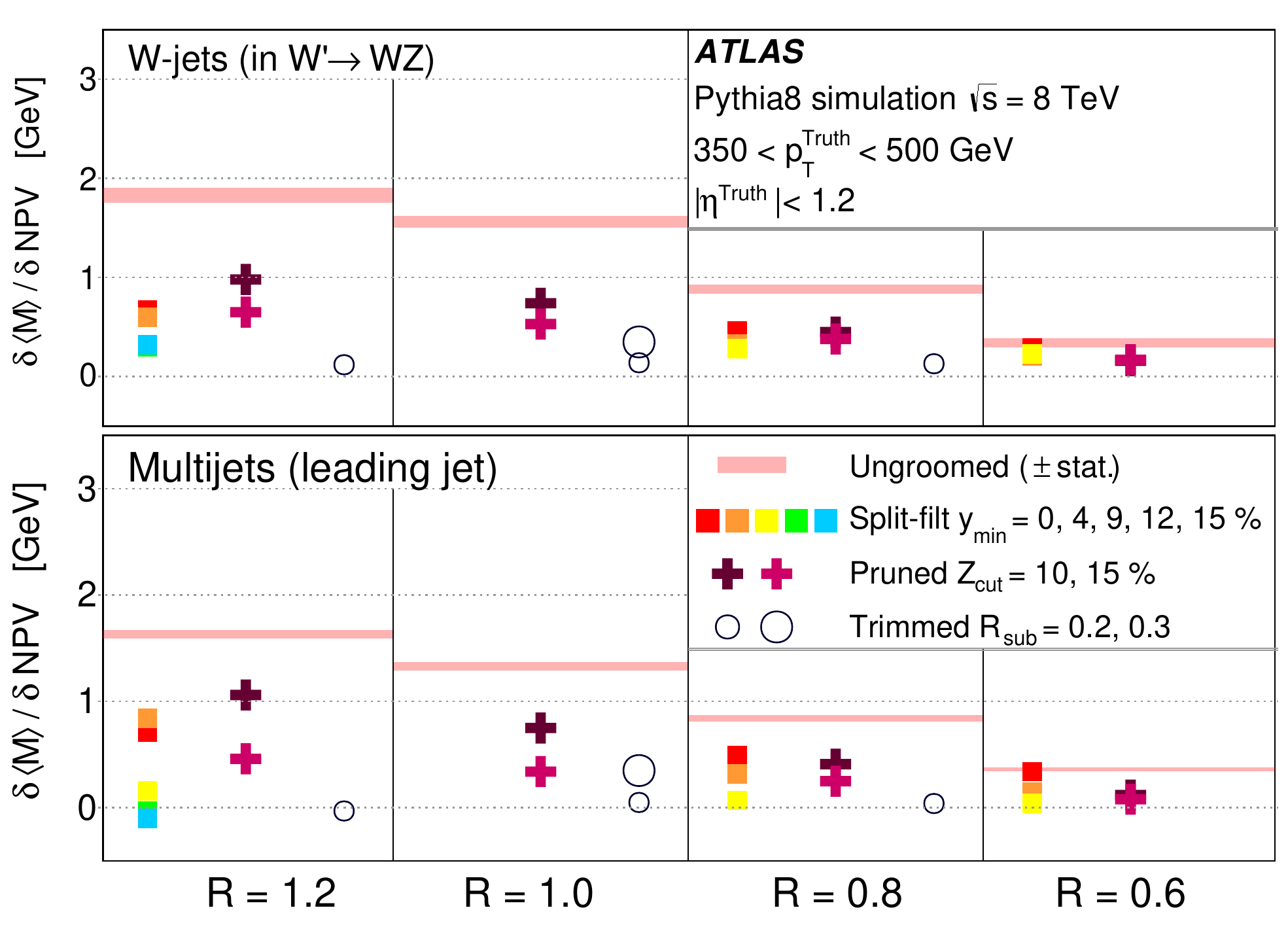}
\caption[Pileup dependence summary of jet mass after grooming.]{A summary of the pileup dependence $\delta \avg{M} / \delta {\rm{NPV}}$ for the 27 jet configurations selected for further study. The top panel shows the dependence for signal $W$-jets, the bottom panel for background multijets, and from left to right shows decreasing values of the initial jet radius parameter, $R$. Each value of $\delta \avg{M} / \delta \rm{NPV}$ is the slope of a straight line fit of $\avg{M}$ versus NPV, an example of which is shown in \figref{part1_trim1b_top}.}
\label{fig:part1_trim1b_bottom}
\end{center}
\end{figure}

The pileup dependence of the mean jet mass obtained with all 27 of the grooming configurations selected for stage two of the optimisation studies is shown in terms of the fitted slope of $\delta \avg{M}/\delta \rm{NPV}$ in \figref{part1_trim1b_bottom} for the \pt\ range 350--500~\GeV. In general, the average masses of jets with larger radii have a more pronounced pileup dependence, and the trimmed jet mass has a weaker pileup dependence than that obtained with the pruning and split-filtering algorithms. For all jet algorithms, the pileup dependence is much reduced with respect to that of ungroomed jets.

\subsection{Performance of substructure variables}\label{sec:substructure_vars}

Substructure variables are introduced in \secref{primer_vars}. A brief description of the variables studied in this analysis are listed below:\\

\begin{itemize}
\item The energy correlation ratios \CTwoBeta and \DTwoBeta, described in detail in \secref{primer_vars}.
\item The N-subjettiness ratios \TauTwo, \TauTwoWTA, \TauTwoOne, and \TauTwoOneWTA are also described in detail in \secref{primer_vars}.
\item Planar flow~\cite{Almeida:2008yp}, $P$, is a measure of how uniformly distributed the energy of a jet is, perpendicular to its axis.
\item The angularity, \Angularity, distribution is expected to peak sharply at values close to zero for a balanced two-body decay, such as that of a $W$ boson, while a broader 
tail is expected for jets initiated by quarks and gluons.
The general formula for the mass-normalised angularity can be found in Ref.~\cite{Almeida:2008yp}.
\item Splitting scales~\cite{ButterworthSplit2002} are calculated, within the jet clustering 
algorithm, and can be calculated for any jet using its constituents.
The splitting scale \SplitOneTwo, is calculated for a jet (re)clustered with the \kt-clustering algorithm, and is the $k_t$ distance between the two proto-jets of the final clustering step.
\item The variable \ZOneTwo~\cite{Thaler2008} is a variant on the original splitting scale \SplitOneTwo which uses the jet mass.
\item The momentum balance~\cite{Butterworth:2008iy}, \MomentumBalance, and mass-drop fraction \MassDrop, are defined at the first de-clustering step that satisfies a minimum mass-drop and momentum balance requirement, and are only available for those jets that are groomed with the split-filtering algorithm.
\item The soft-drop algorithm~\cite{Larkoski:2014sd} declusters the jet, following the path of highest \pt through the clustering history. A condition is defined:

\begin{equation}
z_{\mathrm g}  >  z_{\mathrm{cut}} \times r_{\mathrm g} ^{ \beta}, \label{eq:SD}
\end{equation}

where the fractional momentum of the softest of the two branches is $z_{\mathrm g}=\frac{\min(p_{\mathrm T1}, p_{\mathrm T2})}{p_{\mathrm T1}+p_{\mathrm T2}}$, and the fractional angular separation of the two branches (with respect to the $R$ parameter of the initial jet algorithm, $R_0$) is $r_{\mathrm g}=\frac{ \Delta R_{12} }{ R_0 }$. Nine values of the $z_{\mathrm{cut}}$ parameter between 4\% and 20\% are explored here, given in \Tabref{softdroplevels}. The $\beta$ values chosen here are $-1.0$, $-0.75$, and $-0.5$. The starting condition of \eref{SD} with $z_{\mathrm{cut}} = 4\%$ is applied to the first step in the declustering. If this condition is not satisfied, the algorithm continues to the next step in the jet's clustering history, and so on, checking if the condition is satisfied at any point. If it is not, the `soft-drop-level', \SoftDropLevel is zero. If this condition is satisfied, \SoftDropLevel $= 1$. The algorithm then remains at this point in the clustering history and asks for the same condition with the harder momentum condition, $z_{\mathrm{cut}} = 6\%$. If this condition is not satisfied, the algorithm continues to the next step in the jet's clustering history, and so on.

\begin{table}
\caption{The soft-drop levels \SoftDropLevel are defined as the highest level of balance in the jet history.}
\label{tab:softdroplevels}
\begin{center}
\begin{tabular}{c | c c c c c c c c c}
\hline
\hline
\SoftDropLevel & 1 & 2 & 3 & 4 & 5 & 6 & 7 & 8 & 9 \\
\hline
 $z_{\mathrm{cut}}$ & 4\% & 6\% & 8\% & 10\% & 12\% & 14\% & 16\% & 18\% & 20\% \\
 \hline
 \hline
\end{tabular}
\end{center}
\end{table}

\item The dipolarity~\cite{Hook2011}, $D$, is a measure of the colour flow between two hard centres within a jet. 
\item Jet shape variables are computed in the centre-of-mass frame of a jet, which can increase the separation power between $W$-jets and jets in multijet events.
Sphericity, $S$, aplanarity, $A$, and thrust minor and major, \ThrustMin, \ThrustMaj, already used in a previous ATLAS measurement~\cite{AtlasWtag7tev}, as well
as the ratio of the second to zeroth order Fox--Wolfram moments, \FoxWolfRatio~\cite{Chen2012} are considered.
\item For a jet clustered with a given recombination jet clustering algorithm, the Q-jets technique~\cite{Ellis:2012sn} reclusters the jet many times for each step in the clustering.
Following this, any jet observable, such as the mass, will have a distribution for a given jet. The Q-jets configuration optimised in Ref.~\cite{QjetsPerf2012} is adopted in this study. The high mass in $W$-jets tends to persist during the re-clustering while the mass of QCD jets fluctuates. A sensitive observable to this trend is the coefficient of variation of the mass distribution for a single jet, called the volatility~\cite{Ellis:2012sn,QjetsPerf2012}, \Volatility. The superscript $\alpha$ denotes the rigidity, which controls the sensitivity of the pair selection to the random number generation used in the clustering.  
\end{itemize}

\begin{figure}
\begin{center}
\begin{subfigure}[]{ \label{fig:variable2_c2_1d}
\includegraphics[width=0.6\textwidth]{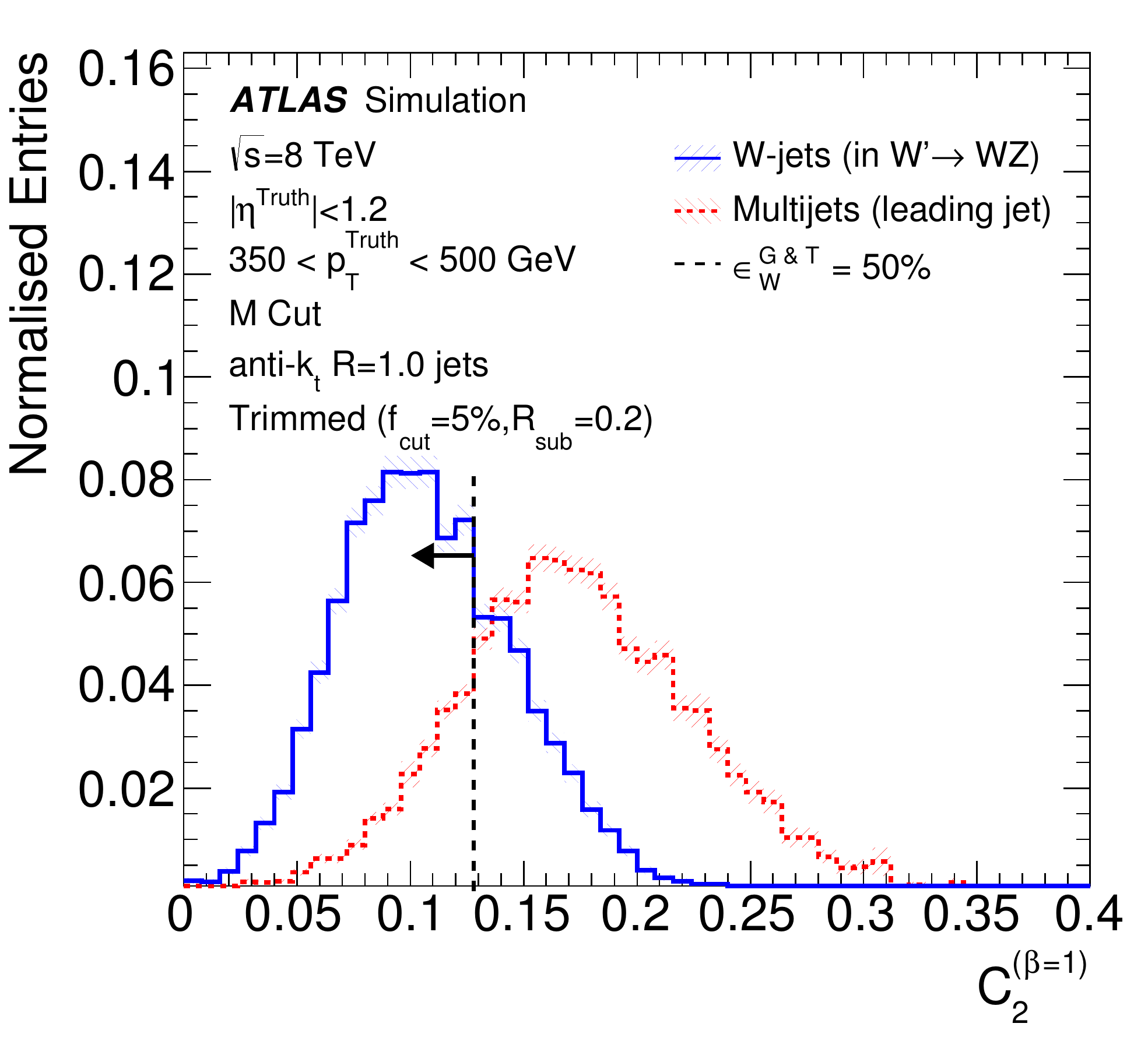}
}
\end{subfigure}
\begin{subfigure}[]{ \label{fig:variable2_c2_2d}
\includegraphics[width=0.99\textwidth]{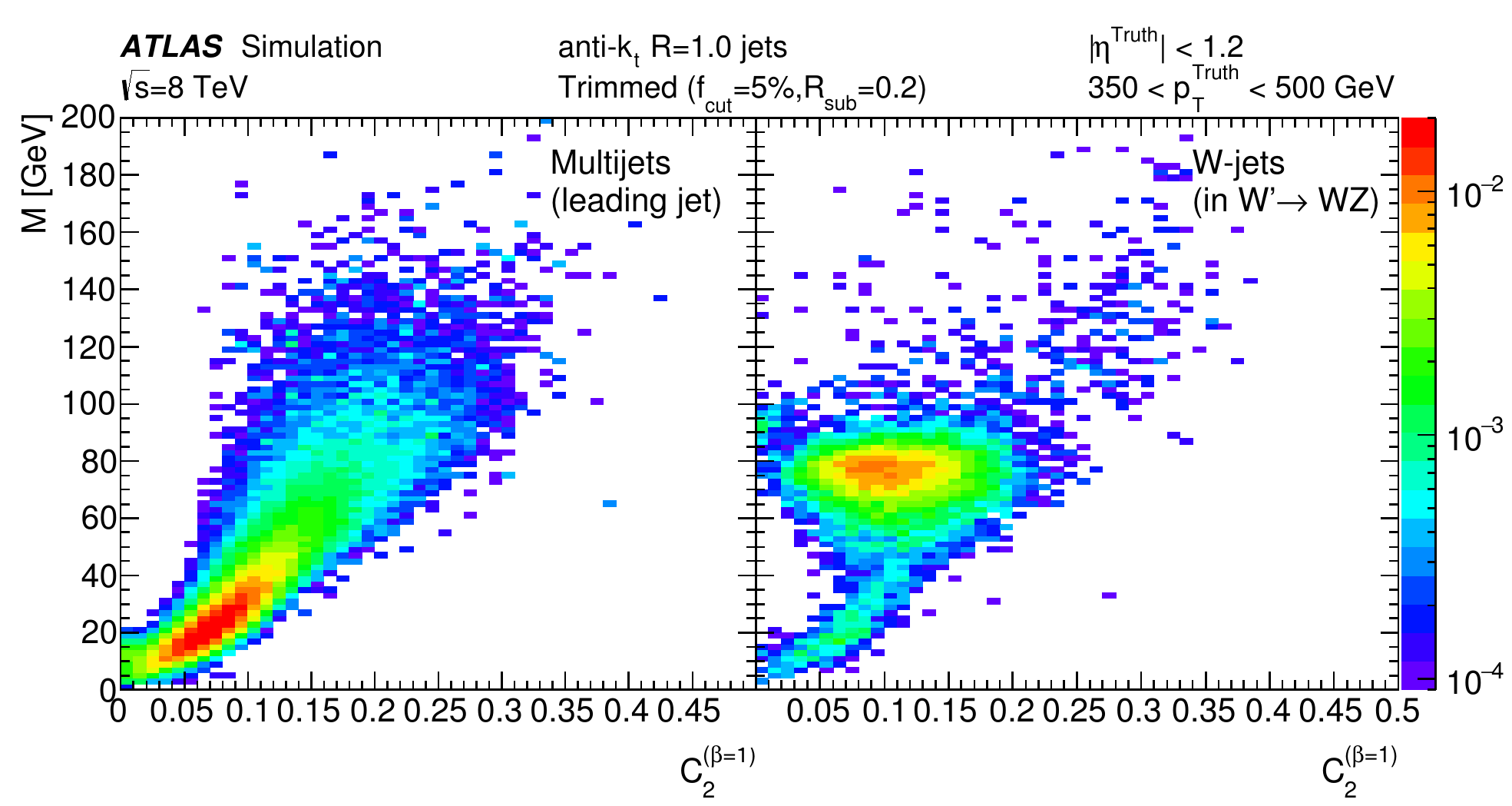}
}
\end{subfigure}
\end{center}
\caption[\CTwoBetaOne, R2-trimmed, $\pt^\text{Truth} = [350,500)$~\GeV.]{The \CTwoBetaOne variable, for R2-trimmed jets: (a) distributions in signal (blue solid line) and background (red dashed) in MC in the range $350 < \pt < 500~\GeV$, obtained after applying the 68\% signal efficiency mass window requirement (discussed in \secref{groomers_perf}); (b) correlation with the leading jet's mass in (left) multijet background and (right) $W$-jet signal events. No truth-matching requirements are made, so the signal events can contain background jets as well as $W$-jets. The vertical line corresponds to the value of the cut providing a combined 50\% efficiency for grooming and tagging (corresponding to a tagging-only efficiency of 50\%/68\% = 73.5\%)}
\label{fig:variable2_c2}
\end{figure}

\begin{figure}
\begin{center}
\begin{subfigure}[]{ \label{fig:variable2_d2_1d}
\includegraphics[width=0.6\textwidth]{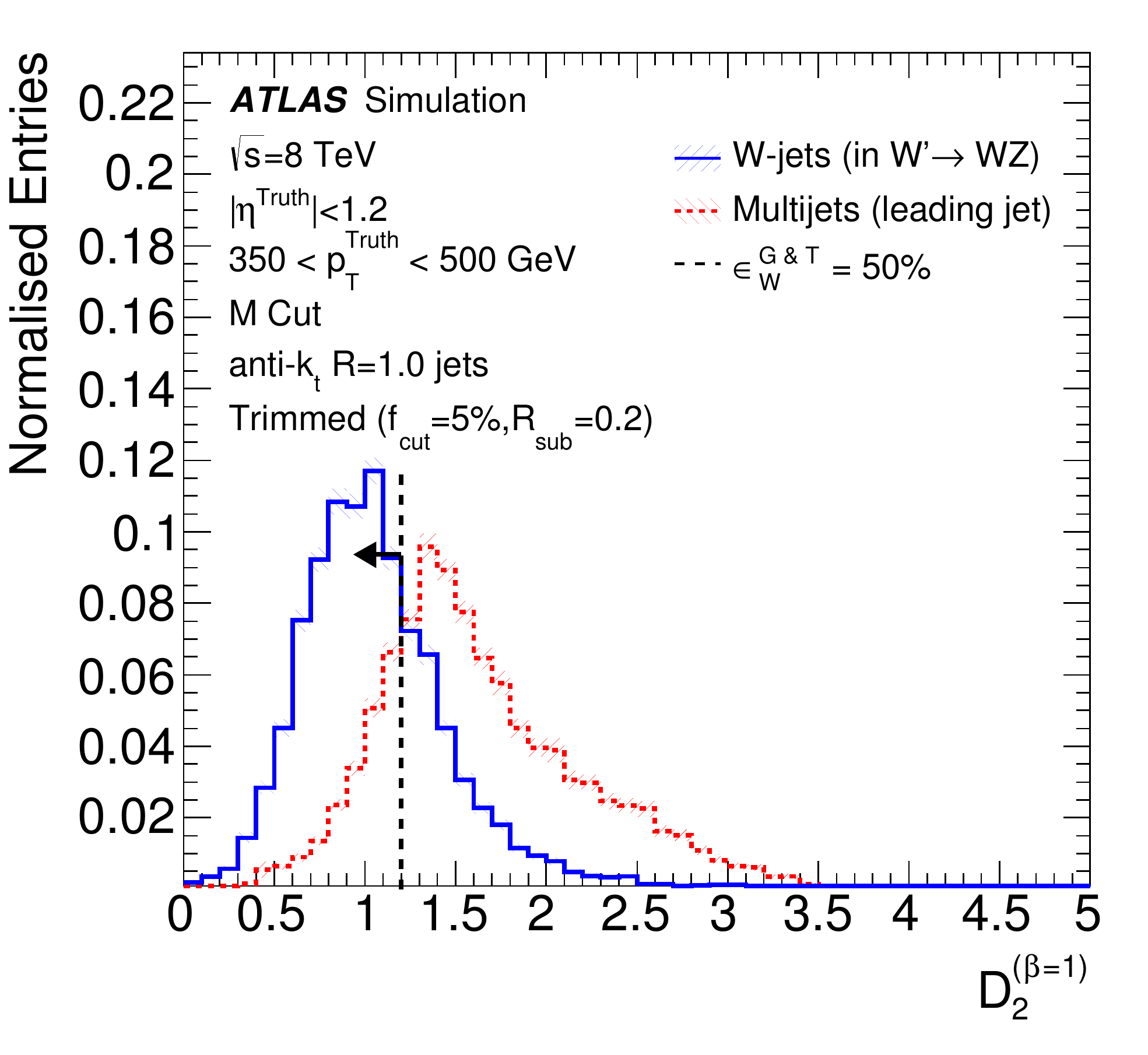}
}
\end{subfigure}
\begin{subfigure}[]{ \label{fig:variable2_d2_2d}
\includegraphics[width=0.99\textwidth]{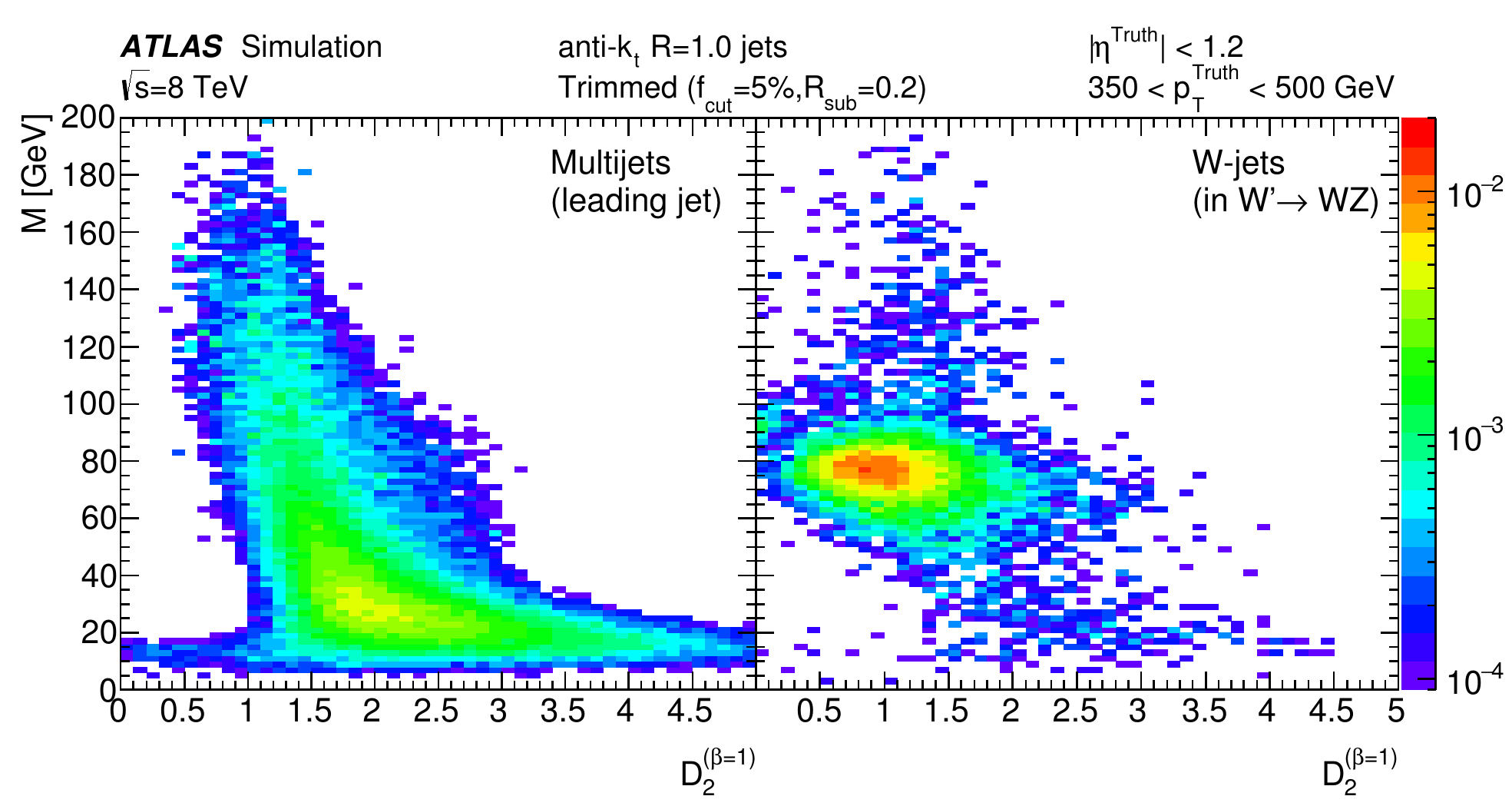} 
}
\end{subfigure}
\end{center}
\caption[\DTwoBetaOne, R2-trimmed, $\pt^\text{Truth} = [350,500)$~\GeV.]{The \DTwoBetaOne variable, for R2-trimmed jets: (a) distributions in signal (blue solid line) and background (red dashed) in MC in the range $350 < \pt < 500~\GeV$, obtained after applying the 68\% signal efficiency mass window requirement (discussed in \secref{groomers_perf}); (b) correlation with the leading jet's mass in (left) multijet background and (right) $W$-jet signal events. No truth-matching requirements are made, so the signal events can contain background jets as well as $W$-jets. The vertical line corresponds to the value of the cut providing a combined 50\% efficiency for grooming and tagging (corresponding to a tagging-only efficiency of 50\%/68\% = 73.5\%)}
\label{fig:variable2_d2}
\end{figure}

\begin{figure}
\begin{center}
\begin{subfigure}[]{ \label{fig:variable2_tau21_1d}
\includegraphics[width=0.6\textwidth]{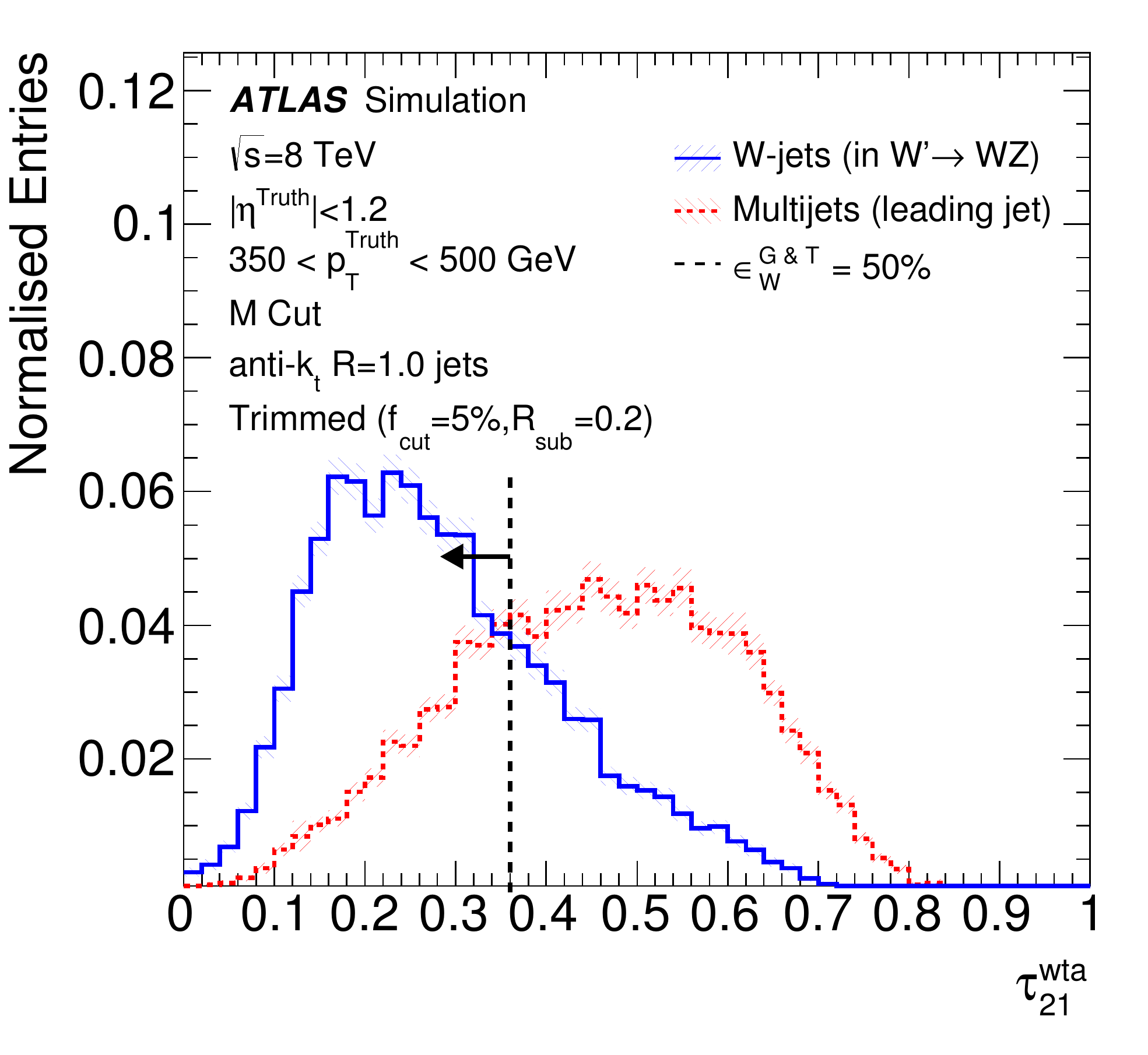}
}
\end{subfigure}
\begin{subfigure}[]{ \label{fig:variable2_tau21_2d}
\includegraphics[width=0.99\textwidth]{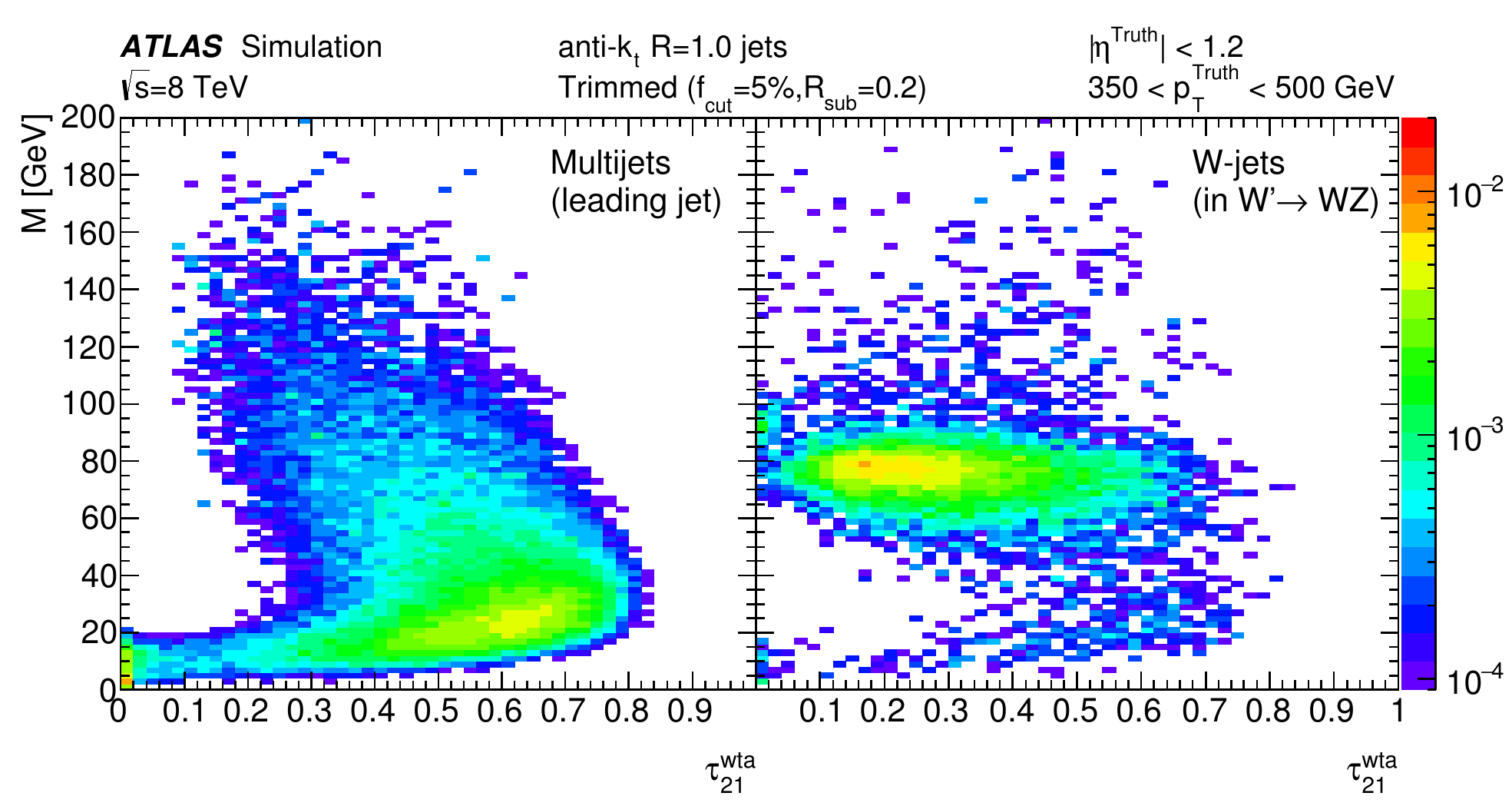}
}
\end{subfigure}
\end{center}
\caption[\TauTwoOneWTA, R2-trimmed, $\pt^\text{Truth} = [350,500)$~\GeV.]{The \TauTwoOneWTA variable, for R2-trimmed jets: (a) distributions in signal (blue solid line) and background (red dashed) in MC in the range $350 < \pt < 500~\GeV$, obtained after applying the 68\% signal efficiency mass window requirement (discussed in \secref{groomers_perf}); (b) correlation with the leading jet's mass in (left) multijet background and (right) $W$-jet signal events. No truth-matching requirements are made, so the signal events can contain background jets as well as $W$-jets. The vertical line corresponds to the value of the cut providing a combined 50\% efficiency for grooming and tagging (corresponding to a tagging-only efficiency of 50\%/68\% = 73.5\%)}
\label{fig:variable2_tau21}
\end{figure}

For all 27 jet collections and grooming algorithms described in \secref{groomers_perf}, the full list of substructure variables described above are computed. The distributions of the three variables \TauTwoOneWTA, \CTwoBetaOne and \DTwoBetaOne are shown in Figs.~\ref{fig:variable2_c2}--\ref{fig:variable2_tau21} for \antiktten jets trimmed with \fcut = 0.05 and \subjetr = 0.2, after applying the 68\% signal efficiency mass window requirement. This grooming algorithm is referred to in the remainder of this paper as `R2-trimming'. At this stage no jet mass calibrations have been applied for any of the grooming configurations. Also shown are the correlations between the jet mass and each of these variables, shown separately for the $W$-jet signal and multijet background, in both cases before applying the 68\% signal efficiency mass window requirement. No truth-matching between the subjets and the quarks from the W decay is required, such that the signal sample contains both full $W$-jets and jets made of fragments of the $W$-decay, generally because the $W$-decay is not completely captured in the $R = 1.0$ jet. The background jets within the signal sample are particularly visible in the low-mass region of \figref{variable2_c2_2d}, where the distributions echo those seen in the background sample.

\begin{figure}
\begin{center}
\includegraphics[width=1.35\textwidth,angle=270]{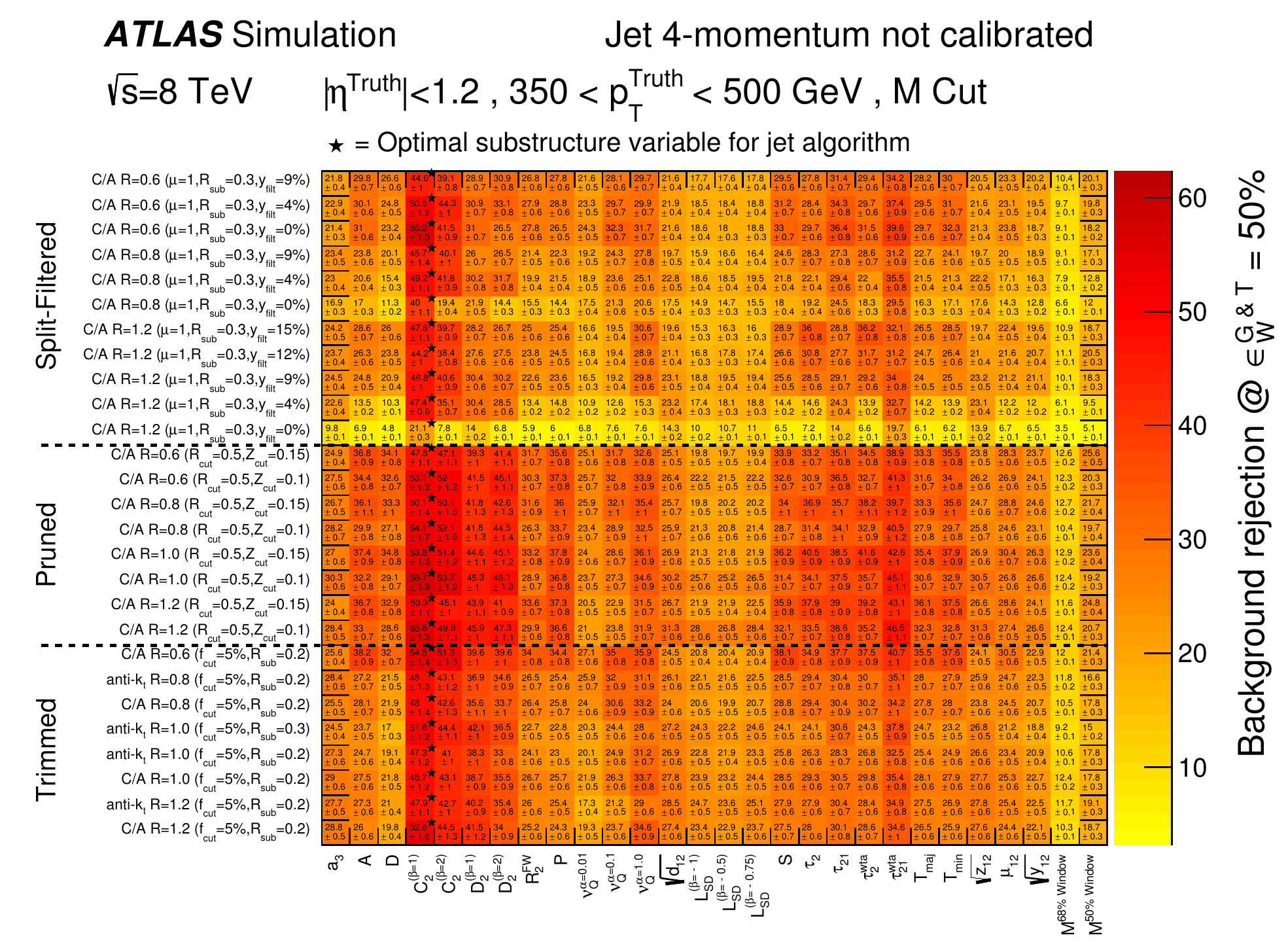}
\end{center}
\vspace{-20 pt}
\caption[Background rejection power for 27 x 26 matrix.]{For jets with 350~$< \pt^\text{Truth} <$~500~\GeV, the background rejection factors corresponding to a 50\% efficiency are shown for all possible combinations between the 27 grooming configurations and 26 substructure variables, after applying the uncalibrated groomed mass window requirement that provides a 68\% signal efficiency. The error shown are the result of the finite Monte Carlo sample size.}
\label{fig:rejection350-500}
\end{figure}

The background rejection power (1 / background efficiency) is shown in \figref{rejection350-500} for the $\epsilon_{W}^{\mathrm{G}\&\mathrm{T}} = 50\%$ efficiency working point for each substructure variable inside the mass window determined by the grooming, and for each of the 27 grooming configurations, for the range $350 < \pt < 500\GeV$. 

\begin{figure}
\begin{center}
\includegraphics[width=\textwidth]{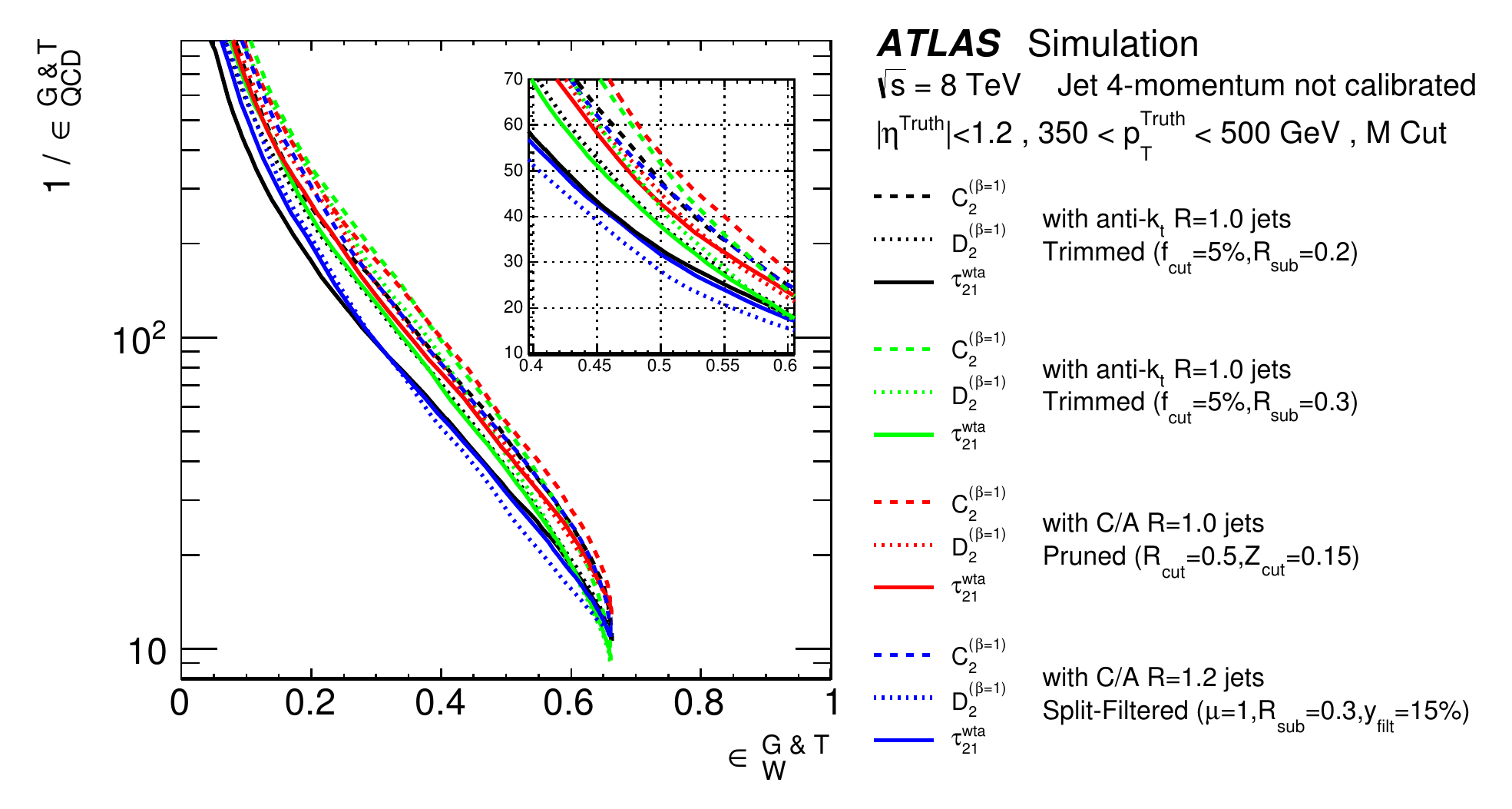}
\end{center}
\caption[ROC curve]{For jets with 350~$< \pt^\text{Truth} <$~500~\GeV, the signal efficiency versus background rejection power ``ROC'' curve for selected tagging variables (combined with the uncalibrated groomed mass window) on a subset of high-performance algorithms is shown. The endpoint at 68\% signal efficiency is a result of the 68\% mass window. The inset enlarges the high-efficiency region.}
\label{fig:roc350-500}
\end{figure}

In addition to calculating the background rejection power at a particular signal efficiency working point, full rejection versus efficiency curves (so-called Receiver Operating Characteristic `ROC' curves) are produced for each combination. An example showing the relationship between the $W$-jet signal efficiency and the multijet background rejection for the range $350 < \pt < 500\GeV$ is shown in \figref{roc350-500}. The maximal efficiency
value for each algorithm is by definition 68\%, since the tagging criteria are applied after requiring the jet mass to be within the mass window defined by the grooming.

\par

\subsection{Summary of grooming and substructure in MC}\label{sec:findings_mc}

Four grooming configurations, given in \Tabref{bestgroomers}, show consistently high performance in all \pt\ bins. The jet $\eta$, mass and energy calibrations are derived for these four using a simulation-based calibration scheme, used as the standard one by ATLAS in previous studies~\cite{Aad:2013gja}. The mass window sizes for calibrated jets, the background efficiencies for $\epsilon_{W}^{\mathrm{G}} = 68\%$ and the $\delta \avg{M} / \delta \rm{NPV}$ in the range $200 < \pt\ < 350\GeV$ are also given in \Tabref{bestgroomers}. 

\begin{table}\renewcommand{\arraystretch}{1.5}
\begin{center}
\begin{tabular}{  c | c | c | c }
\hline
\hline
 Grooming configuration & $\epsilon_{W}^{\mathrm{G}} = 68\%$ mass range & $\epsilon_{\mathrm{QCD}}^{\mathrm{G}}$ & $\delta \avg{M} / \delta {\rm NPV}$ \\
\hline
\antiktten trimmed \fcut = 0.05, \subjetr = 0.2 &  61--93~\GeV & 11\% & 0.1--0.2 ~\GeV  \\
\antiktten trimmed \fcut = 0.05, \subjetr = 0.3 & 65--99~\GeV  & 16\% & 0.5--0.6 ~\GeV \\
\camktten pruned \Zcut = 0.15, \rcut = 0.5      & 59--111~\GeV & 16\%  & 0.9--1.1 ~\GeV \\
\camkttwelve split-filt \MomentumBalance = 0.15, \subjetr = 0.3 & 63--103~\GeV  & 13\% & 0.1--0.3 ~\GeV \\
\hline
\hline
\end{tabular}
\caption[Performance of the four best grooming algorithm.]{The four favoured grooming configurations along with their mass windows (derived using calibrated jets), background efficiencies, and pileup dependence for $\epsilon_{W}^{\mathrm{G}} = 68\%$ in the range $200 < \pt\ < 350\GeV$.}
\label{tab:bestgroomers}
\end{center}
\end{table}

Since the first algorithm in \Tabref{bestgroomers} is the only one of the four with negligible pileup dependence across all \pt\ ranges (the central \pt range only is shown in \figref{part1_trim1b_bottom}), it is adopted for all successive studies.

The best substructure variables for use with R2-trimmed jets at the $\epsilon_{W}^{\mathrm{G}\&\mathrm{T}}=50\%$ working point, providing background efficiencies $\epsilon_{\mathrm{QCD}}^{\mathrm{G}\&\mathrm{T}} \sim 2\%$ (background rejection power $\sim$ 50, in terms of \figref{rejection350-500}) for jets with $\pt{} > 350\GeV$, are given in \Tabref{bestvars}. Studies of the R2-trimmed grooming configuration and the three preferred substructure variables are described 
in the next section, where the results obtained from Monte Carlo simulations are compared to data.

\begin{table}\renewcommand{\arraystretch}{1.5}
\begin{center}
\begin{tabular}{ c  c | c | c | c }
\hline
\hline
\multicolumn{2}{c |}{\multirow{2}{*}{Variable}} & \multicolumn{3}{c }{Tagging criteria in \pt\ range}\\\cline{3-5}
&& 200--350 ~\GeV & 350--500 ~\GeV & 500--1000 ~\GeV \\
\hline
\hline
\multicolumn{2}{c | }{$\epsilon_{W}^{\mathrm{G}} = 68\%$ mass range} & 61--93~\GeV & 71--91~\GeV & 73--91~\GeV \\[1ex]
\multirow{3}{*}{$\epsilon_{W}^{\mathrm{G}\&\mathrm{T}} = 50\%$}& \CTwoBetaOne & $<$ 0.18 & $<$ 0.13 & $<$ 0.10 \\
& \DTwoBetaOne & $<$ 1.14 & $<$ 1.23 & $<$ 1.35 \\
& \TauTwoOneWTA & $<$ 0.32 & $<$ 0.36 & $<$ 0.40 \\
\hline
 \hline
\end{tabular}
\caption[Mass windows and tagging criteria for R2-trimmed jets.]{The mass windows for calibrated R2-trimmed jets that provide $\epsilon_{\mathrm{W}}^{\mathrm{G}} = 68\%$, and the requirements on the three substructure variables that result in the lowest background efficiencies $\epsilon_{\mathrm{QCD}}^{\mathrm{G}\&\mathrm{T}}$, when combined with the mass windows to provide $\epsilon_{W}^{\mathrm{G}\&\mathrm{T}} = 50\%$.}
\label{tab:bestvars}
\end{center}
\end{table}

\section{Detailed studies of selected techniques in data}
\label{sec:datamc}

This section describes a comparison of the $W$-jet and multijet tagging efficiencies measured using three tagging variables \CTwoBetaOne, \DTwoBetaOne and \TauTwoOneWTA computed for the leading R2-trimmed jet in data and MC. 

In data, a relatively pure sample of boosted, hadronically decaying $W$ bosons can be obtained from
decays of top quark pairs in the lepton-plus-jets decay channel: {$\ttbar \to W^{+}b W^{-}\bar{b} \to \ell \nu q\bar{q} b\bbar$}. The selection requirements detailed in \secref{objects} are applied to events in data and MC, where relevant. The composition of the data and MC samples introduced in \secref{mc} is discussed in \secref{defs_and_procs}. Details of the event topology differences between the \ttbar final state examined in this section and the $W^{\prime}$ final state used in the preliminary optimisation studies are given in \secref{wprime_ttbar_comparison}. The systematic uncertainties are discussed in \secref{sysJSS}, and the distributions of mass and substructure variables in data and MC are presented in \secref{datamc_distributions}. The signal and background efficiency estimation procedures and their uncertainties are detailed in \secref{sysEff}. A summary of the signal and background tagging efficiencies measured in data and compared to MC is given in \secref{results}.

In all the following studies, events are categorised according to the leading, reconstructed R2-trimmed jet \pt\ in three ranges: $[200, 250]$, $[250, 350]$, and $[350, 500]\GeV$. This characterisation differs from that used in the first stage of the optimisation in \secref{partOne}, which uses ungroomed \camkttwelve truth jets and different ranges; the selection is extended only to $500\GeV$ here because there are insufficient data above $500\GeV$ in the 2012 dataset. The lowest \pt range used in the preliminary optimisation stage, $[200, 350]\GeV$, is now divided in two, since the 2012 dataset has an abundance of top-decay events in this range.

\subsection{Sample compositions and definitions}
\label{sec:defs_and_procs}

Signal $W$-jets are extracted from \ttbar{} events in data and in the MC samples detailed in \secref{mc}. The \ttbar{} production cross-section is scaled to match the value obtained from NNLO calculations~\cite{TopXsecTheory}. An additional reweighting is then applied to the \ttbar{} MC using the generator-level \pt of the top quark and the \pt of the \ttbar{} system to reproduce the \pt-dependence of the measured cross-section~\cite{Top7TeV}.

The dominant backgrounds to the \ttbar{} event topology come from \ttbar production where there is only partial reconstruction of the $W$ boson decay, with or without contamination from radiation outside of the top quark decay (such as hard gluon emission, non-tagged $b$-jets). Generator-level information from the \ttbar{} and $Wt$ samples is used to distinguish the cases where the $W$ candidate jet is matched to a genuine $W$ boson or to other jets (referred to as top quark background events). An event is categorised as belonging to the $W$ signal when both partons from the $W$ boson decay are within $\Delta R = 1.0$ of the jet axis; otherwise, the event is labelled as non-$W$ background. 

The leading non-top background process is production of $W$ bosons in association with jets. The $W$+jets contribution is estimated using a data-driven charge asymmetry method~\cite{ATLAS:2012an}. \Alpgen+~\Pythia MC samples provide the event kinematics, and the relative flavour contributions and overall normalisation are determined from data. The flavour fractions are found using a control region in which there is no $b$-tagged jet requirement and instead of requiring a large-$R$ jet, events are required to have exactly two small-$R$ jets.  The relative contributions from each jet flavour are found using the charge asymmetry and the flavour fractions are fixed for $W$+jets events in the signal region before the $b$-tagged jet requirement is applied. Finally, an overall normalisation is obtained by scaling the simulated $W$+jets charge asymmetry to match the charge asymmetry in data, after other charge-asymmetric backgrounds are accounted for using MC.

The contribution from multijet events to the sample composition is estimated by using loose lepton identification criteria and deriving the contribution of non-prompt leptons using the matrix method~\cite{ATLAS-CONF-2014-058,TOPQ-2011-02}. This method relies on the fact that the tight lepton identification criteria selects primarily prompt leptons, while loose leptons that do not satisfy the tight criteria are primarily from backgrounds. The probabilities for a non-prompt lepton from multijet production which satisfies the loose/tight identification criteria are measured from data 
in control regions dominated by multijet events, with prompt-lepton contributions subtracted based on MC. The corresponding probabilities for a lepton from prompt sources (such as $W$ bosons) which satisfies the loose/tight identification criteria are derived from MC samples, corrected using data-to-MC correction factors derived from $Z\to \ell\ell$ events. Once the fraction of events satisfying the different identification criteria is known, an event weight is calculated and applied to data events with the loosened lepton identification criteria to provide an estimate of the multijet contribution.

\subsection{Event topology effects in Monte Carlo simulations}
\label{sec:wprime_ttbar_comparison}

The preliminary MC-based optimisation studies in \secref{partOne} use a signal composed of well-isolated $W$-jets from the hypothetical process $W^{\prime} \rightarrow WZ \rightarrow qq\ell\ell$ provided by \Pythia and a background sample of jets initiated by light quarks or gluons, also provided by \Pythia. In the following sections, efficiencies are measured in data, so the \ttbar final state is used as a source of $W$-jets. As described in \secref{mc}, the main \ttbar{} signal processes are provided by either \PowhegBox+~\Pythia or \Mcatnlo+~\Herwig and the multijet background is provided by \Pythia or by \Herwigpp.

Despite the backgrounds in both event topologies being \Pythia multijets, they are different in that the background efficiencies obtained in data include a leading-jet minimum \pt\ requirement of 450~\GeV\ in order to ensure full efficiency with respect to the trigger used. With this selection, the lower \pt ranges, [200, 250] and [250, 350]$\GeV$, are composed entirely of sub-leading jets, and the highest \pt\ bin, [350, 500]$\GeV$, is a mixture of leading and sub-leading jets. Jets softer than the sub-leading jet are not considered. In the background sample used for the studies in \secref{partOne} there is no comparison with data, thus there are no trigger requirements and the leading jet is always shown. A higher average jet mass is observed in the leading + sub-leading jet selection than with the leading-jet selection. This in turn leads to a higher background efficiency for the studies summarised in \secref{datamc} than for those in \secref{findings_mc}. These differences are relevant in that leading and sub-leading jets have different flavour compositions (light-quark versus gluon). Gluon-initiated jets have higher average mass than quark-initiated jets~\cite{Gallicchio:2012ez}.

The signal event topologies are more obviously different, with the $W^{\prime}$ process producing potentially more isolated $W$-jets than those found in the \ttbar final state. The $W$ bosons produced in the $W^{\prime}$ decay are also generally longitudinally polarised, making them potentially easier to distinguish from multijet background than $W$-jets from top decays, which are produced in both the longitudinal and transverse modes~\cite{Cui2010,Khachatryan:2014vla}.

\begin{figure}
\begin{center}
\includegraphics[width=0.9\textwidth]{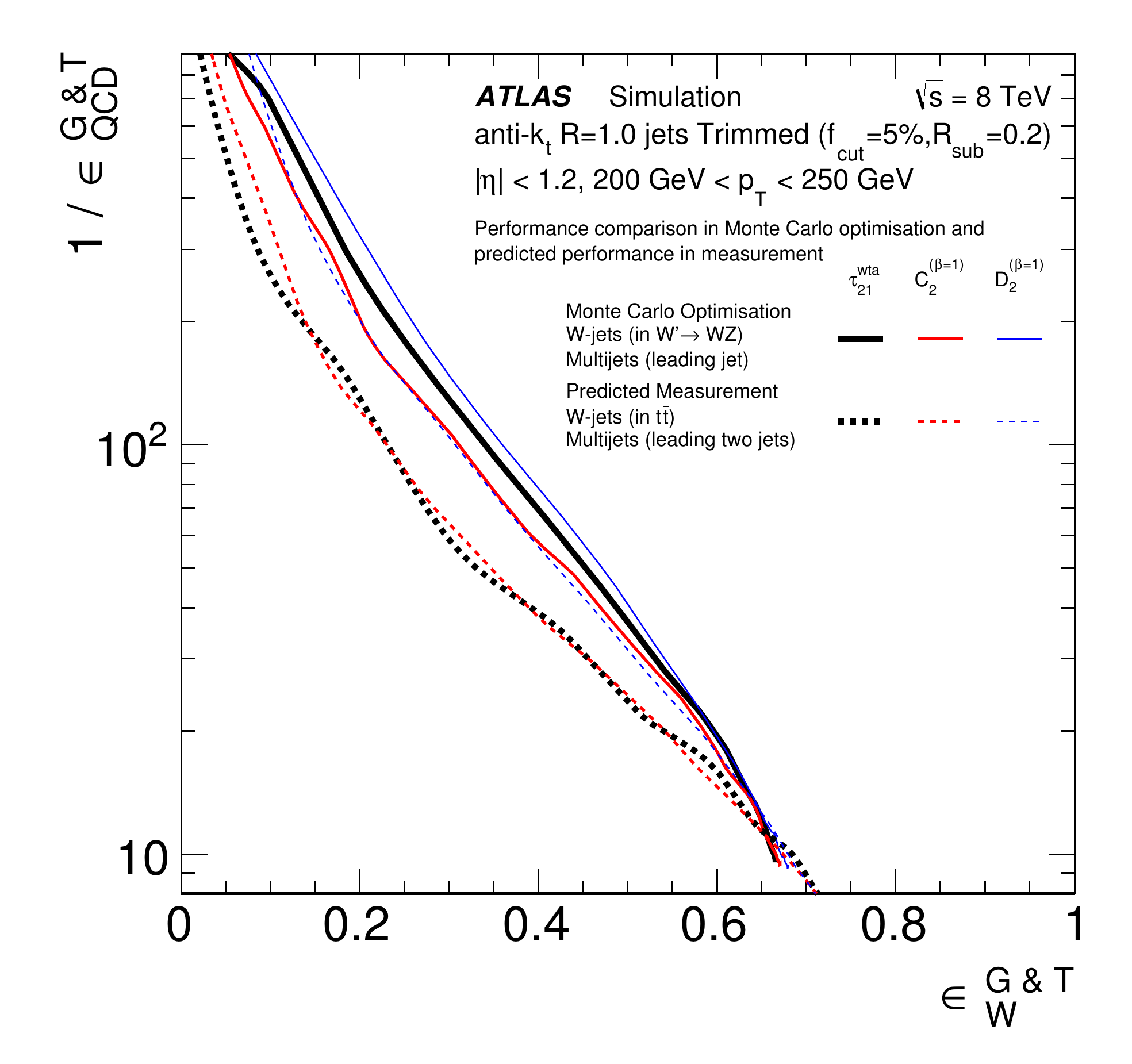}
\end{center}
\vspace{-20 pt}
\caption[Topologies ROC.]{Signal versus background efficiency curves for different event topologies. The solid lines show the curves obtained for the $W^\prime$ signal efficiencies and the leading jet from the \Pythia multijet background. The dashed lines show the curves obtained for \PowhegBox+~\Pythia \ttbar signal efficiencies and the leading+sub-leading jets from the \Pythia multijet background. }
\label{fig:topologies_roc_fulldiffs}
\end{figure}

\begin{figure} 
\begin{center} 
\includegraphics[width=0.9\textwidth]{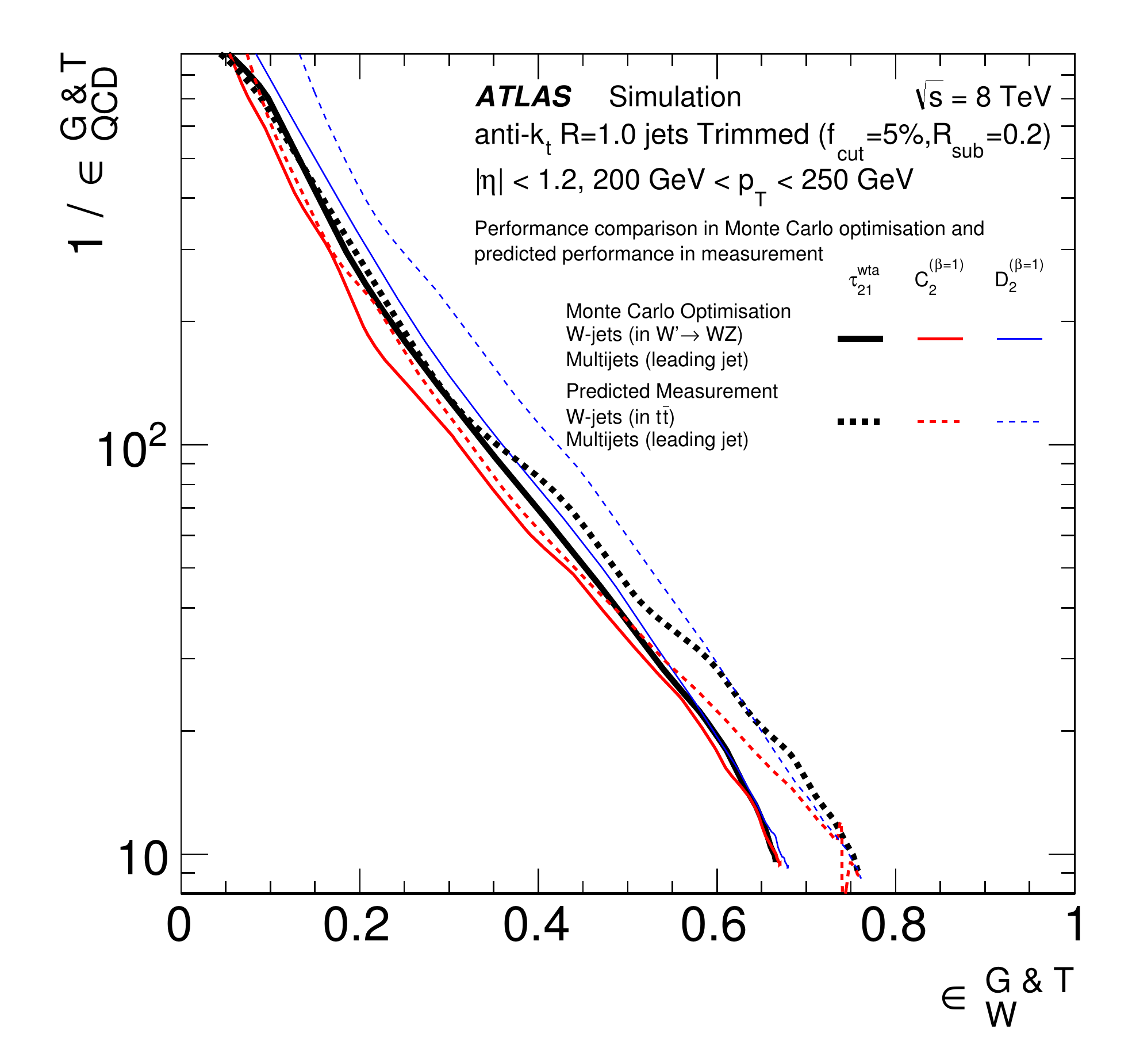} 
\end{center} 
\vspace{-20 pt} 
\caption[ROC placeholder- no background differences.]{Signal versus background efficiency curves for different event topologies. The solid lines show the curves  obtained from the $W^\prime$  signal efficiencies and the \Pythia background efficiencies calculated in \secref{findings_mc}. The dashed lines show the curves obtained with \PowhegBox+~\Pythia \ttbar signal efficiencies and the same \Pythia background efficiencies, thus removing the differences in background efficiencies seen in \figref{topologies_roc_fulldiffs}.} 
\label{fig:topologies_roc_signaldiffs} 
\end{figure}

The signal efficiency versus background rejection curves in the two different event topologies, including the differences in both signal and background, are shown in \figref{topologies_roc_fulldiffs}. The curves for tagging $W$-jets from the $W^{\prime}$ against a leading-jet background indicate better performance in this event topology, with the magnitude of the difference depending on the substructure variable used for tagging. Figure~\ref{fig:topologies_roc_signaldiffs} shows the curves again, but this time the leading jet from the \Pythia multijet background is used in both cases, thus removing the differences in background efficiencies, and isolating the differences resulting from the different signal event topologies. With identical background compositions, the performance is generally slightly better in the \PowhegBox \ttbar{} sample.

\begin{figure}[pt!]
\begin{center}
\begin{subfigure}[]{ \label{fig:mass_diff_mcs_200a}
\includegraphics[width=0.45\textwidth]{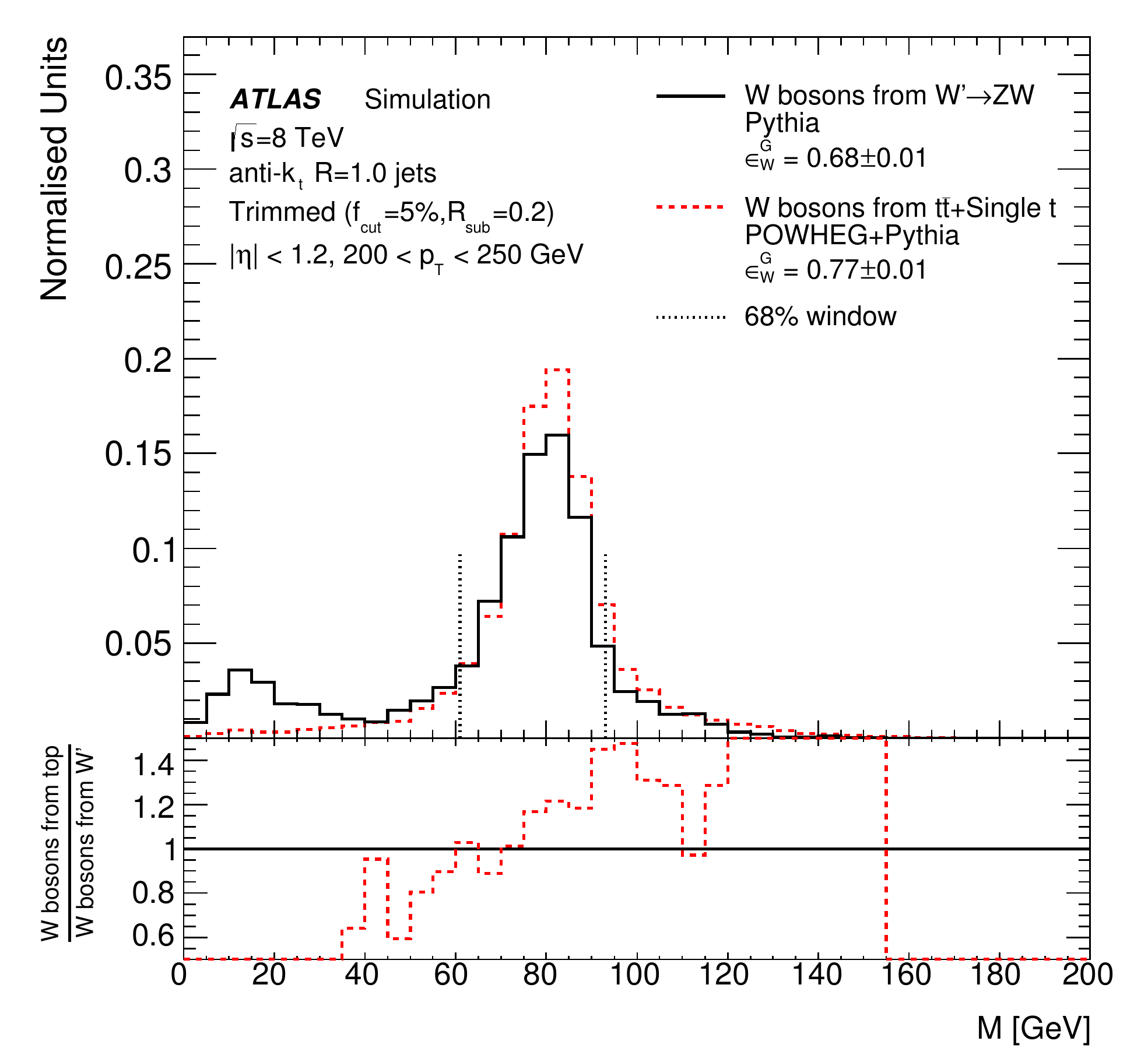}
}
\end{subfigure}
\begin{subfigure}[]{ \label{fig:mass_diff_mcs_350a}
\includegraphics[width=0.45\textwidth]{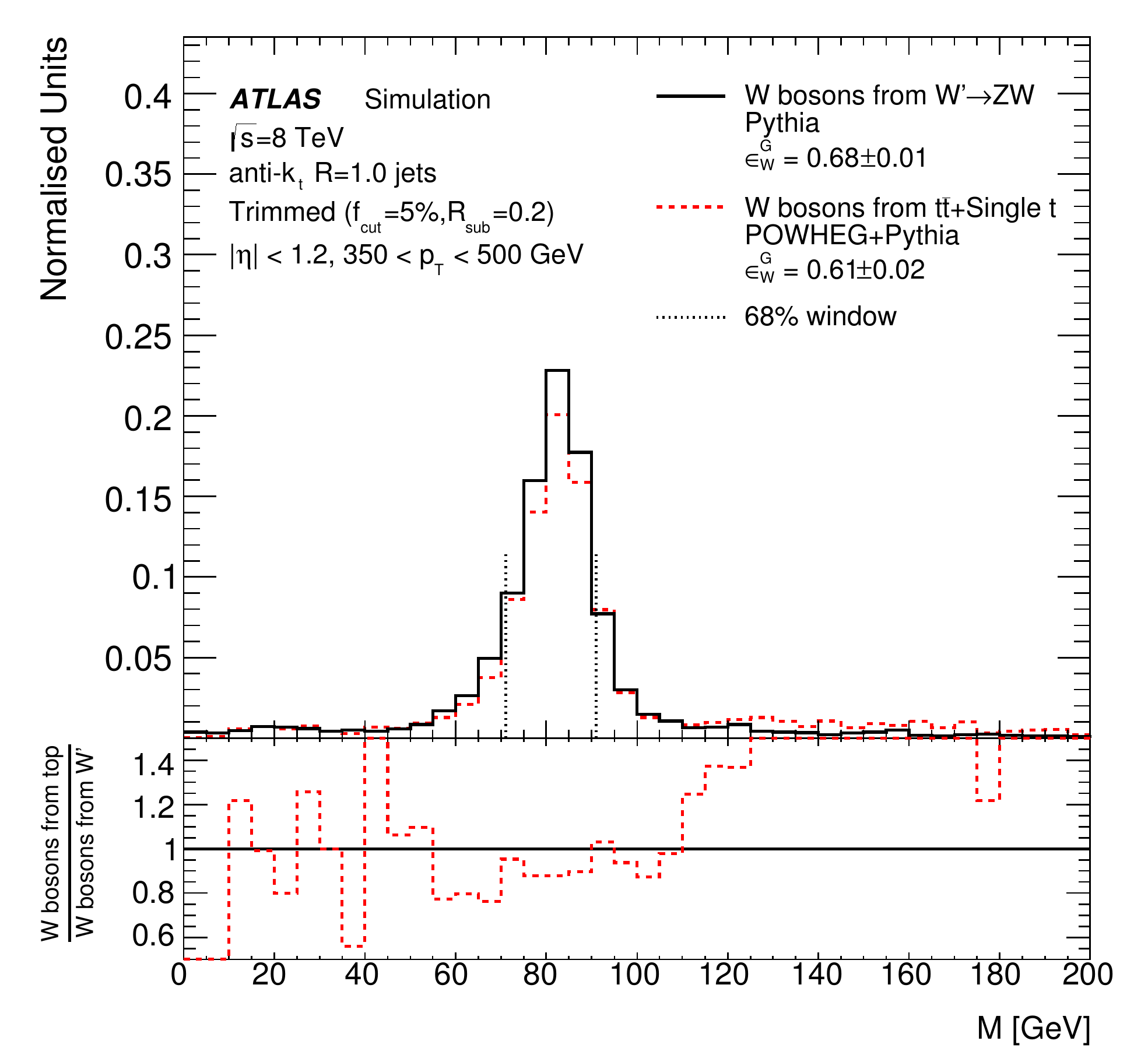}
}
\end{subfigure}
\begin{subfigure}[]{ \label{fig:mass_diff_mcs_200b}
\includegraphics[width=0.45\textwidth]{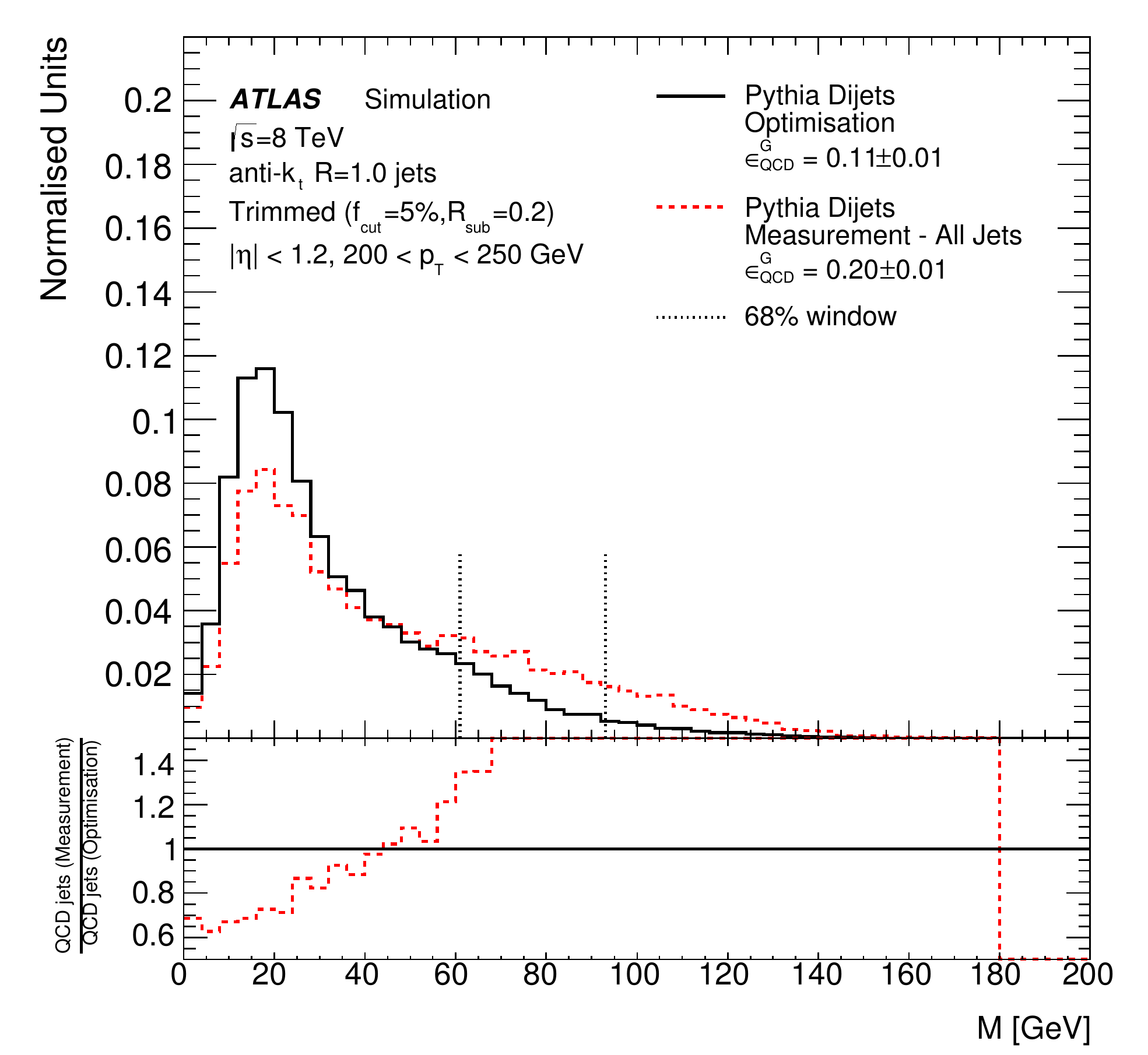}
}
\end{subfigure}
\begin{subfigure}[]{ \label{fig:mass_diff_mcs_350b}
\includegraphics[width=0.45\textwidth]{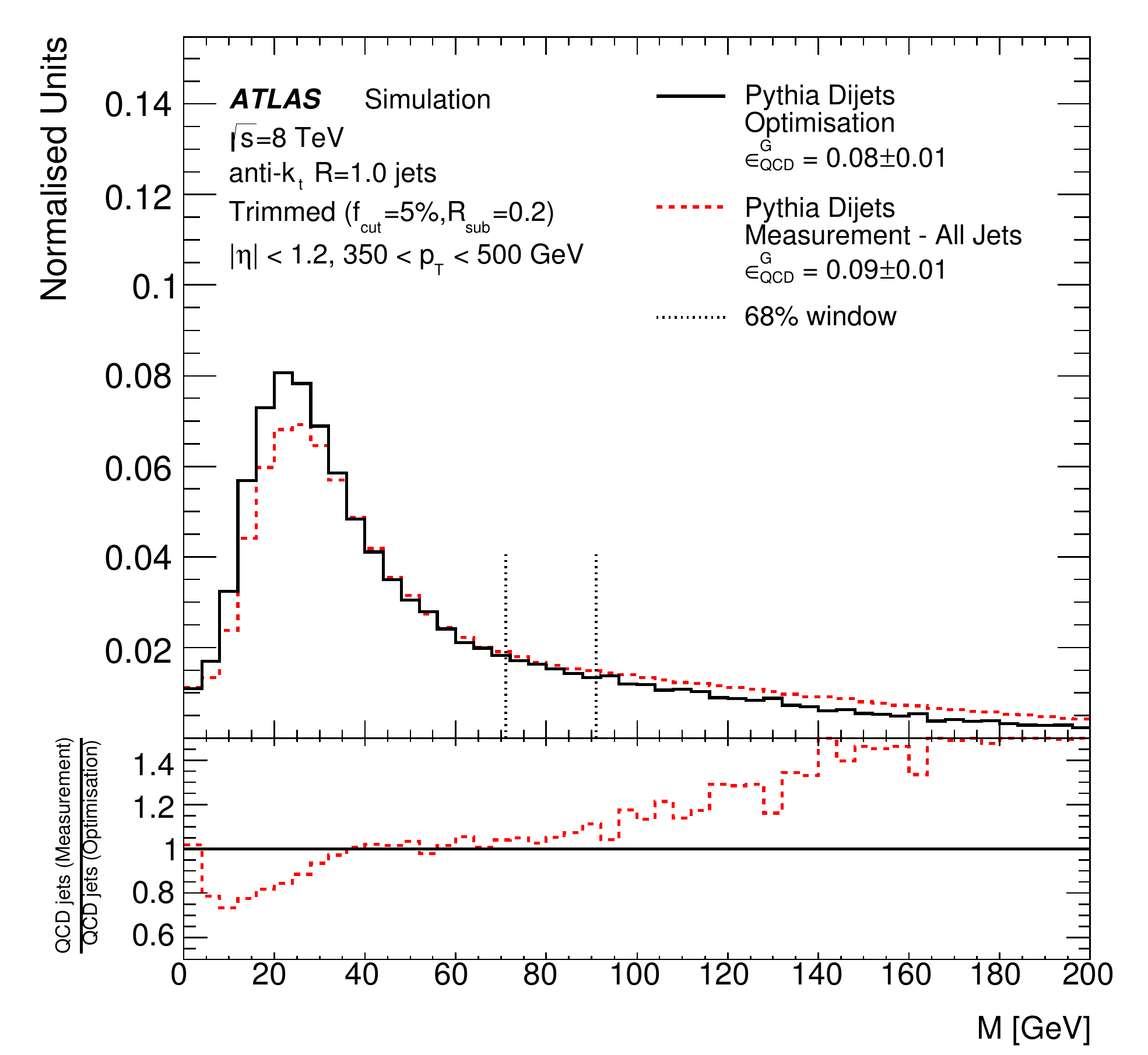}
}
\end{subfigure}
\end{center}
\vspace{-20 pt}
\caption[R2-trimmed jet mass with different MCs.]{The R2-trimmed jet mass distributions for signal $W$-jet candidates in the range (a) $200 < \pt\ < 250\GeV$, and (b) $350 < \pt\ < 500\GeV$, and multijet background candidates (c,d) in the same ranges. The $W$-jets are taken from the processes $W^{\prime}\rightarrow WZ$ (solid black), and \ttbar{} events provided by \PowhegBox (dotted red). Two kinds of \Pythia multijets are shown: the solid black line is for the leading jets only, and the dotted red line is for the leading and sub-leading jets. The ratios between the models is shown at the bottom. The inclusion of sub-leading jets, which are more likely to be initiated by gluons, results in higher-mass jets. The vertical lines represent the signal mass window.}
\label{fig:mass_diff_mcs}
\end{figure}

The mass distributions for the different signal and background samples are compared in \figref{mass_diff_mcs} for the lowest and highest \pt ranges. The signal distributions also include the R2-trimmed leading-jet mass from \ttbar events provided by \Mcatnlo+~\Herwig. The mass shape differences are less pronounced at higher $\pt$, although the difference in $\epsilon_{W}^{\mathrm{G}}$ for the different signal event topologies is still a non-negligible 10\% even in the highest \pt\ range. 

\subsection{Systematic uncertainties}
\label{sec:sysJSS}

The sources of systematic uncertainty that are common to both the signal and background efficiency measurements include the jet mass scale (JMS), jet mass resolution (JMR), jet energy scale (JES), jet energy resolution (JER) and jet substructure variable (JSS). 

The uncertainty on the JER is taken from previous studies~\cite{Aad:2012ag} and is parameterised as a function of \pt{}.  The size of JER uncertainty is approximately 10\% for the \pt\ ranges presented here. The uncertainty on JMR is also taken from previous studies~\cite{Aad:2013gja}, where it was determined from the data/MC variations in the widths of the $W$-jet mass peaks in \ttbar events, and is fixed at 20\%. The JMS, JES and JSS are varied up and down by $\pm 1 \sigma$, using the standard deviation derived from the double-ratio method; this is described in detail below using the JSS as an example.

The systematic uncertainty on the JSS is needed in order to derive the full systematic uncertainties on the signal and background efficiencies. Uncertainties are derived using in-situ methods by comparing the measured calorimeter jet energy, mass and substructure variables to the same quantities measured by well-calibrated and completely independent detectors in both data and MC, using the double ratio: 
\begin{equation}
\langle{X^{\rm jet} / X^{\rm ref}}\rangle_{\rm data} / \langle{X^{\rm jet} / X^{\rm ref}}\rangle_{\rm MC}\ ,
\label{eq:doubleRatioLargeR}
\end{equation}
where $X$ denotes a jet variable. In this case, track-jets are used as reference objects, since tracks from charged hadrons are well-measured and are independent of the calorimeter. In addition, the use of track-jets, where tracks are required to come from the hard scattering vertex, suppresses pileup effects. A geometrical matching in the $\eta$--$\varphi$ plane is applied to associate track-jets with calorimeter-jets. This approach was widely used in the measurement of the jet mass and substructure properties of jets in the 2011 data~\cite{Aad:2013gja}. Performance studies have also shown that there is excellent agreement between the measured positions of clusters and tracks in data, indicating no systematic misalignment between the calorimeter and the inner detector. This technique achieves a precision of around 3--7\% in the central detector region, which is dominated by systematic uncertainties arising from the inner-detector tracking efficiency and MC modelling uncertainties of the charged and neutral components of jets. 

The double ratio of \equref{doubleRatioLargeR} is computed for two different MC generators, \Pythia and \Herwigpp, and the largest disagreement between data and each of the MC generators is taken as a modelling uncertainty. The total uncertainty is then obtained by adding in quadrature this modelling uncertainty to the tracking efficiency uncertainty. Specific uncertainties for tracks inside the core of dense jets are not needed here, because only jets with $\pT < 1\TeV$ are considered.
The scale uncertainties for the jet energy, mass and substructure variables  are derived in ranges of the \pt, $\eta$, and $M/\pt$ of the reconstructed calorimeter jet.

\begin{figure}[pt!]
\begin{center}
\begin{subfigure}[]{\label{double_ratio_c2_a}
\includegraphics[width=0.45\textwidth]{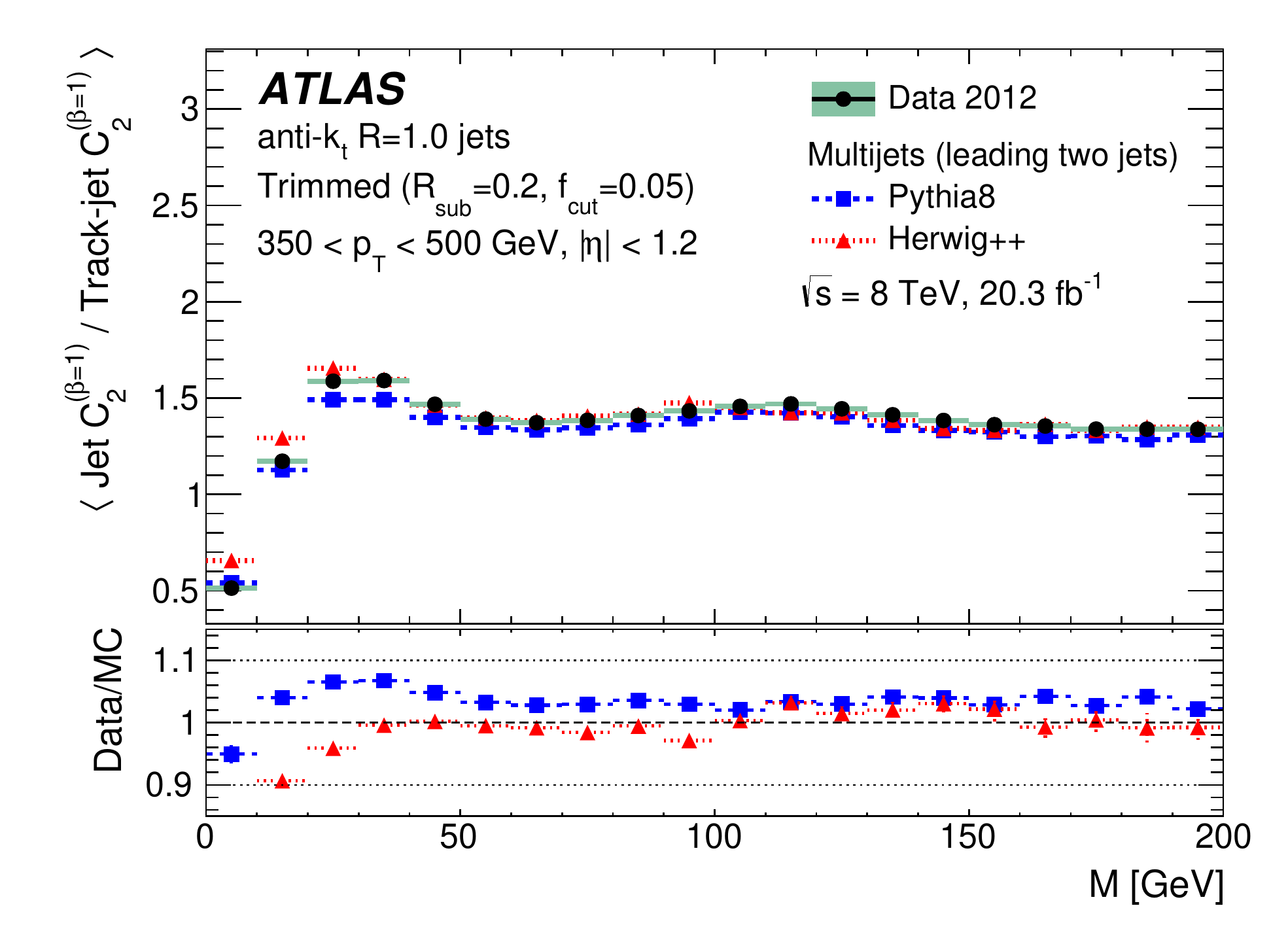}
}
\end{subfigure}
\begin{subfigure}[]{\label{double_ratio_c2_}
\includegraphics[width=0.45\textwidth]{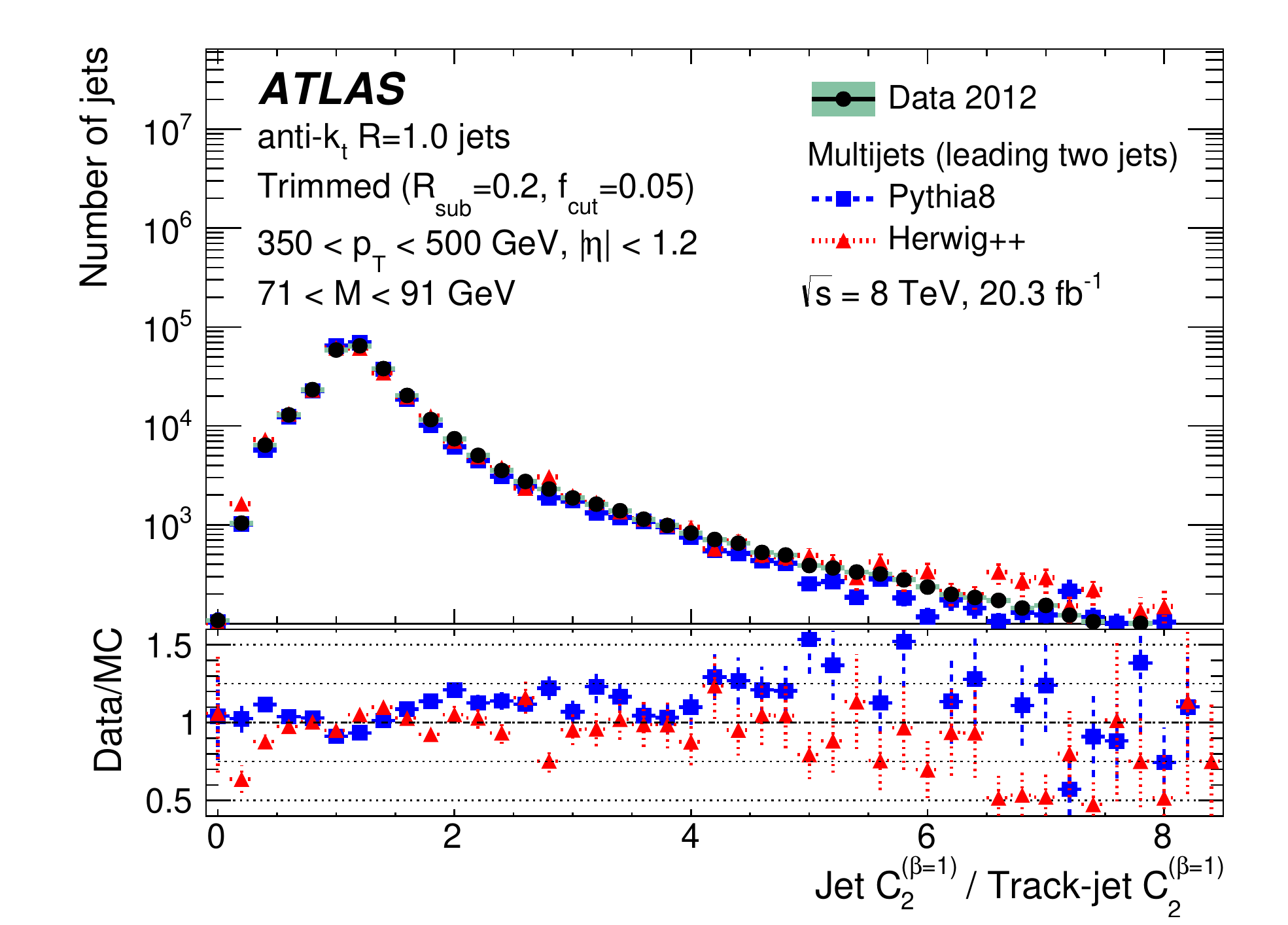}
}
\end{subfigure}
\begin{subfigure}[]{\label{double_ratio_d2_a}
\includegraphics[width=0.45\textwidth]{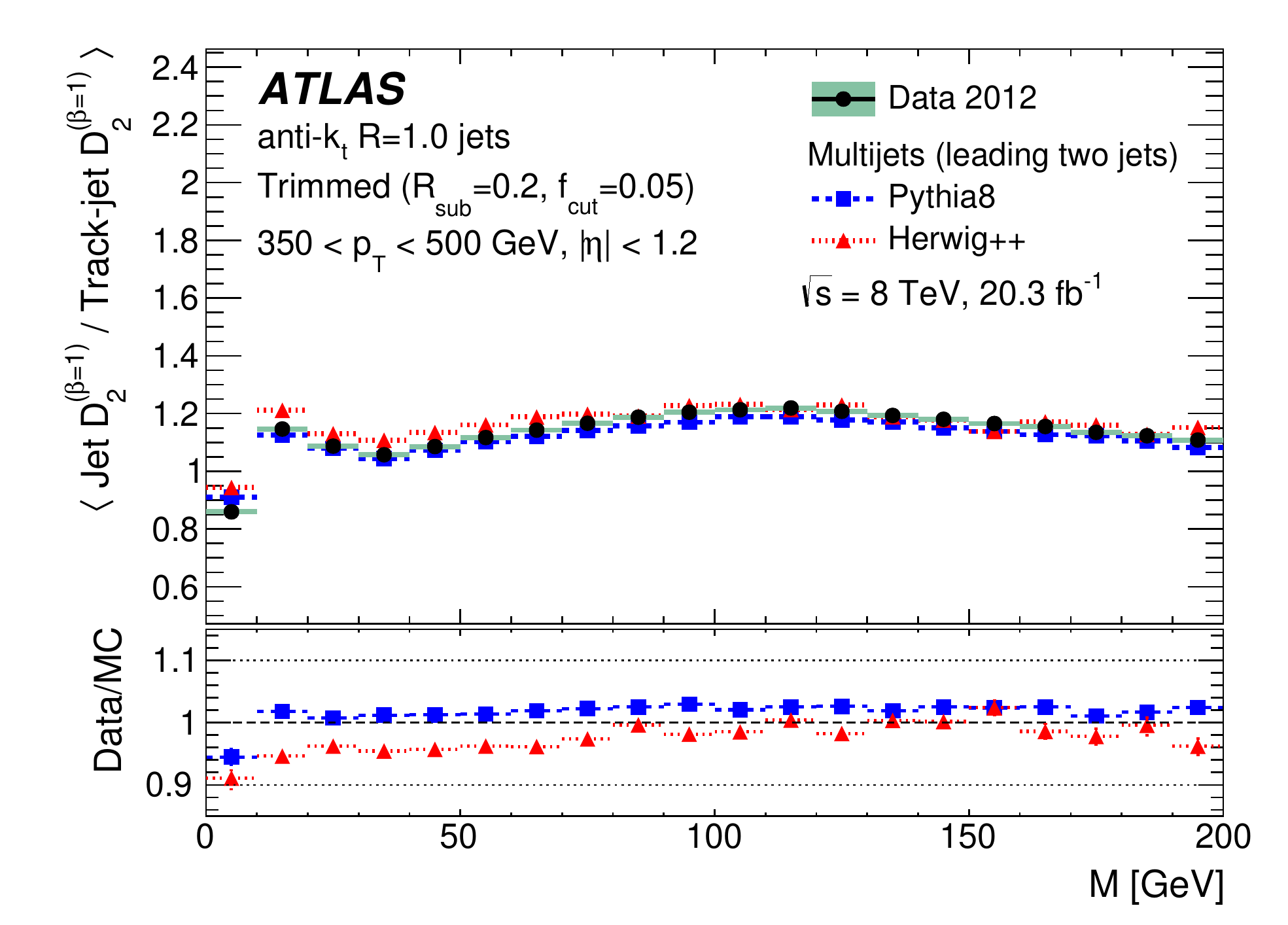}
}
\end{subfigure}
\begin{subfigure}[]{\label{double_ratio_d2_b}
\includegraphics[width=0.45\textwidth]{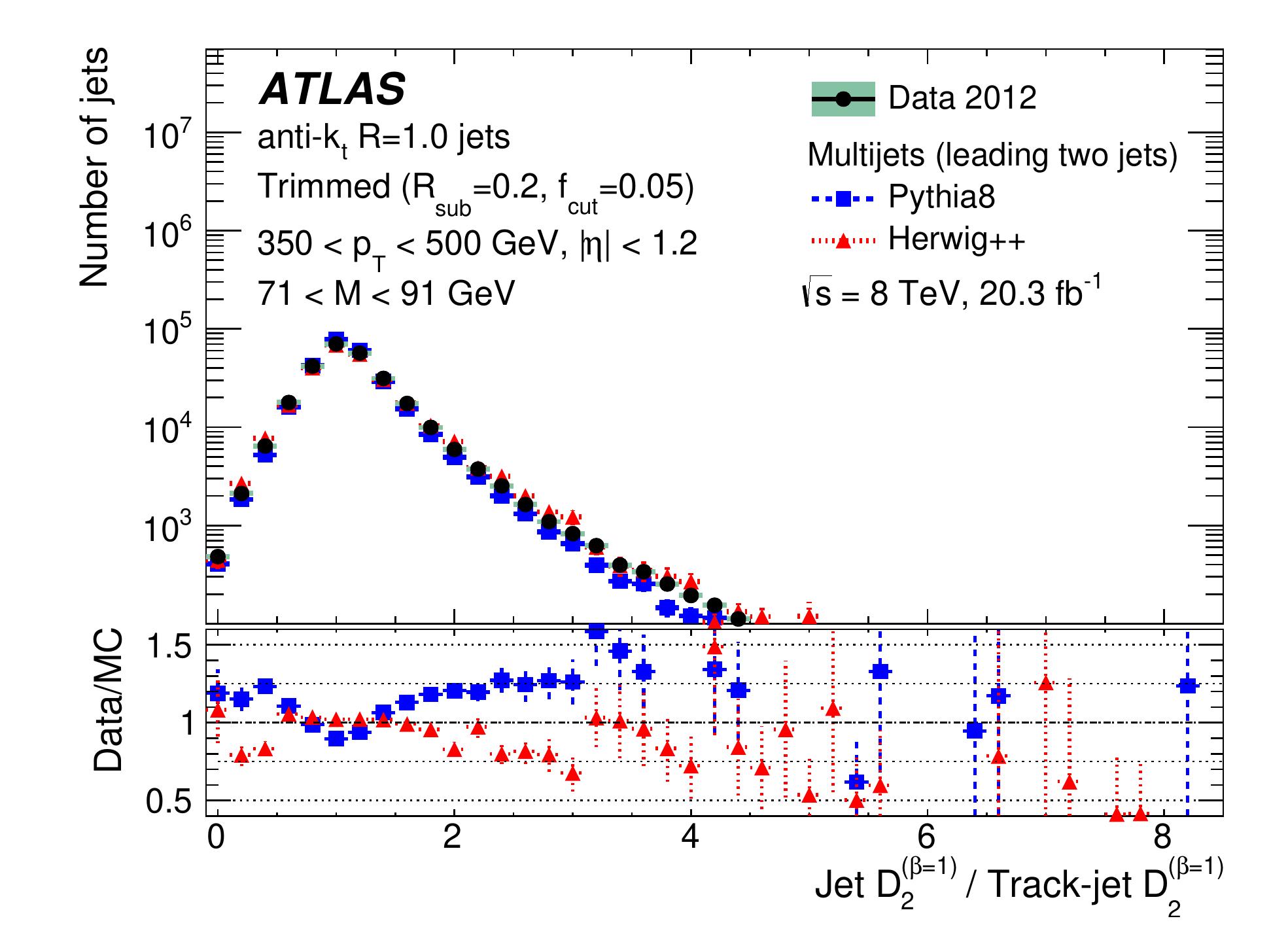}
}
\end{subfigure}
\begin{subfigure}[]{\label{double_ratio_tau21_a}
\includegraphics[width=0.45\textwidth]{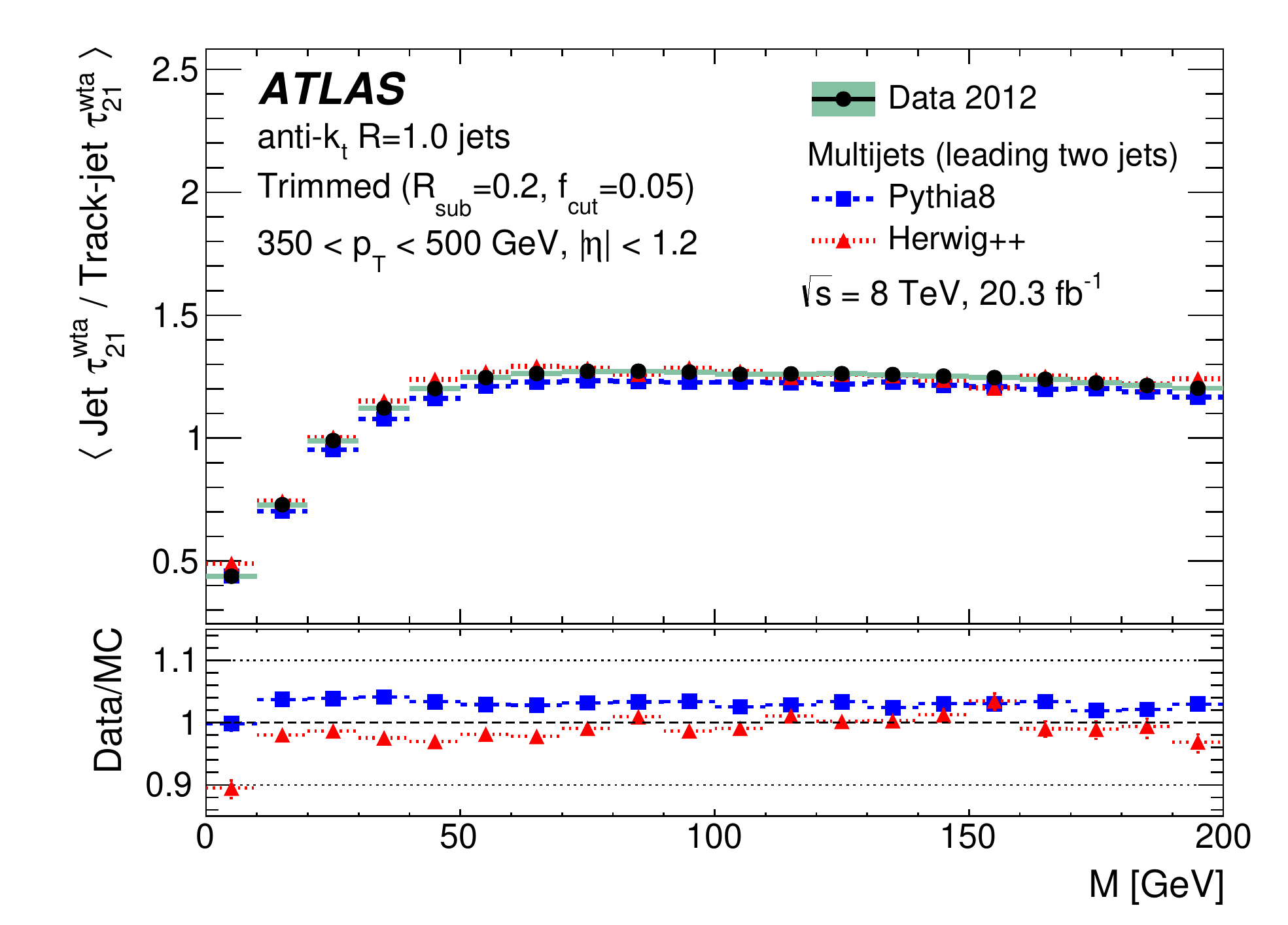}
}
\end{subfigure}
\begin{subfigure}[]{\label{double_ratio_tau21_b}
\includegraphics[width=0.45\textwidth]{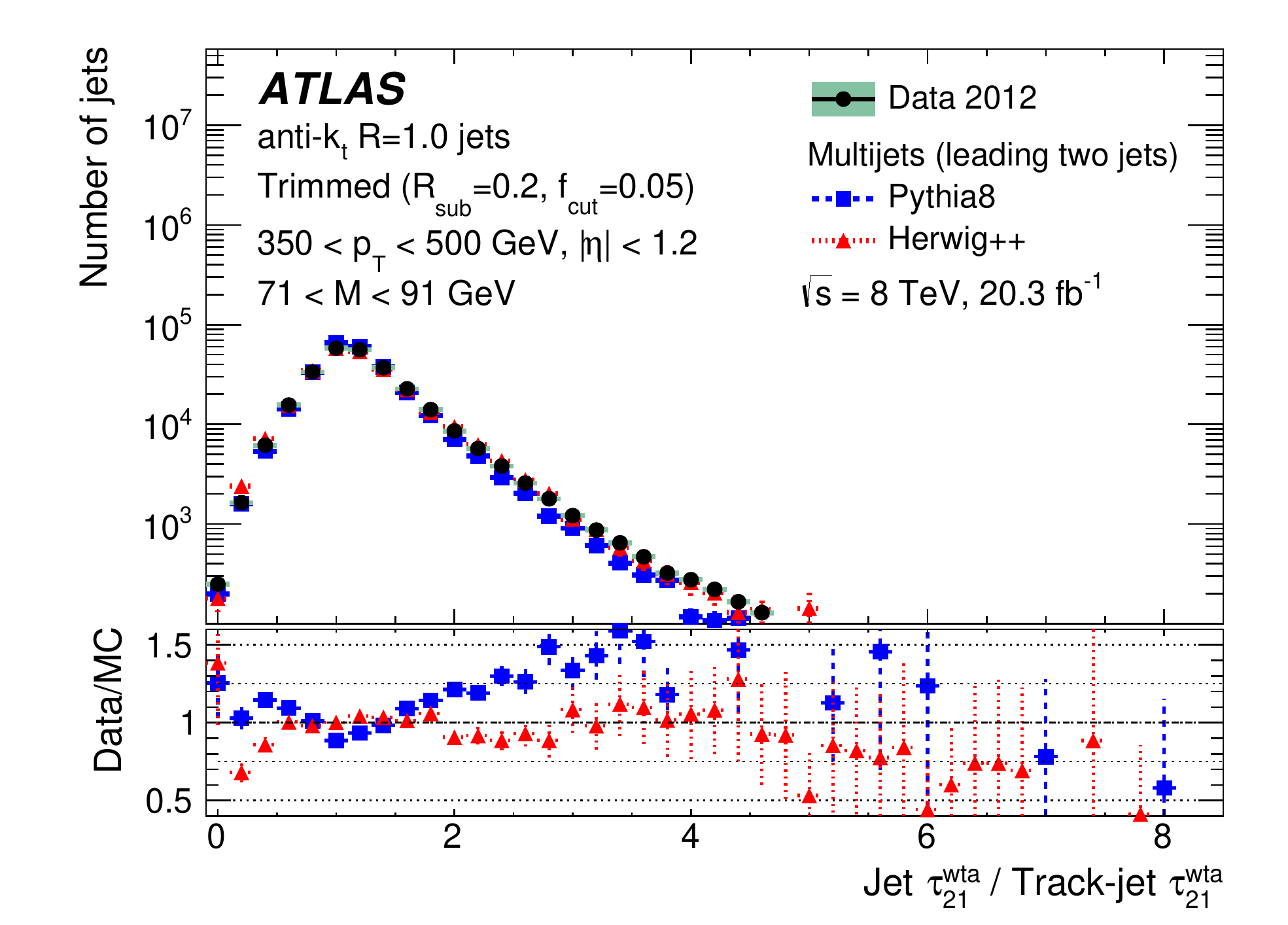}
}
\end{subfigure}
\end{center}
\vspace{-20 pt}
\caption[Mean calo/track ratios as a function of mass. 
]{Left: Distributions of the mean calorimeter-jet / track-jet ratios as a 
function of the R2-trimmed jet mass for three tagging variables. 
Right: distribution of these ratios for the three variables in data 
compared to the \Pythia and \Herwigpp models. (a), (b): 
\CTwoBetaOne, (c), (d): \DTwoBetaOne and (e), (f): \TauTwoOneWTA. The distributions are shown for R2-trimmed jets in the central calorimeter region, $|\eta| < 1.2$ and in the range $350 < p_{\rm T} < 500\GeV$. The data/MC comparisons (the `double-ratios') for \Pythia (blue dashed) and \Herwigpp (red dotted) are shown in the lower panel of each plot.}
\label{fig:JMS_double_ratio_vs_m}
\end{figure}

\figref{JMS_double_ratio_vs_m} shows a set of six representative distributions for \CTwoBetaOne, \DTwoBetaOne and \TauTwoOneWTA in the range $350 < \pt < 500\GeV$. The mean values of the single-ratio $X^{\rm jet} / X^{\rm ref}$ distributions are shown as a function of the jet mass, along with the distributions of $X^{\rm jet} / X^{\rm ref}$ themselves within the relevant $\epsilon_{W}^{\mathrm{G}}\sim68\%$ mass window. 

Large discrepancies between data and MC are observed for low-mass jets, while for masses around 80$\GeV$ the data/MC agreement is within 5\%. 
In the distributions of $X^{\rm jet} / X^{\rm ref}$ it is noticed that 
while the tails of the ratio distributions show discrepancies between data and the MC, the agreement is good for values of the ratio close to one, which represents the large majority of events.
In summary, the scale uncertainty of the three jet substructure variables ranges between 1\% and 5\% in the different kinematic regions.

Additional, sub-dominant systematic uncertainties come from MC sources listed in \Tabref{sys_eff_D2} and described in \secref{sysEff} in terms of the uncertainty on the final measured signal and background efficiencies. The full systematic uncertainty on the mass and substructure variables are obtained by adding each of the scale, resolution, statistical and MC uncertainties in quadrature.

\subsection{Mass and substructure distributions in \ttbar events}
\label{sec:datamc_distributions}

\begin{figure}[pt!]
\begin{center}
\includegraphics[width=0.65\textwidth,angle=0]{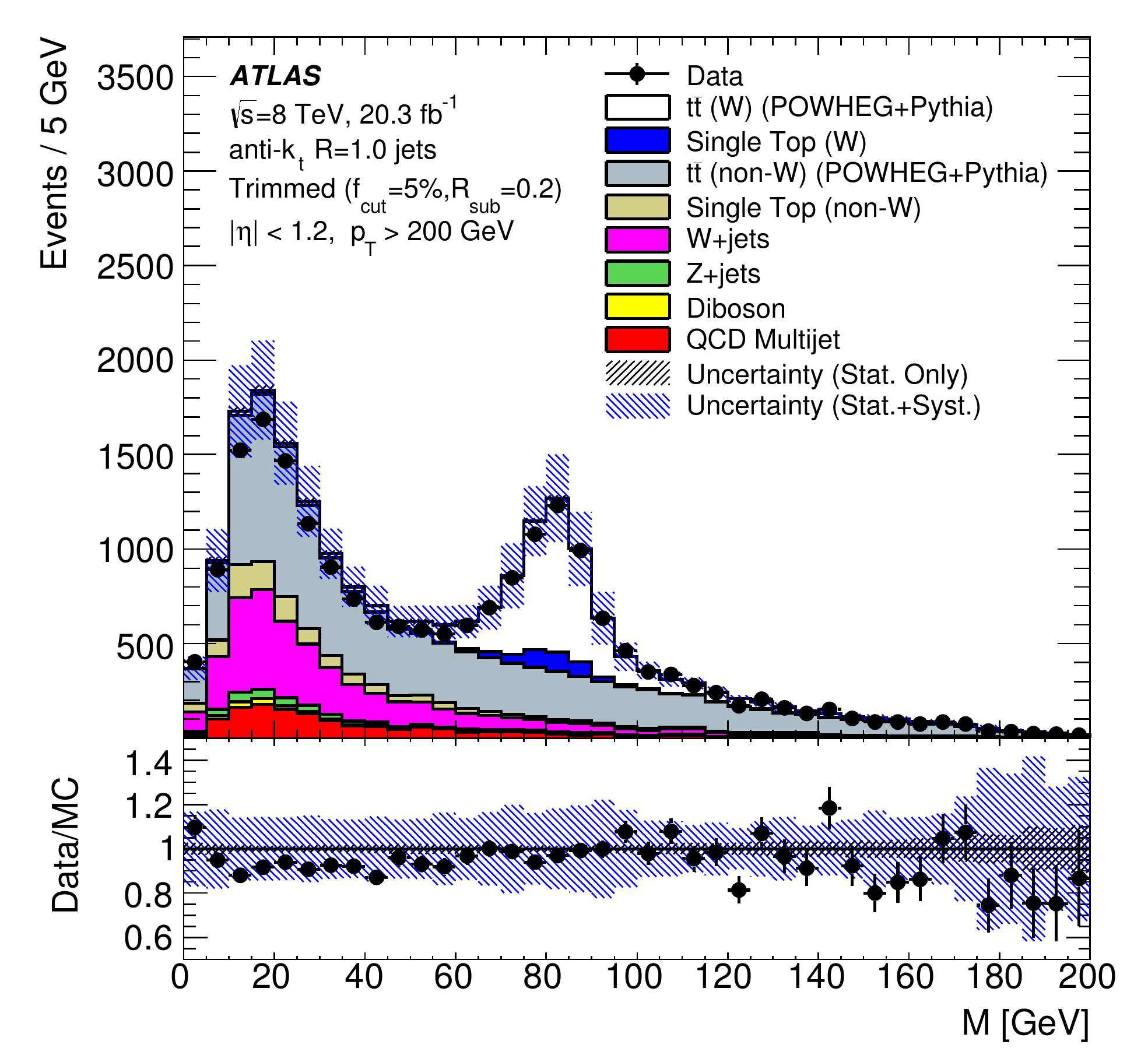}
\end{center}
\vspace{-20 pt}
\caption[Data and \PowhegBox+~\Pythia distributions in pre-tag sample.]{Distribution of the $W$ candidate jet mass for selected lepton+jets \ttbar events in data and \PowhegBox~+~\Pythia MC  for the combined electron and muon channel. Data points are shown with statistical uncertainties, and the combined MC is shown with full systematic and statistical uncertainties. The lower panel shows the data/MC ratio, with the statistical uncertainty on the MC given in the black forward-slashed band, and the full systematic uncertainty given in the blue, back-slashed band.}
\label{fig:tt_pretag1}
\end{figure}

\begin{figure}[pt!]
\begin{center}
\begin{subfigure}[]{\label{pretag_c2}
\includegraphics[width=0.4\textwidth,angle=0]{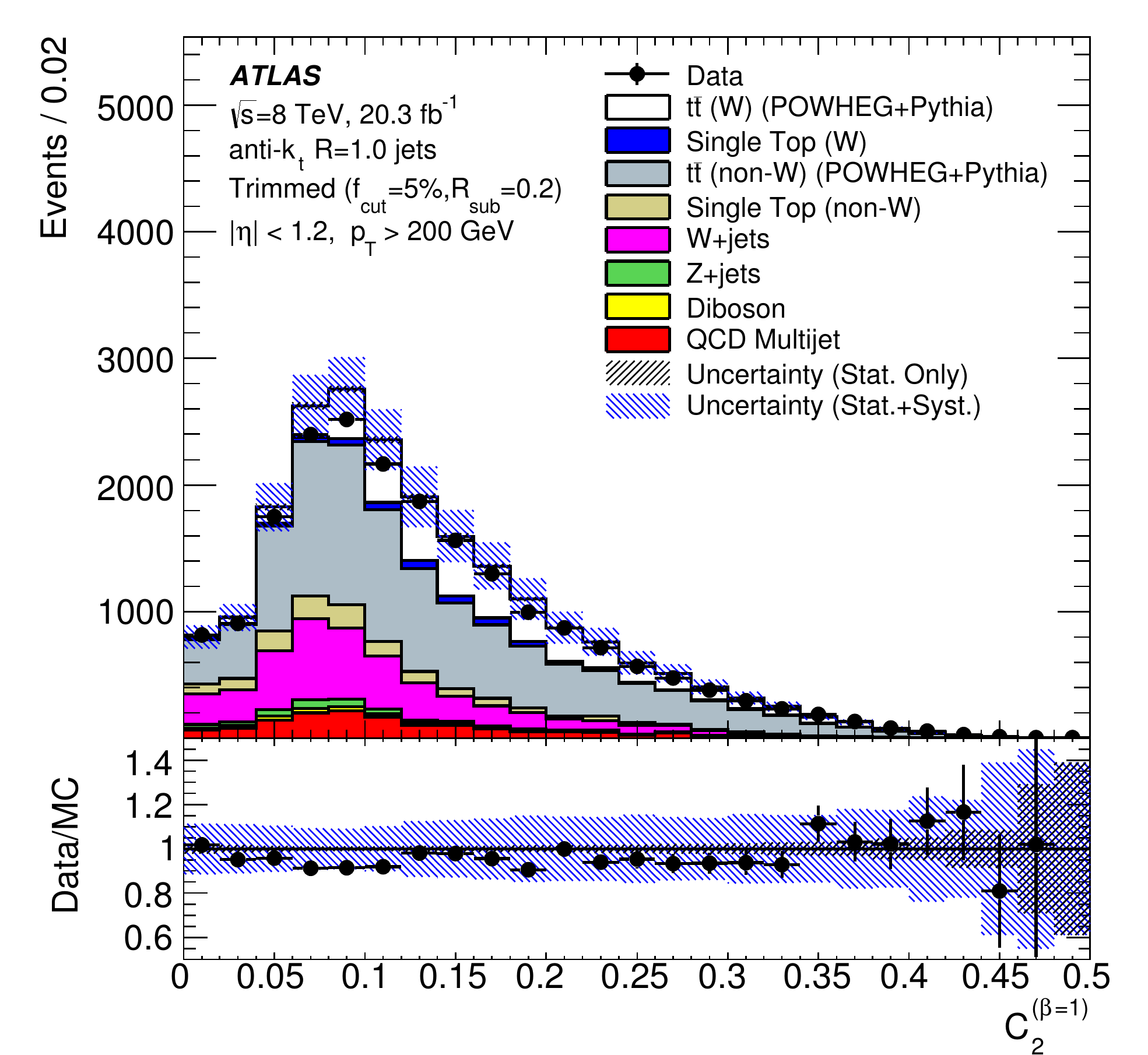}
}
\end{subfigure}
\begin{subfigure}[]{\label{posttag_c2}
\includegraphics[width=0.4\textwidth,angle=0]{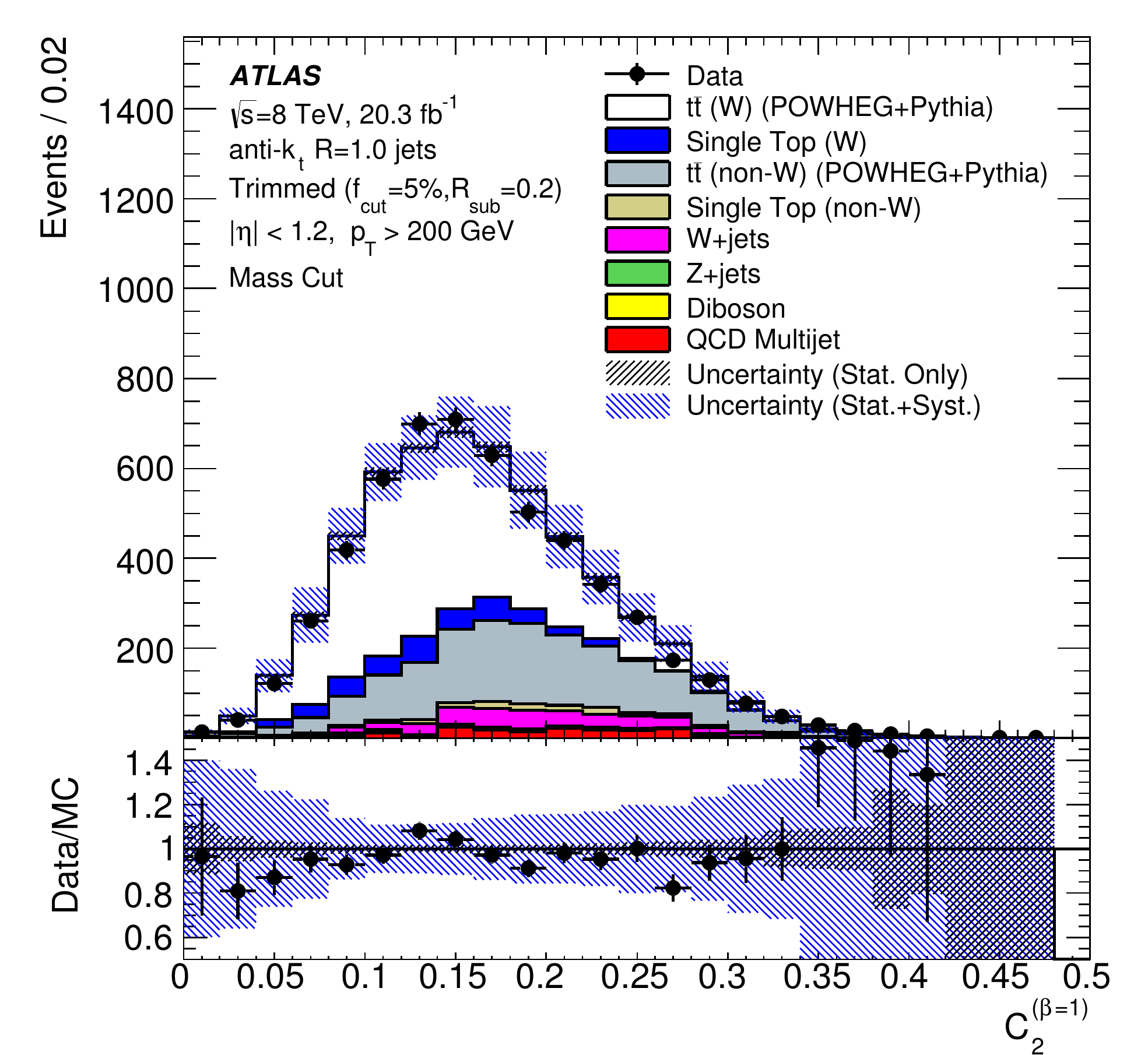}
}
\end{subfigure}
\begin{subfigure}[]{\label{pretag_d2}
\includegraphics[width=0.4\textwidth,angle=0]{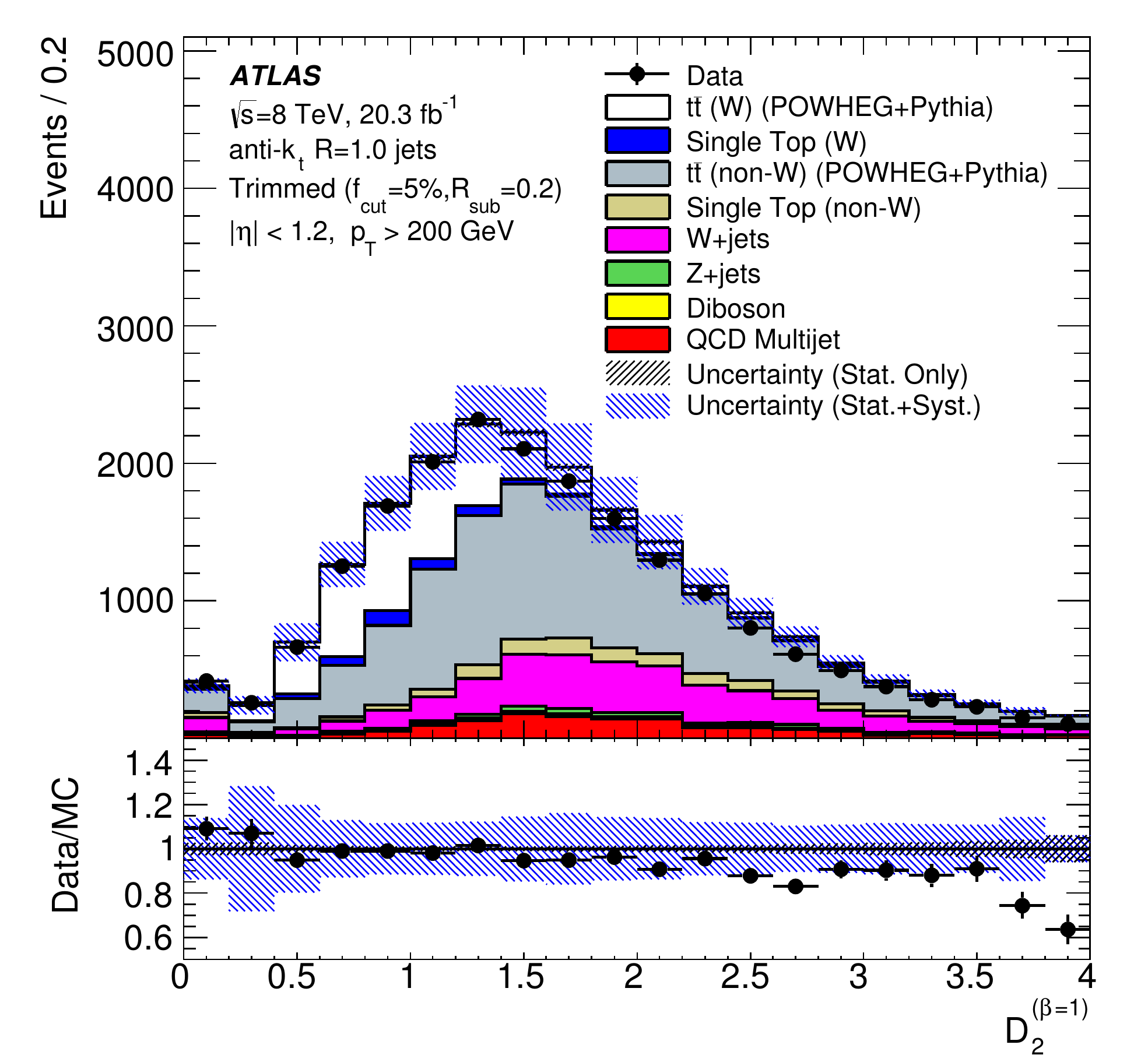}
}
\end{subfigure}
\begin{subfigure}[]{\label{posttag_d2}
\includegraphics[width=0.4\textwidth,angle=0]{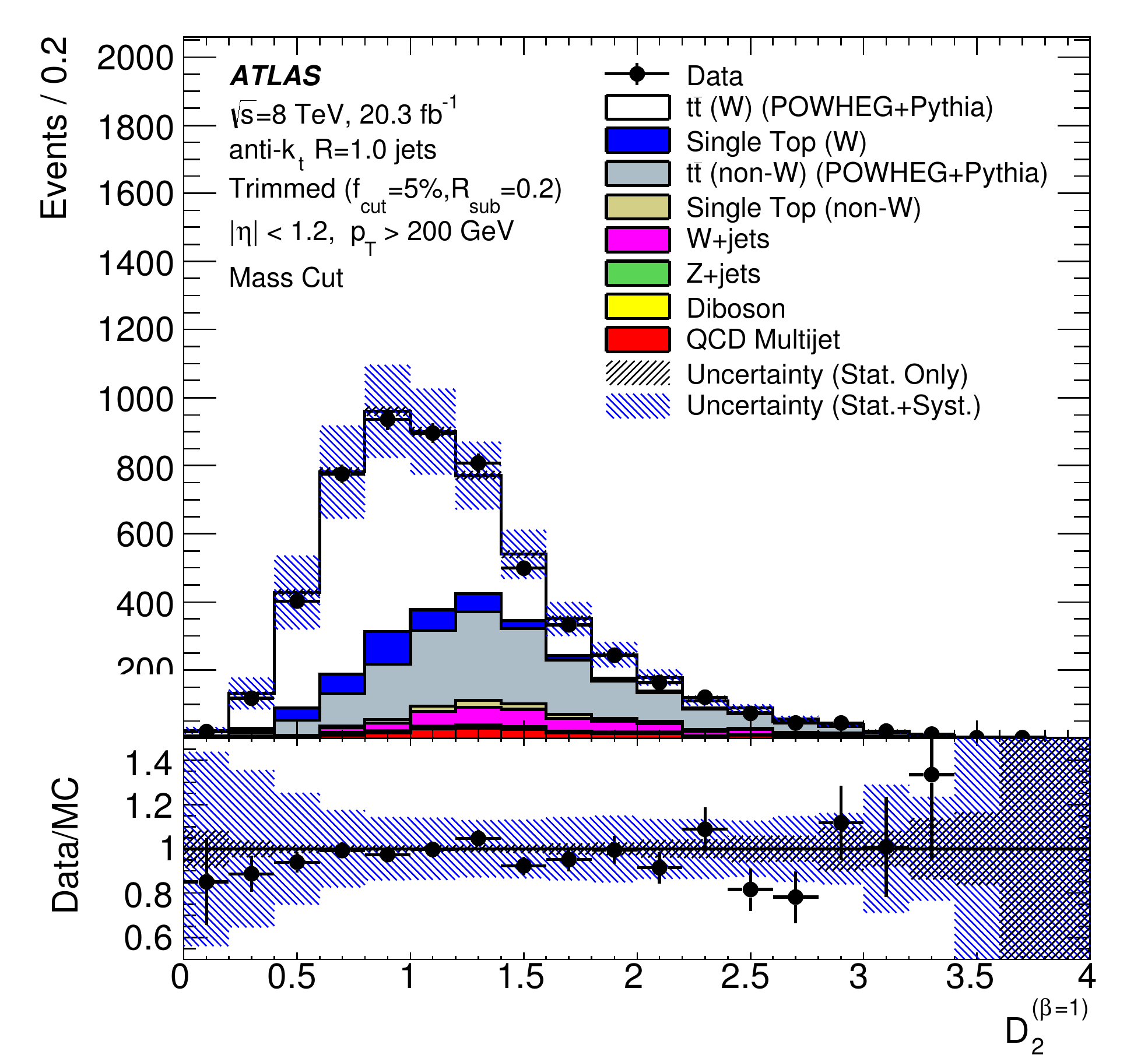}
}
\end{subfigure}
\begin{subfigure}[]{\label{pretag_tau21}
\includegraphics[width=0.4\textwidth,angle=0]{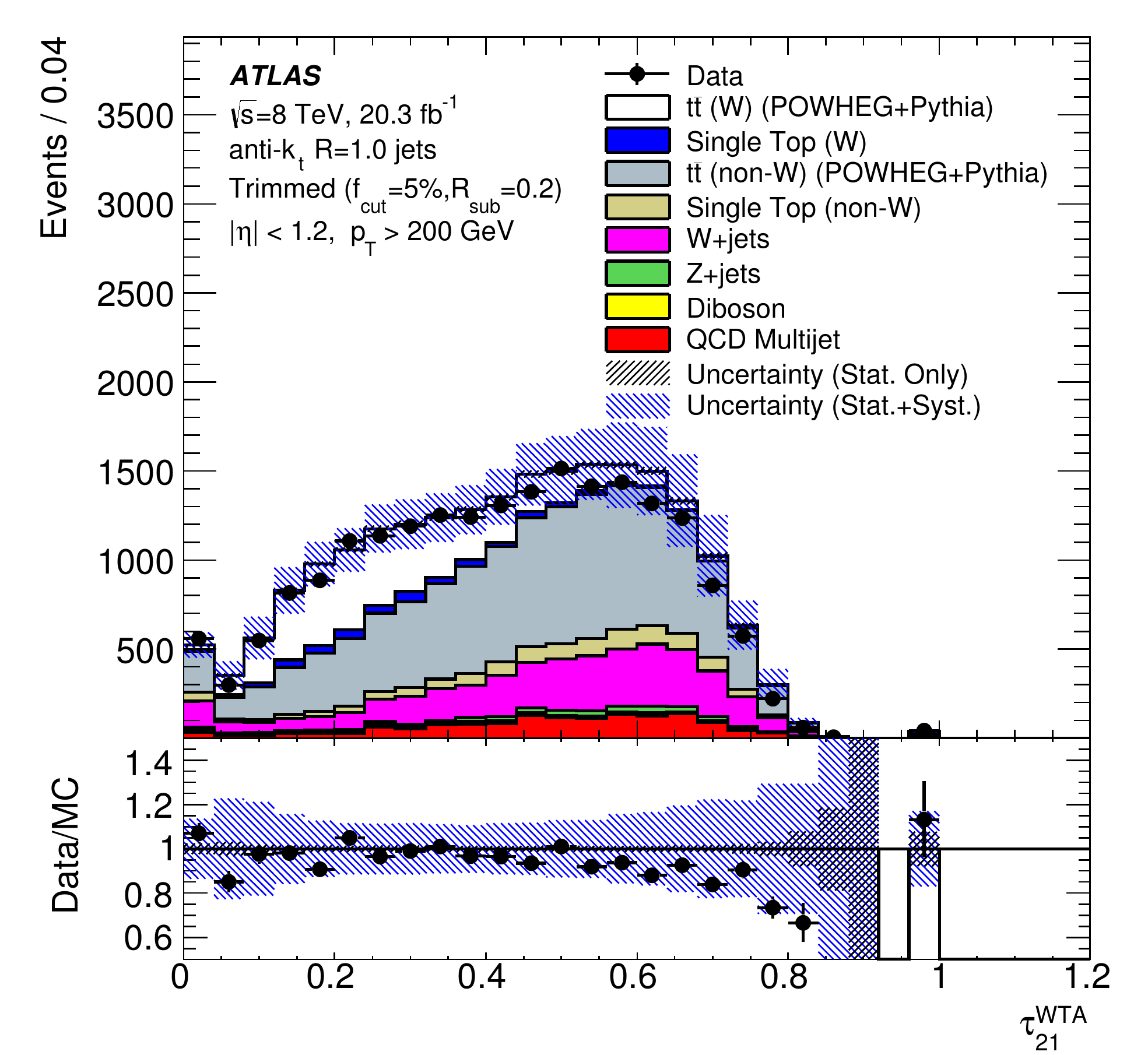}
}
\end{subfigure}
\begin{subfigure}[]{\label{posttag_tau21}
\includegraphics[width=0.4\textwidth,angle=0]{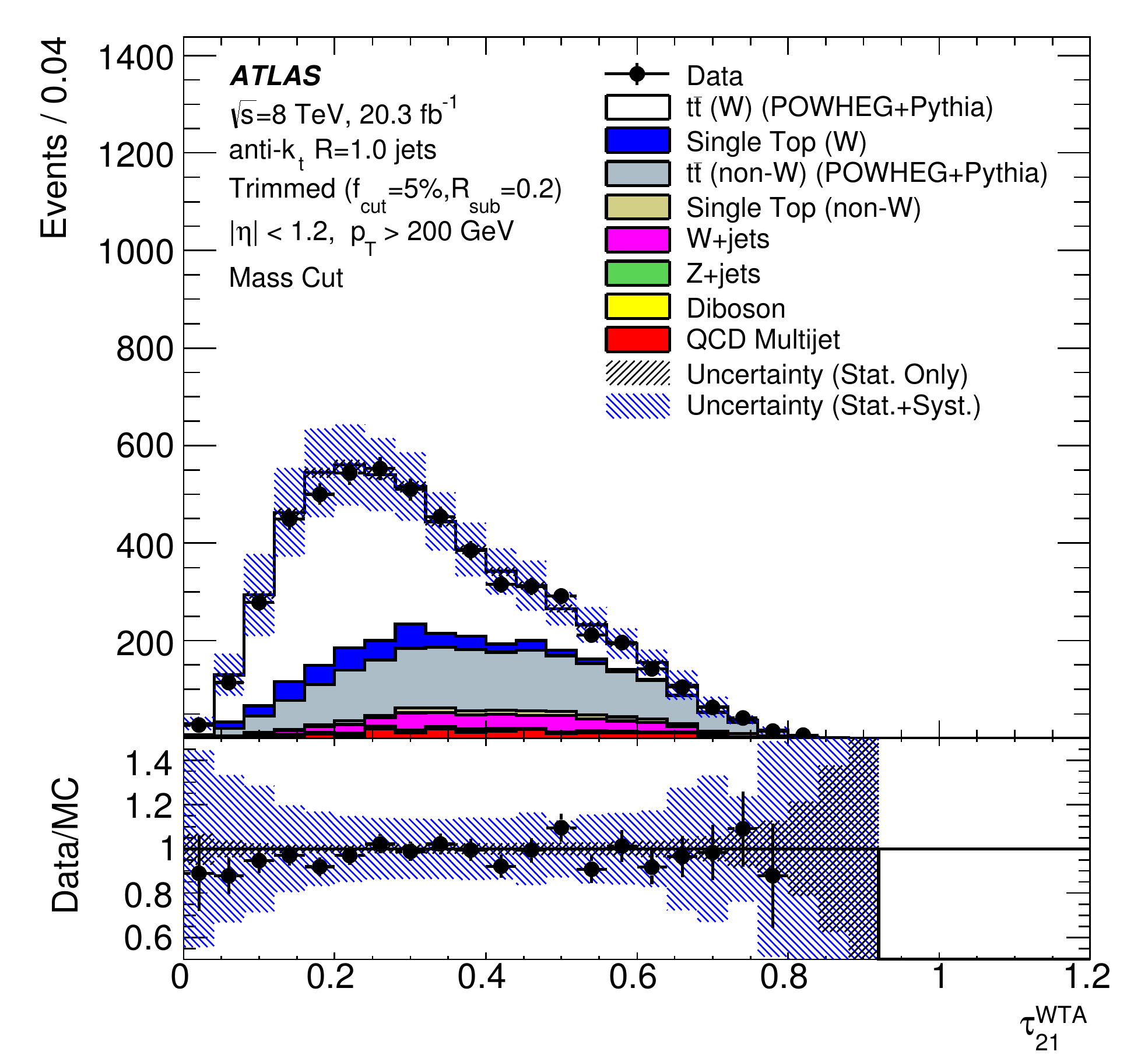}
}
\end{subfigure}
\end{center}
\vspace{-20 pt}
\caption[Data and \PowhegBox+~\Pythia distributions pre-tag and in mass window.]{Distributions of the $W$ candidate jet substructure variables before (left) and after (right) the $\epsilon_{W}^{\mathrm{G}} = 68\%$ mass window for selected lepton+jets \ttbar events in data and \PowhegBox~+~\Pythia MC for the combined electron and muon channel. (a), (b): \CTwoBetaOne, (c), (d): \DTwoBetaOne and (e), (f) \TauTwoOneWTA. Data points are shown with statistical uncertainties, and the combined MC is shown with full systematic and statistical uncertainties. The lower panels show the data/MC ratios, with the statistical uncertainty on the MC given in black forward-slashed bands, and the full systematic uncertainty given in the blue, back-slashed bands.}
\label{fig:tt_postgroom}
\end{figure}

The jet mass distribution for the leading R2-trimmed jets in events satisfying the pre-selection criteria in \secref{objects} are shown in \figref{tt_pretag1}. The data and events in \PowhegBox+~\Pythia and \Mcatnlo+~\Herwig simulations agree within the uncertainties detailed in \secref{sysJSS}. Distributions of the three tagging variables \CTwoBetaOne, \DTwoBetaOne and \TauTwoOneWTA are shown for the same pre-selection criteria, before and after making the relevant $\epsilon_{W}^{\mathrm{G}} = 68\%$ mass window requirements for the \pt range in question, in \figref{tt_postgroom}. These variables are used to define medium and tight tagging criteria, where the medium working point provides a signal efficiency of $\epsilon_{W}^{\mathrm{G}\&\mathrm{T}} = 50\%$ and the tight working point provides $\epsilon_{W}^{\mathrm{G}\&\mathrm{T}} = 25\%$. 

\begin{figure}[pt!]
\begin{center}
\begin{subfigure}[]{\label{fig:c2_tagged_fail}
\includegraphics[width=0.4\textwidth]{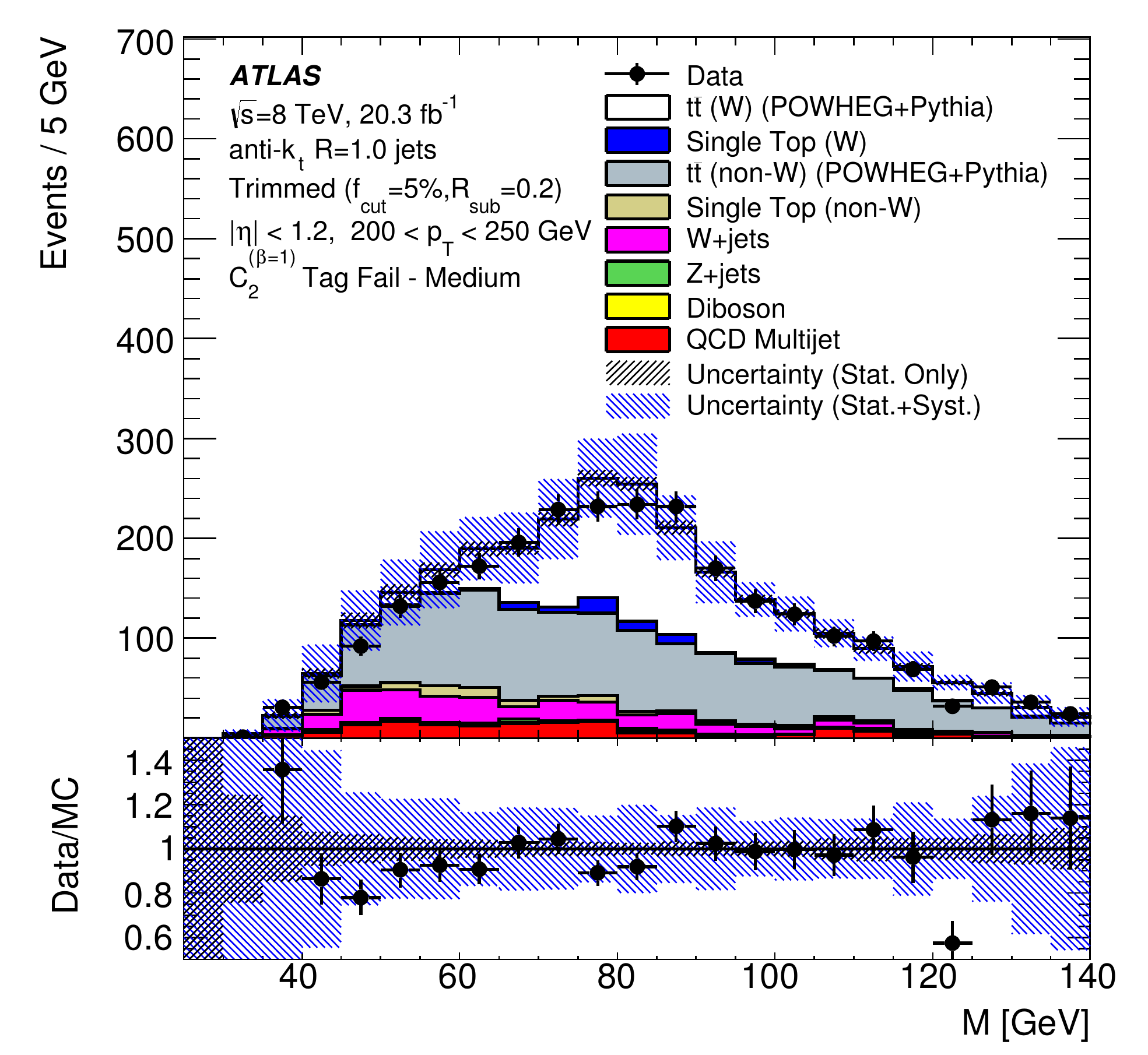}
}
\end{subfigure}
\begin{subfigure}[]{\label{fig:c2_tagged_pass}
\includegraphics[width=0.4\textwidth]{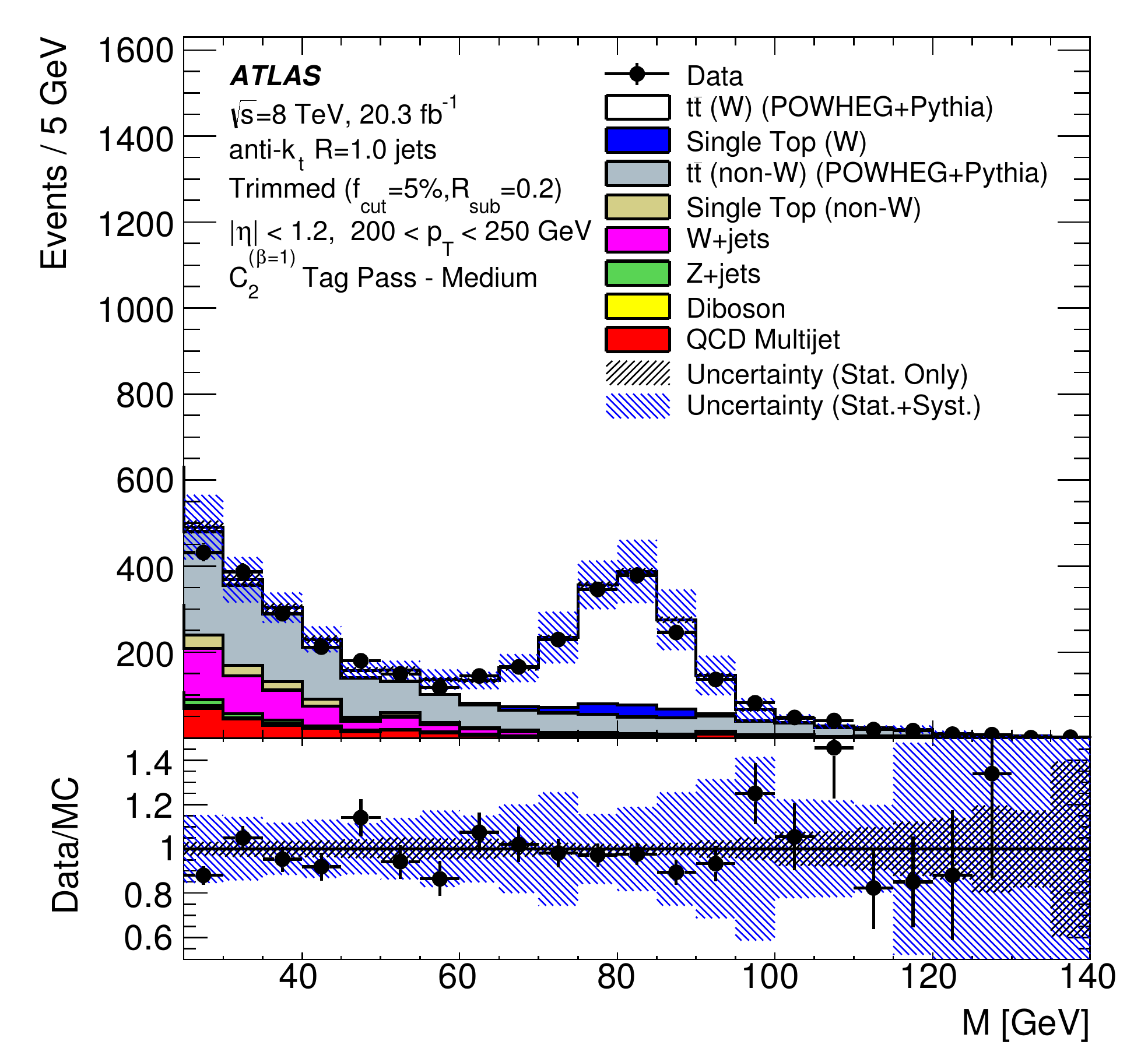}
}
\end{subfigure}
\begin{subfigure}[]{\label{fig:d2_tagged_fail}
\includegraphics[width=0.4\textwidth]{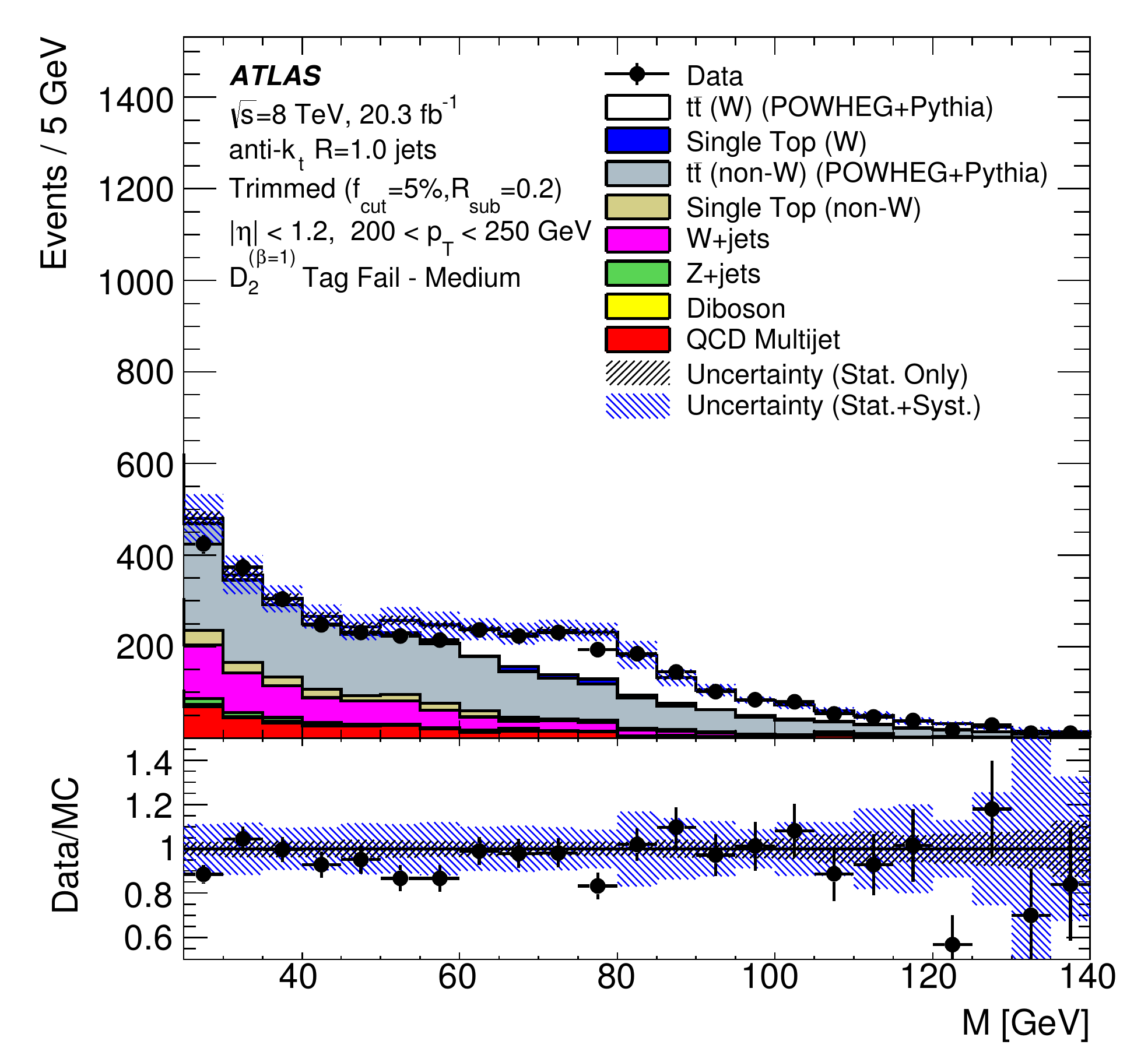}
}
\end{subfigure}
\begin{subfigure}[]{\label{fig:d2_tagged_passed}
\includegraphics[width=0.4\textwidth]{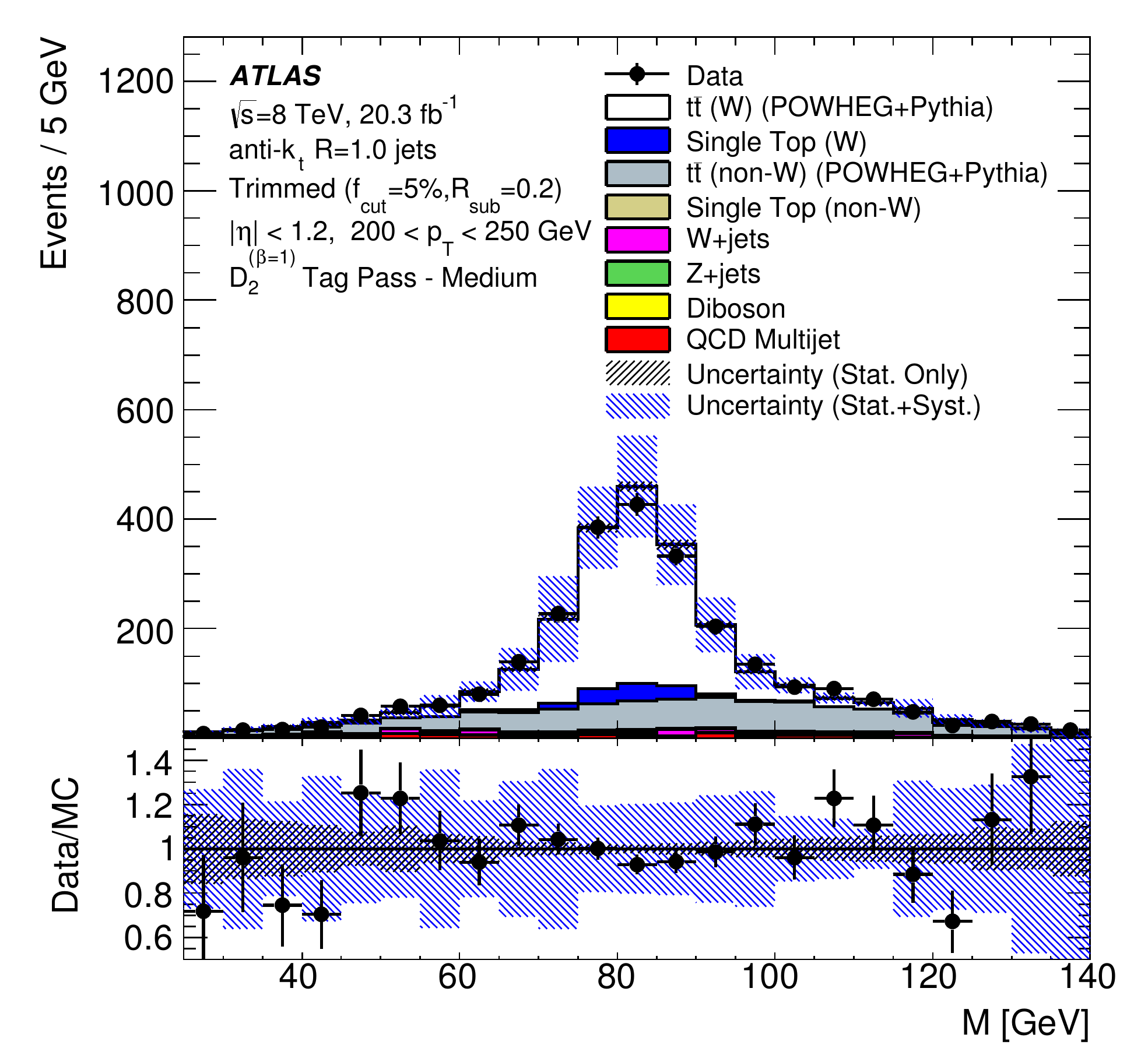}
}
\end{subfigure}
\begin{subfigure}[]{\label{fig:tau21_tagged_failed} 
\includegraphics[width=0.4\textwidth]{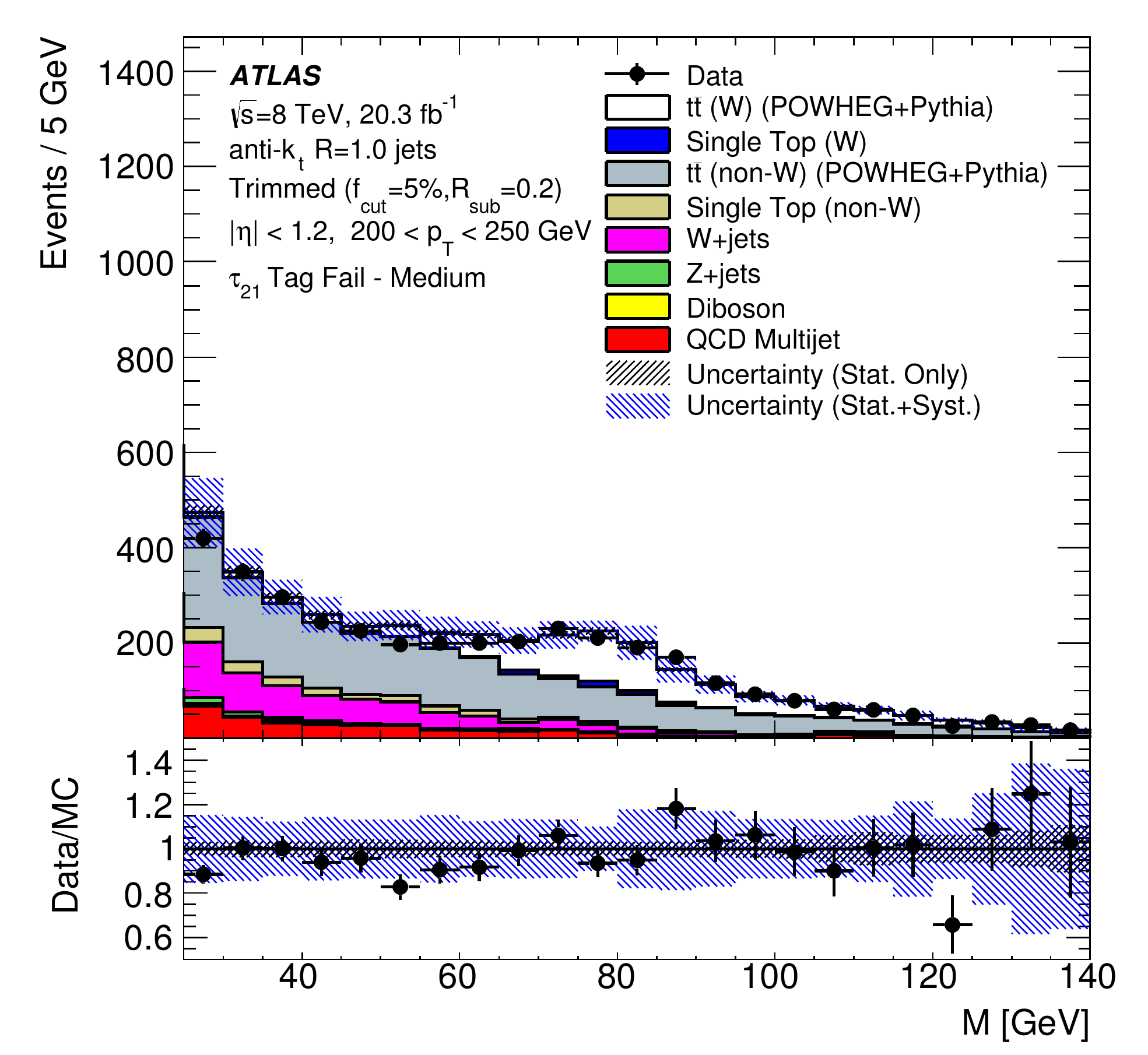}
}
\end{subfigure}
\begin{subfigure}[]{\label{fig:tau21_tagged_passed}
\includegraphics[width=0.4\textwidth]{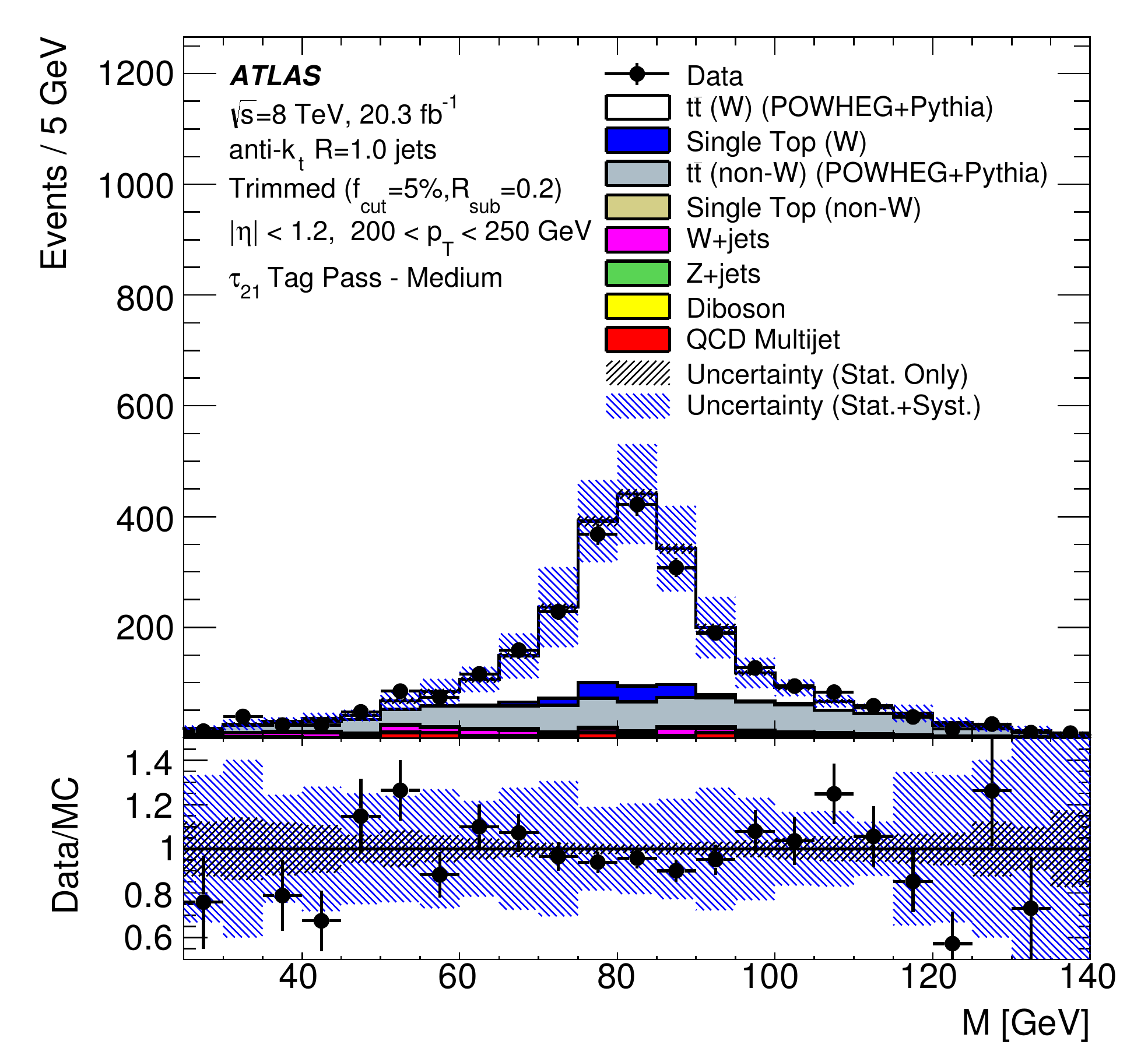}
}
\end{subfigure}
\end{center}
\vspace{-20 pt}
\caption[Data and \PowhegBox+~\Pythia mass distributions in fail/ pass substructure requirement sample.]{The distribution of the $W$ candidate mass for R2-trimmed jets failing (left) and passing (right) the selection corresponding to $\epsilon_{W}^{\mathrm{G}\&\mathrm{T}} = 50\%$ for the combined electron and muon channel in the \pt range 200--250~\GeV, without application of the mass cut. In (a), (b) the variable used for selection is \CTwoBetaOne, in (c) and (d) it is \DTwoBetaOne, and in (e), (f) it is \TauTwoOneWTA. Data points are shown with statistical uncertainties, and the combined MC is shown with full systematic and statistical uncertainties. The lower panels show the data/MC ratios, with the statistical uncertainty on the MC given in black forward-slashed bands, and the full systematic uncertainty given in the blue, back-slashed bands.}
\label{fig:tt_jm_tagged_d2beta1}
\end{figure}

The jet mass distributions of the $W$ boson candidates satisfying or failing to satisfy the medium signal efficiency requirement for each of the three substructure variables are shown in \figref{tt_jm_tagged_d2beta1}. The mass distribution for jets failing the \CTwoBetaOne tagger (\figref{c2_tagged_fail}) is notably different from the mass distributions for jets that fail the \DTwoBetaOne and/or \TauTwoOneWTA taggers, with a significantly higher mass peak and a low-mass tail that is conspicuous in its absence. This effect can be understood by referring back to \figref{variable2_c2_2d}: the correlation between the mass and \CTwoBetaOne is strong for background jets with low masses, while there is no clear correlation in the signal mass region. This means that the \CTwoBetaOne variable performs well when combined with a mass window, but is not very effective without the mass constraint.

\subsection{Signal and background efficiencies and uncertainties}
\label{sec:sysEff}

Background efficiencies are measured in a multijet-enriched sample of data, using the large-$R$ trigger and event selection described in \secref{objects}. 

The systematic uncertainties on the background efficiency measurements in multijet events are summarised in \Tabref{sys_mistag_D2}. The uncertainties are propagated coherently through to the measurement and then added together in quadrature. The background efficiency uncertainty due the JSS uncertainty can be as large as $\sim 25\%$ for jets with $\pt > 500\GeV$ and is about 15--20\% in the lower \pt{} ranges for the scale uncertainty on \DTwoBetaOne . The background efficiency uncertainties from the JMS are, in general, larger than those from the JES and are of the order of 6--10\% and 2--9\%, respectively. The impact of JER and JMR uncertainties is much smaller than that of the scale uncertainties.

Signal efficiencies are extracted from data by performing a template fit to the mass distributions of jets that satisfy or fail to satisfy the requirement on the given tagging variable. The signal template is constructed using the \PowhegBox~+~\Pythia\ \ttbar\ events, requiring that both partons from the $W$ boson decay in the event record are within $\Delta R = 1.0$ of the jet axis. The mass templates for the background are composed of decays of $W$ bosons from top quarks, where not all the decay products fall inside the jet cone, and the other non-$W$ backgrounds are also estimated using \PowhegBox~+~\Pythia. The normalisations of both templates are allowed to float.

The statistical uncertainty on the efficiency measurement in data includes the statistical uncertainty of the templates. For most sources of systematic uncertainty, a variation of the fit is performed with templates modified by $\pm 1\sigma$. In the case of the JMS, this variation is between $\pm 0.5\sigma$ and $\pm 1.0\sigma$; this reduction in the uncertainty with respect to that obtained with the standard double-ratio technique is made possible by fitting the mass distributions in data to a number of different templates. The templates are obtained by shifting the jet mass up and down by fractions (0.25 -- 1.0) of $\sigma$. The $\chi^{2}/ndf$ fit quality of each template is calculated, and a parabolic fit performed to the $\chi^{2}/ndf$ as a function of the fraction of $\sigma$. The fraction of $\sigma$ that results in a one unit shift from that which minimises $\chi^{2}/ndf$ is used as the uncertainty on the JMS for the signal efficiency calculation.

The full set of contributions to the systematic uncertainty on the signal efficiency is summarised in \Tabref{sys_eff_D2}, after applying the mass and \DTwoBetaOne medium tagging requirements. As in the background efficiency uncertainty estimate, the JSS contributes the largest uncertainty on this efficiency, varying between and 3\% and 5\% for the \DTwoBetaOne scale. The contribution from the JMR is $\sim 3\%$. The contribution from JER is less significant than JMR, being negligible in the lowest \pt{} bin and $\sim 1\%$ for jets with $250 < \pt < 500\GeV$. The contribution from JMS variations is also $\sim1\%$ (symmetrised as a result of the profiling technique) and increases to $\sim 10\%$ in the highest \pt{} range ($350 < \pt < 500\GeV$). The uncertainty from the JES is around 2--4\%.

In addition to the scale and resolution uncertainties, two other types of uncertainty are considered for the signal efficiency measurement: (a) \ttbar modelling -- initial-state radiation (ISR), final-state radiation (FSR), and generator uncertainty; (b) the normalisation of the main background sources -- multijet, $W$+jets, partial-$W$ and non-$W$ in single top and \ttbar.

The generator uncertainty is taken into account as the difference between the signal efficiency measurement using the \Mcatnlo+~\Herwig mass templates for the signal instead of the default \PowhegBox+~\Pythia ones. These uncertainties are between 1\% and 3\%. The modelling uncertainty of the QCD radiation is estimated using \Acermc~\cite{acer} v3.8 plus \Pythia v6.426 MC samples by varying the parameters controlling the ISR and FSR in a range consistent with a previous ATLAS measurement~\cite{TOPQ-2011-21}. The resulting uncertainties on the signal efficiency increase with jet \pt\ and are 2--6\%. The normalisation uncertainties for the main background sources are evaluated using a $\pm 1\sigma$ variation of the cross-section. The normalisation uncertainties are negligible with respect to the scale and resolution uncertainties, and for the \ttbar signal and $W$+jets background they are $<$ 1\%.
 
\begin{table}\renewcommand{\arraystretch}{1.5}
\begin{center}
\begin{tabular}{l|c|c|c|c}
\hline
\hline
\multirow{2}{*}{Source} & \multicolumn{4}{c}{\pt{} range [\GeV]}\\\cline{2-5}
 & 200--250 & 250--350 & 350--500 & 500--1000 \\
\hline
\hline
JES & +3.0 / $-$5.6 & +7.9 / $-$8.3 & +8.8 / $-$5.0 & +2.5 / $-$4.3 \\
JMS & $-$9.3 / +8.6 &  $-$9.7 / $ -$9.6 & $-$6.0 / $-$6.7 & $-$8.0 / +5.7 \\
JER & +1.0 & $-$1.6 & +0.5 & +0.8 \\
JMR & $-$2.0 & +1.8 &  +1.0 & +0.1 \\
JSS (\DTwoBetaOne)  & $-$13.2 / +15.9 & $-$15.7 / +19.7 & $-$17.6 / +22.7 & $-$19.4 / +23.8 \\
\hline
Total & +16.7 / $-$19.1 & +20.3 / $-$23.5 & +20.6 / $-$24.2 & +21.2 / $-$24.8 \\
\hline
\hline
\end{tabular}
\caption[Systematics on background efficiency in \Pythia for mass+\DTwoBetaOne.]{Relative systematic uncertainties (in \%) on the background efficiency from the different sources, for jets in the \Pythia multijet sample after tagging with the R2-trimmed mass and medium \DTwoBetaOne requirement that results in a signal efficiency $\epsilon_{W}^{\mathrm{G}\&\mathrm{T}}\approx50\%$. Uncertainties on scales (JMS, JES and JSS indicate the mass, energy and substructure scale uncertainties) can be in both directions, and so result in pairs of efficiency uncertainties. The mass and energy resolution uncertainties are denoted JMR and JER respectively. The contributions from each source are added in quadrature to get the total uncertainty on $\epsilon_{\mathrm{QCD}}^{\mathrm{G}\&\mathrm{T}}$.}
\label{tab:sys_mistag_D2}
\end{center}
\end{table}

\begin{table}\renewcommand{\arraystretch}{1.5}
\begin{center}
\begin{tabular}{l|c|c|c}
\hline
\hline
\multirow{2}{*}{Source} & \multicolumn{3}{c}{\pt{} range [\GeV]}\\\cline{2-4}
 & 200--250 & 250--350 & 350--500  \\
\hline
\hline
JMS           &       +1.1     &       +1.1     &       +9.6     \\
JES        &       $-$3.5 / +3.6 &       $-$1.7 / +2.5 &       +1.6 / $-$2.3        \\
JER                         &       $-$0.1    &       +1.0     &       +1.0     \\
JMR             &       +2.7     &       +3.7     &       +4.3     \\
JSS (\DTwoBetaOne)      &       +4.3 / $-$2.9&       +4.2 / $-$4.5&       +5.1 / $-$4.8\\
MC generator           &       $-$0.9    &       +1.9     &       $-$3.2    \\
ISR/FSR        &       +1.6 / $-$2.2     &       +2.7 / $-$4.0     &       +4.4 / $-$5.6     \\
Multijet normalisation        &       $-$0.4 / +0.4 &       $-0.3$ / +0.3 &       +0.1 / $-$0.1        \\
Single-top normalisation        &       $-$0.1 / +0.1 &       $-$0.1 / +0.1 &       $-$0.1 / +0.1        \\
$t\bar{t}$ normalisation       &       0.6 / $-$0.5 &       +0.6 / $-$0.6 &       +0.5 / $-$0.5        \\
$W$+jets normalisation       &       $-$0.3 / +0.3 &       $-$0.4 / +0.4 &       $-$0.5 / +0.4 \\
MC statistics         &       $-$1.0    &       $-$1.5    &       $-$3.5    \\
\hline
Total           &       +6.6 / $-$5.4 &       +7.3 / $-$6.6 &       +13.1 / $-$13.2 \\
\hline
\hline
\end{tabular}
\caption[Systematics on efficiency in data for mass+\DTwoBetaOne.]{Relative systematic uncertainties (in \%) on the $W$-jet tagging efficiency from different sources after tagging with the R2-trimmed mass and medium \DTwoBetaOne requirement that results in a signal efficiency $\epsilon_{W}^{\mathrm{G}\&\mathrm{T}}\approx50\%$. The uncertainties on scales (JMS, JES and JSS indicate the mass, energy and substructure scale uncertainties) and normalisations can be in both directions, and so result in pairs of efficiency uncertainties, but here the JMS is symmetrised as part of the profiling technique described in the text. The contributions from each source are added in quadrature to get the total uncertainty on $\epsilon_{\mathrm{QCD}}^{\mathrm{G}\&\mathrm{T}}$. The mass and energy resolution uncertainties are denoted JMR and JER respectively, and ISR/FSR indicate the uncertainties from the modeling of the initial/final state radiation.}
\label{tab:sys_eff_D2}
\end{center}
\end{table}

\subsection{Summary of $W$ boson tagging efficiencies in data and MC}
\label{sec:results}

\begin{figure}[pt!]
\begin{center}
\begin{subfigure}[]{\label{fig:Eff_c2_a}
\includegraphics[width=.45\textwidth]{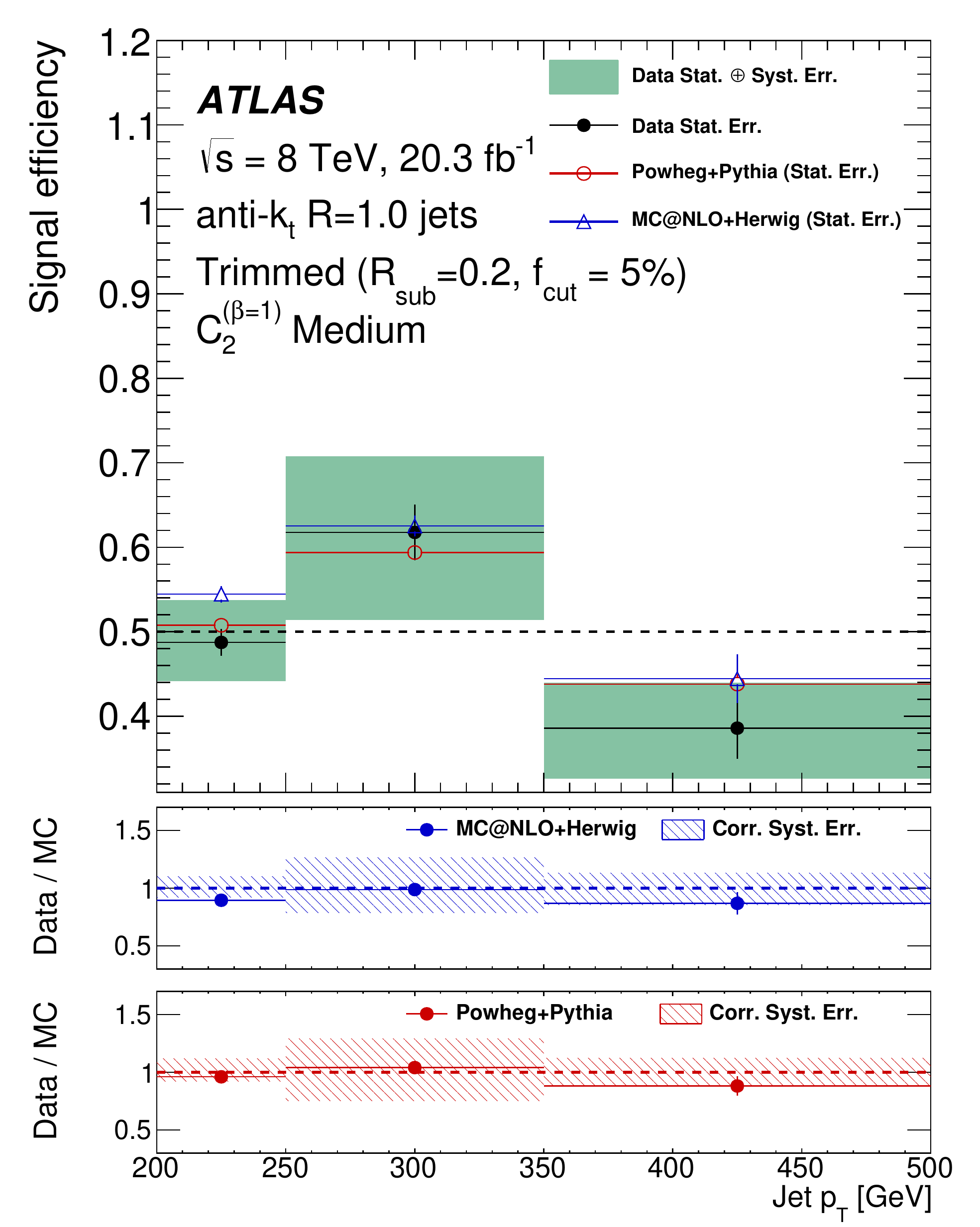}
}
\end{subfigure}
\begin{subfigure}[]{\label{fig:Eff_c2_b}
\includegraphics[width=.45\textwidth]{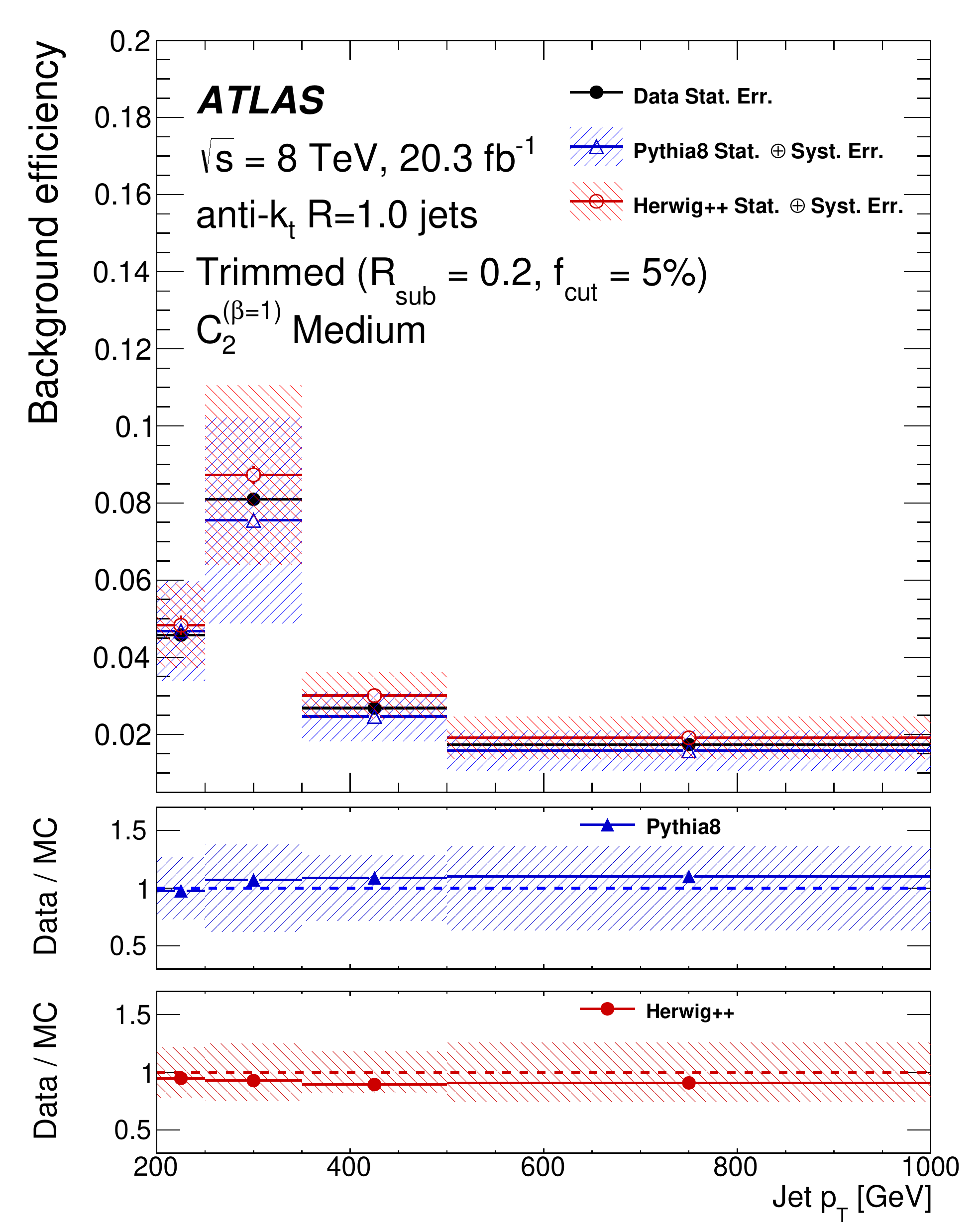}
}
\end{subfigure}
\begin{subfigure}[]{\label{fig:Eff_c2_c}
\includegraphics[width=.45\textwidth]{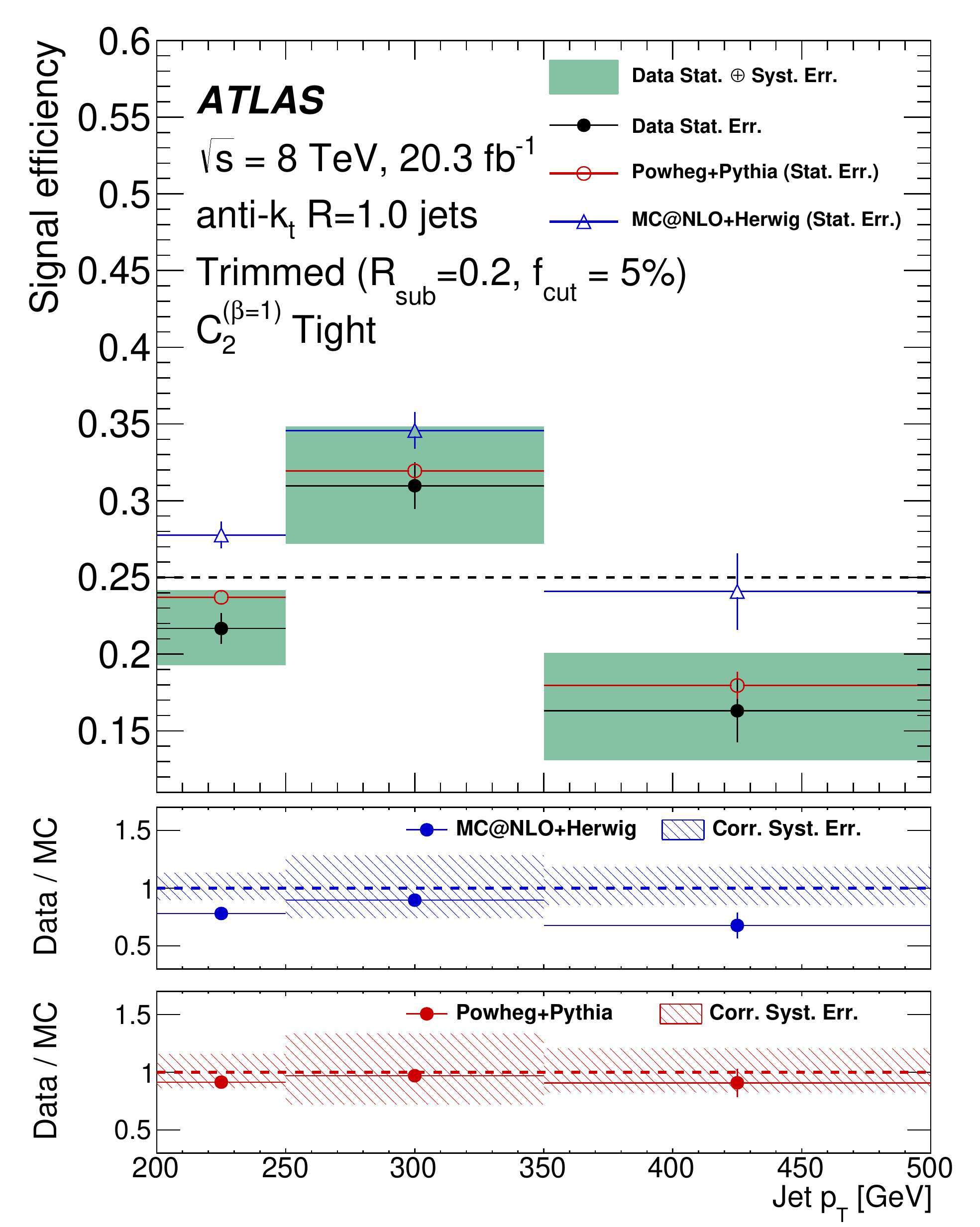}
}
\end{subfigure}
\begin{subfigure}[]{\label{fig:Eff_c2_d}
\includegraphics[width=.45\textwidth]{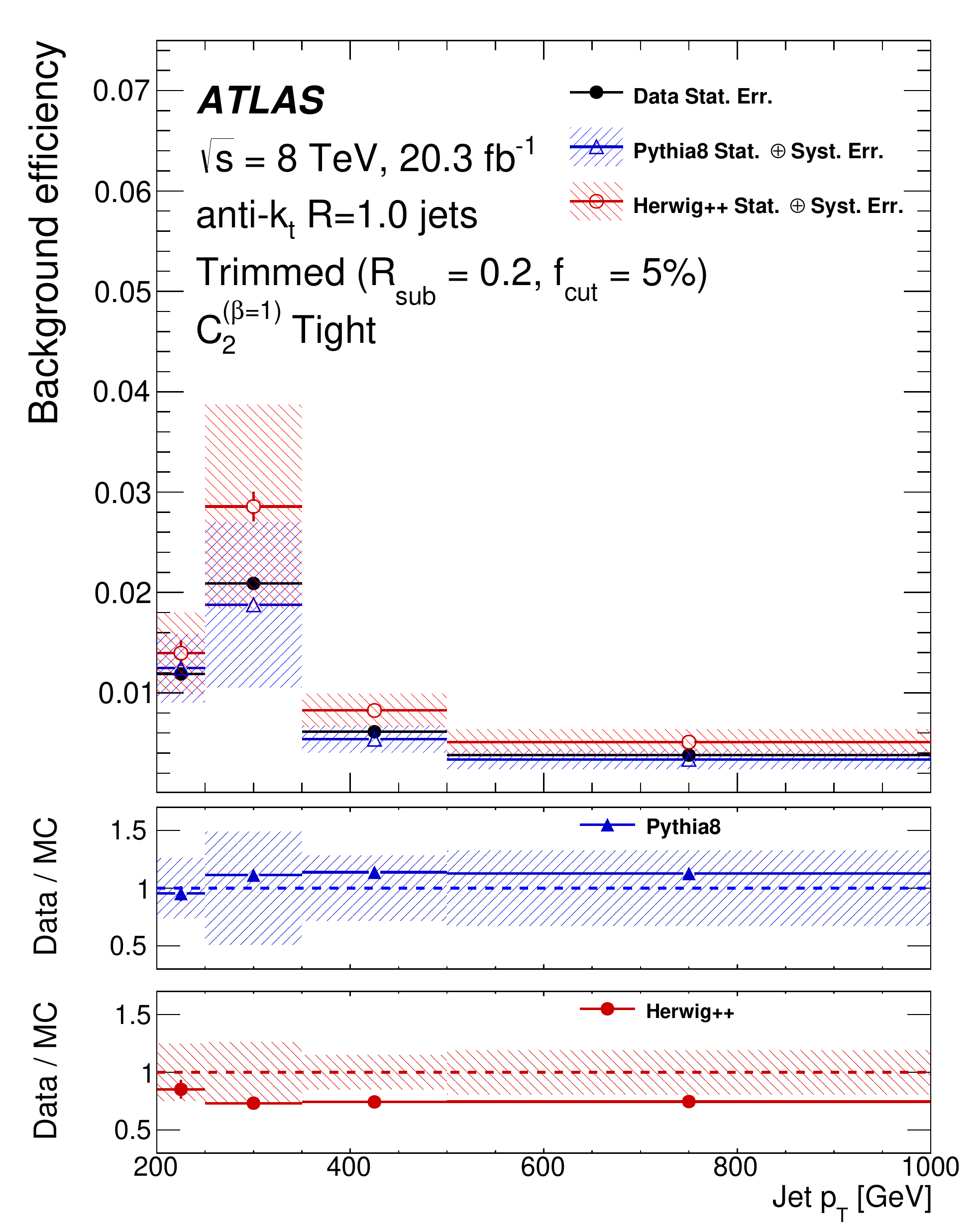}
}
\end{subfigure}
\end{center}
\vspace{-20 pt}
\caption[Signal and background efficiency with \CTwoBetaOne.]{$W$ boson tagging efficiencies in ranges of jet \pt\ for (left) signal $W$-jets in \ttbar{} events and (right) multijet background. The $\epsilon_{W}^{\mathrm{G}\&\mathrm{T}} \sim$ 50\% working points obtained with the combined mass window and \CTwoBetaOne requirements are shown in (a) and (b), and the $\sim$ 25\% working points are shown in (c), (d). The deviations from 50\% and 25\% in (a) and (c) respectively are due to the optimisations being based on $W$-jets in a different $W^{\prime} \rightarrow WZ$ topology, as discussed in the text. The lower panels show ratios of the efficiency measured in data to the efficiency in two different MC simulations.}
\label{fig:Eff_c2}
\end{figure}

\begin{figure}[pt!]
\begin{center}
\begin{subfigure}[]{\label{fig:Eff_d2_a}
\includegraphics[width=.45\textwidth]{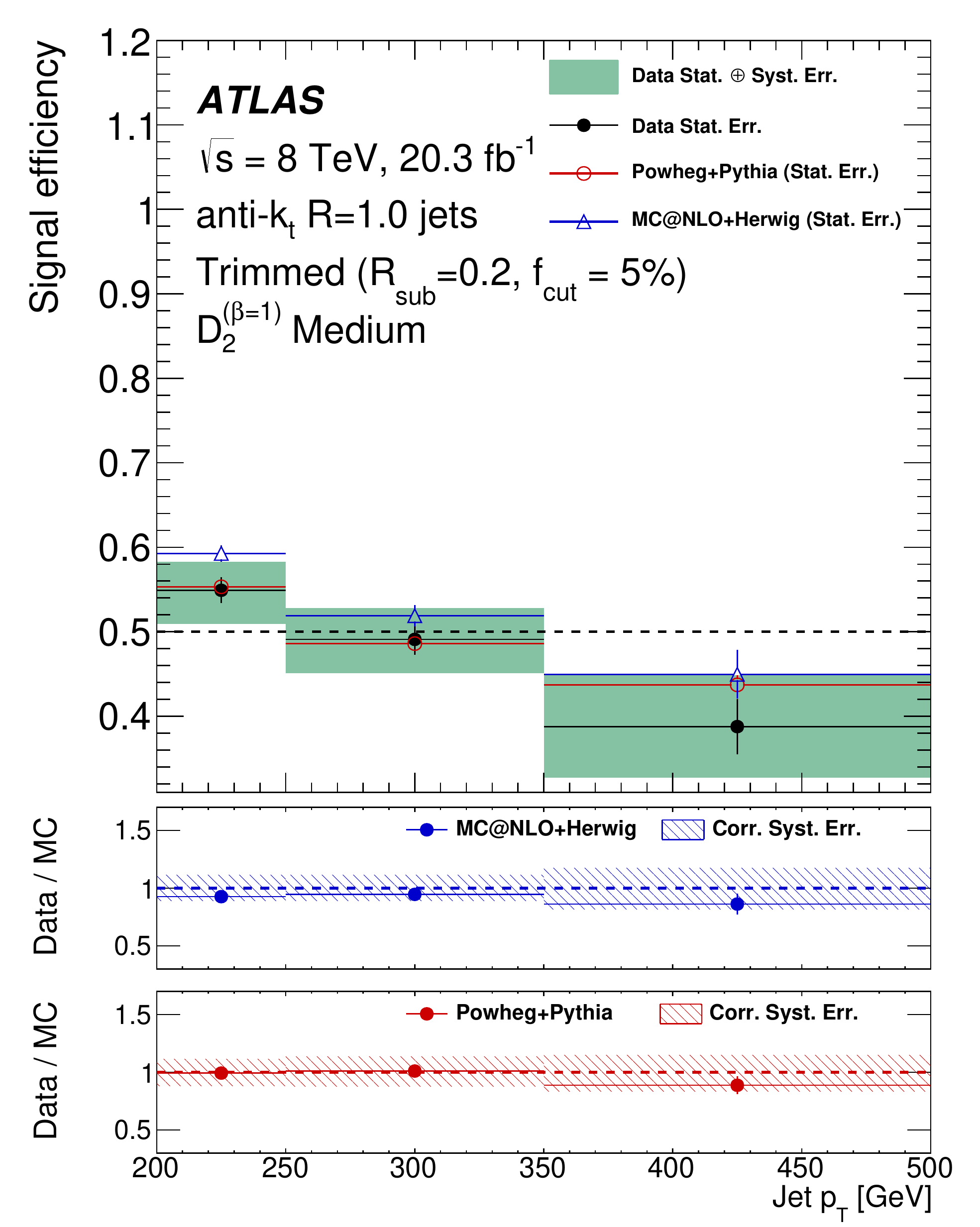}
}
\end{subfigure}
\begin{subfigure}[]{\label{fig:Eff_d2_b}
\includegraphics[width=.45\textwidth]{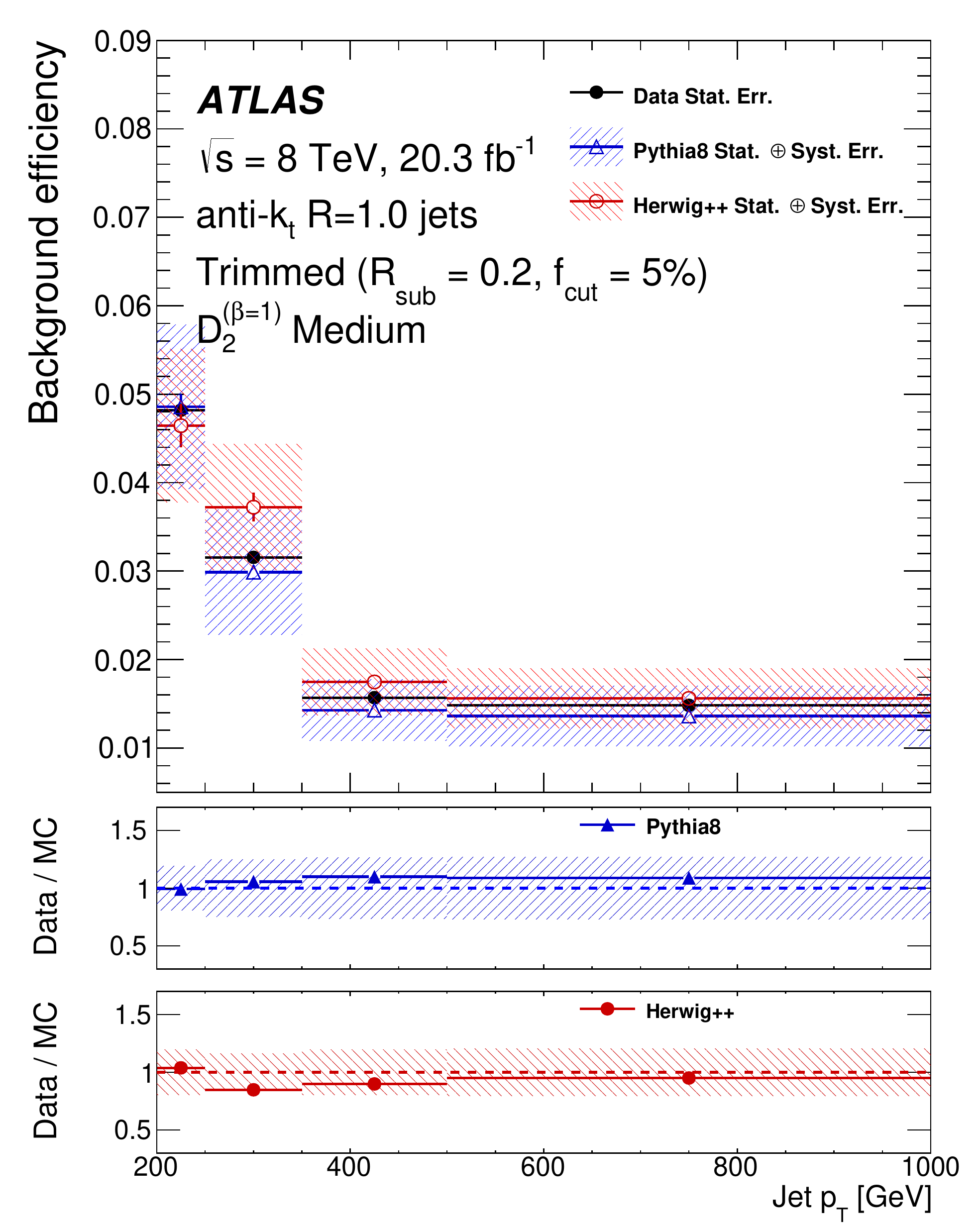}
}
\end{subfigure}
\begin{subfigure}[]{\label{fig:Eff_d2_c}
\includegraphics[width=.45\textwidth]{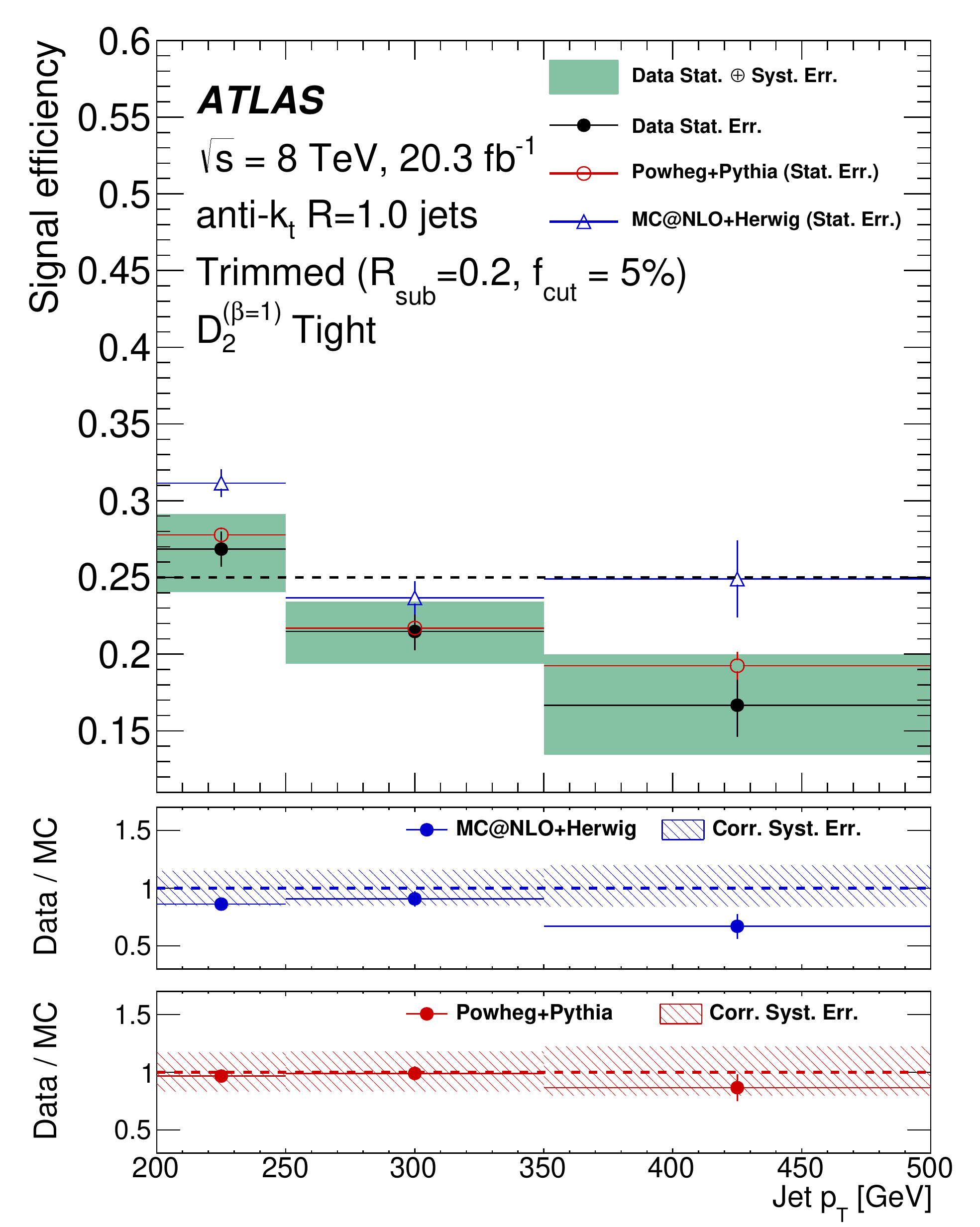}
}
\end{subfigure}
\begin{subfigure}[]{\label{fig:Eff_d2_d}
\includegraphics[width=.45\textwidth]{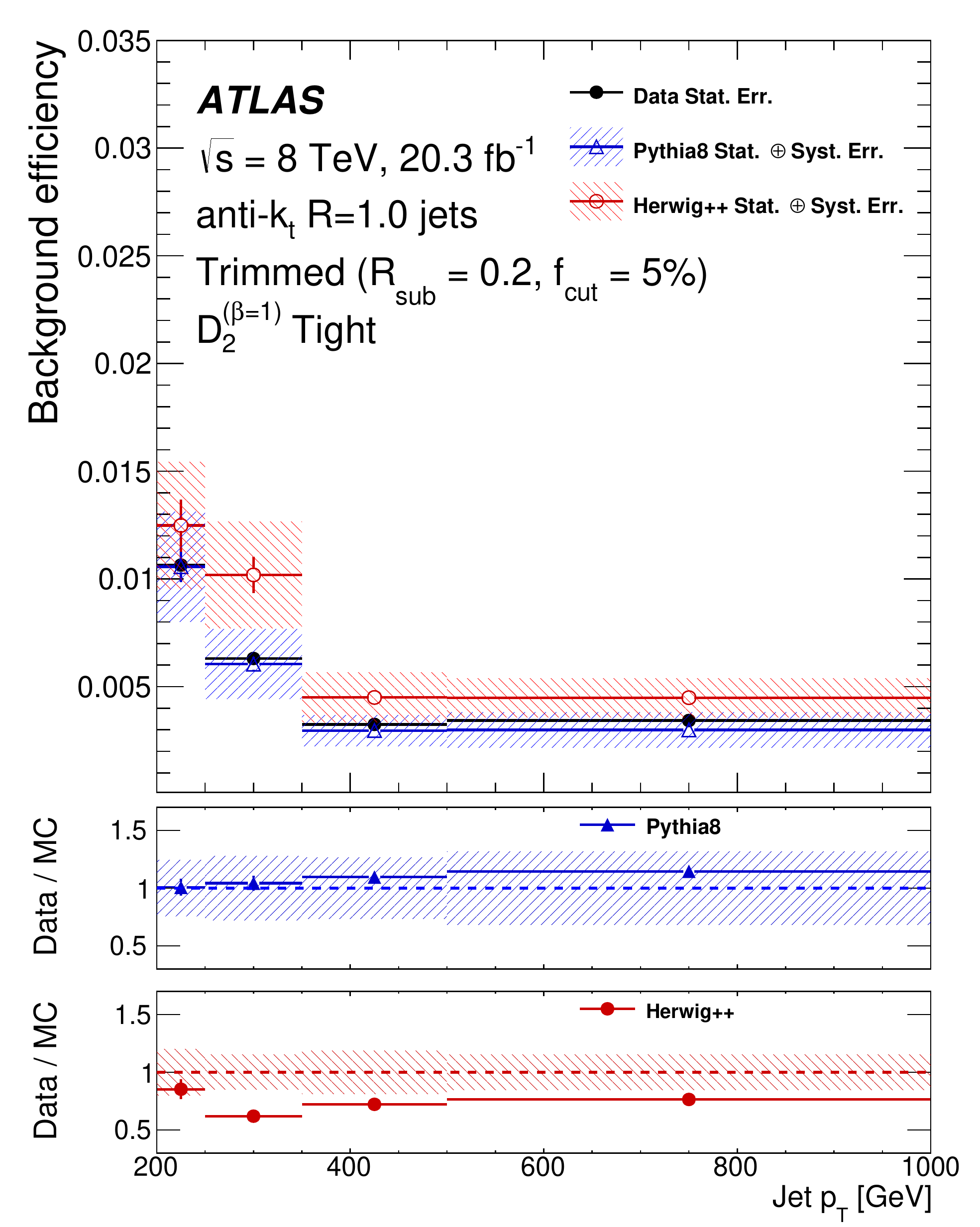}
}
\end{subfigure}
\end{center}
\vspace{-20 pt}
\caption[Signal and background efficiency with \DTwoBetaOne.]{$W$ boson tagging efficiencies in ranges of jet \pt\ for (left) signal $W$-jets in \ttbar{} events and (right) multijet background. The $\epsilon_{W}^{\mathrm{G}\&\mathrm{T}} \sim$ 50\% working points obtained with the combined mass window and \DTwoBetaOne requirements are shown in (a) and (b), and the $\sim$ 25\% working points are shown in (c), (d). The deviations from 50\% and 25\% in (a) and (c) respectively are due to the optimisations being based on $W$-jets in a different $W^{\prime} \rightarrow WZ$ topology, as discussed in the text. The lower panels show ratios of the efficiency measured in data to the efficiency in two different MC simulations.}
\label{fig:Eff_d2}
\end{figure}

\begin{figure}[pt!]
\begin{center}
\begin{subfigure}[]{\label{fig:Eff_t21_a}
\includegraphics[width=.45\textwidth]{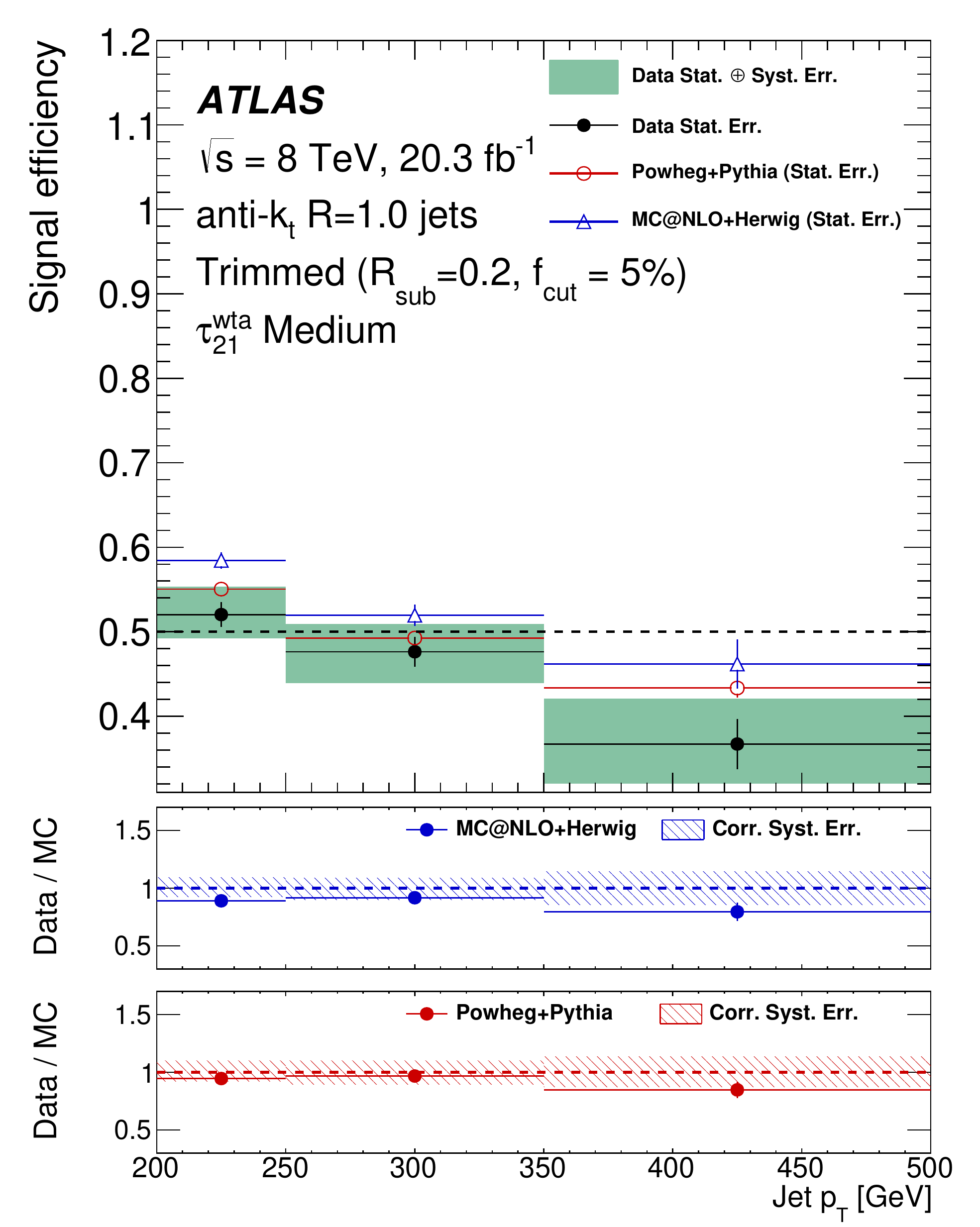}
}
\end{subfigure}
\begin{subfigure}[]{\label{fig:Eff_t21_b}
\includegraphics[width=.45\textwidth]{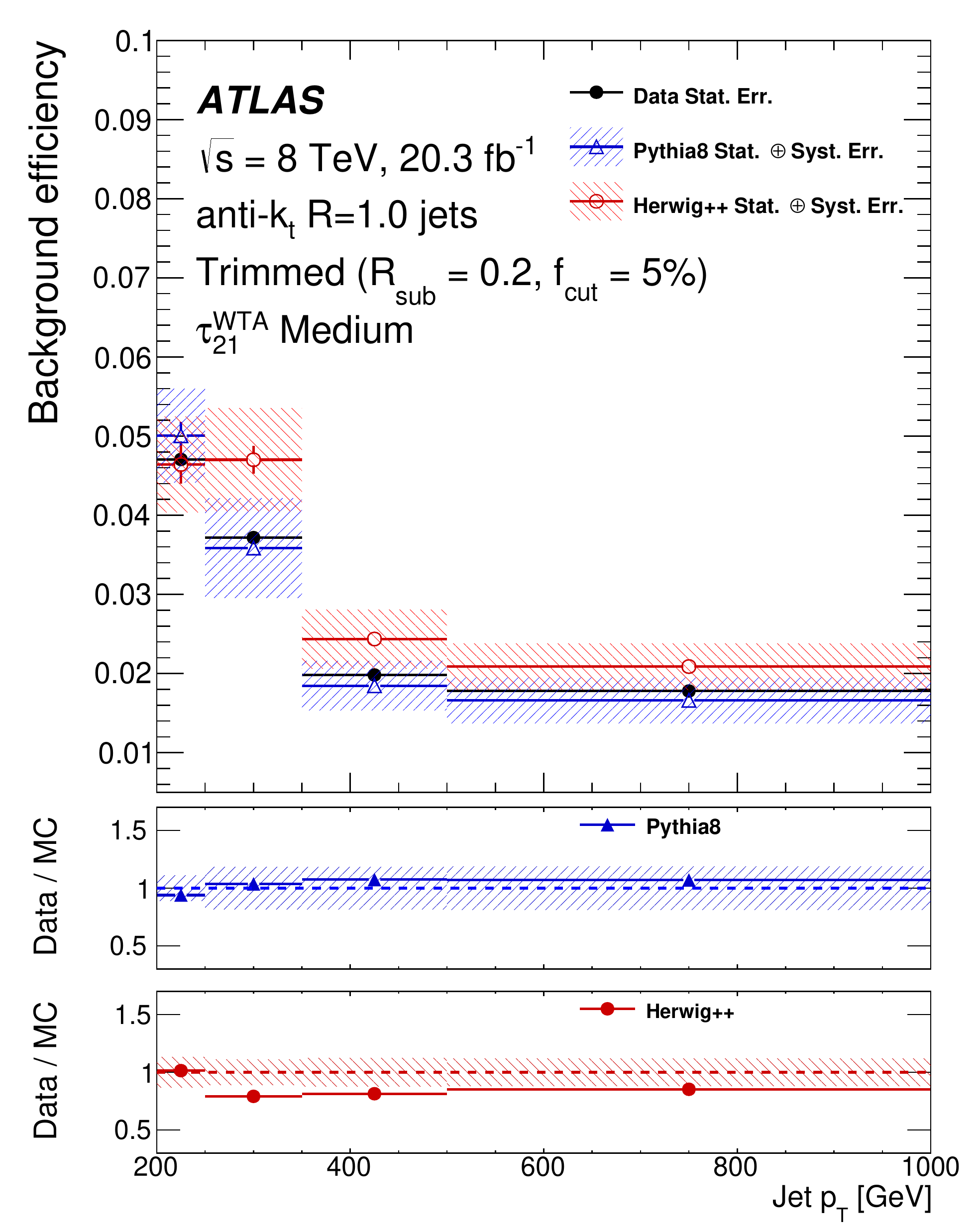}
}
\end{subfigure}
\begin{subfigure}[]{\label{fig:Eff_t21_c}
\includegraphics[width=.45\textwidth]{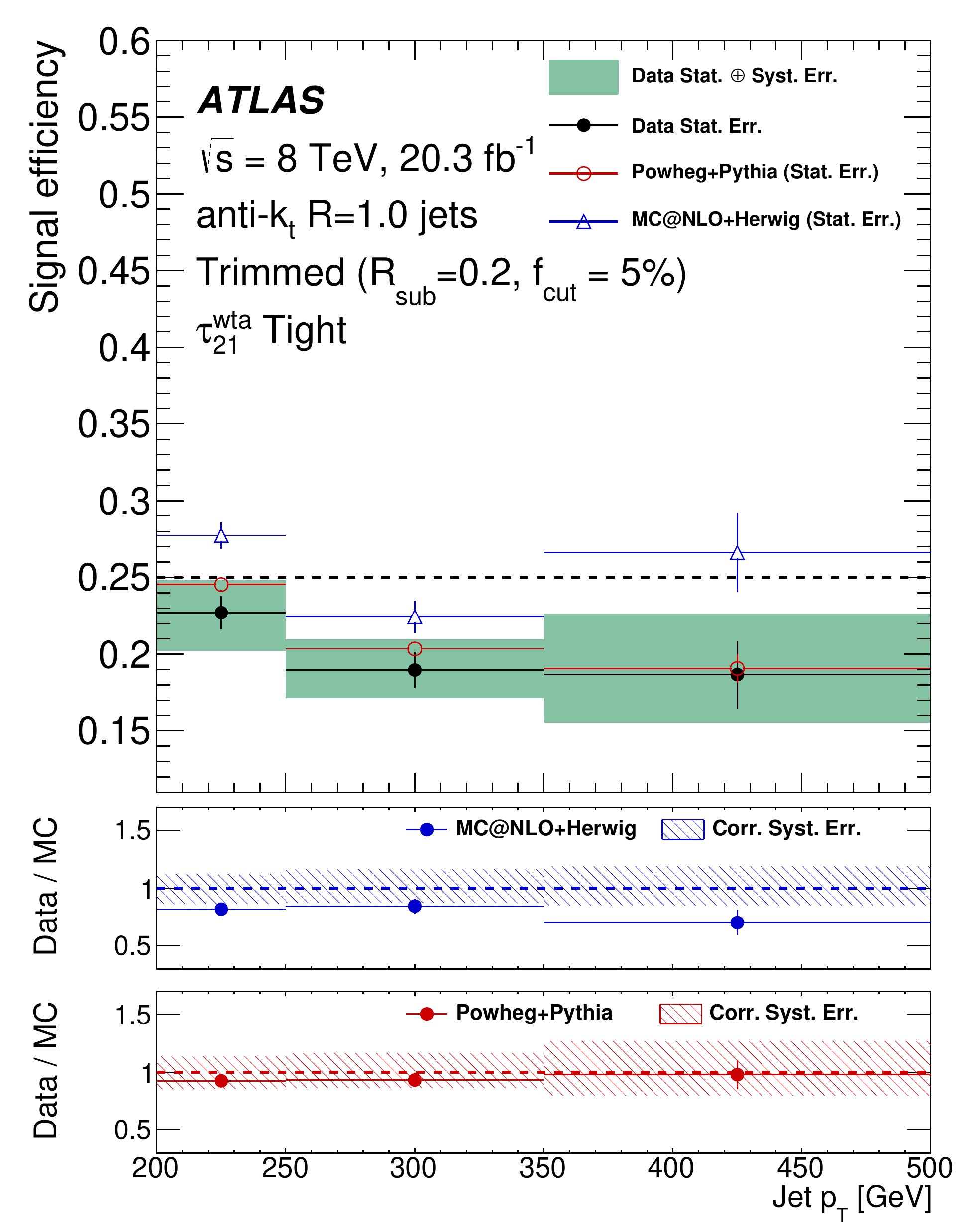}
}
\end{subfigure}
\begin{subfigure}[]{\label{fig:Eff_t21_d}
\includegraphics[width=.45\textwidth]{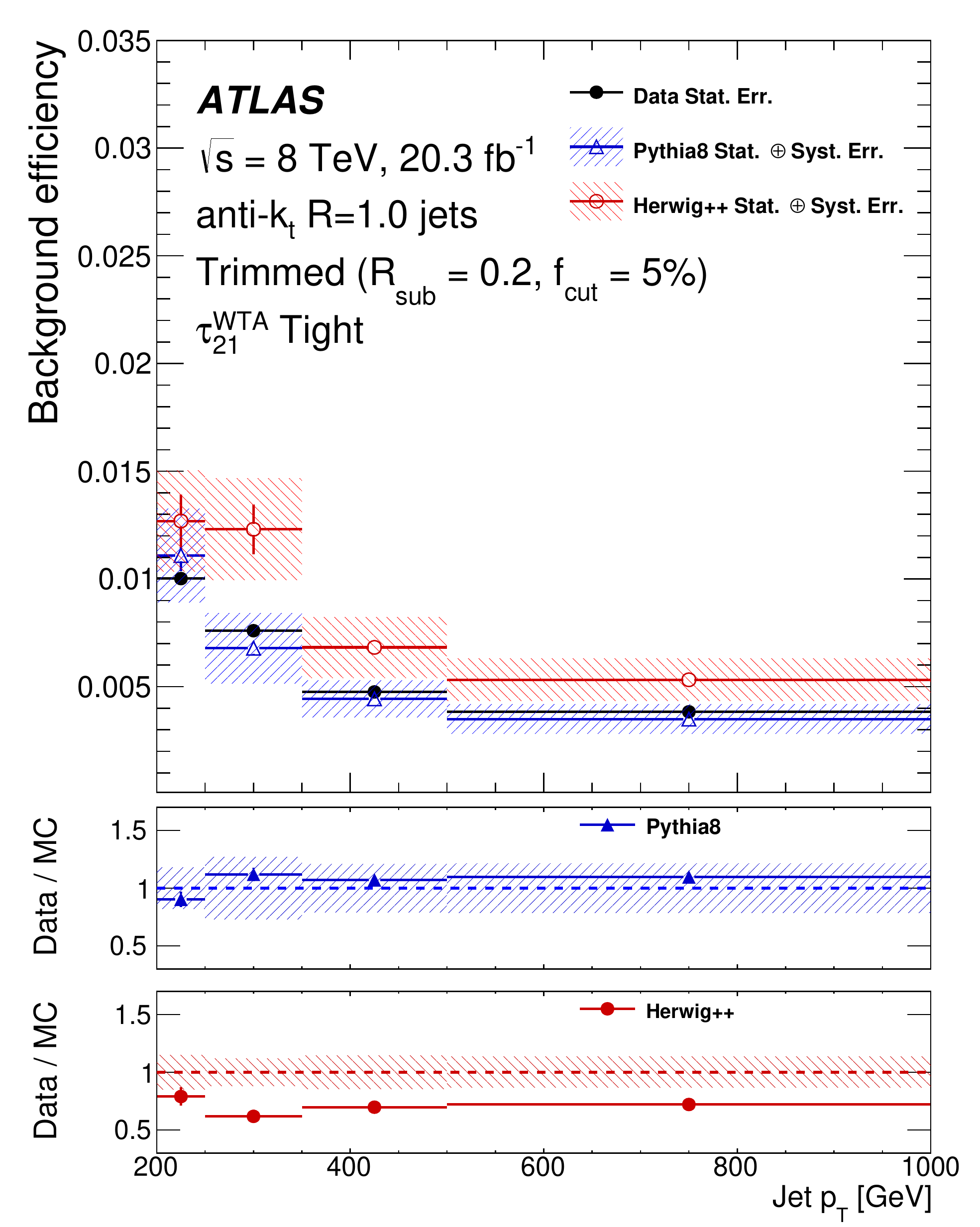}
}
\end{subfigure}
\end{center}
\vspace{-20 pt}
\caption[Signal and background efficiency with \TauTwoOneWTA.]{$W$ boson tagging efficiencies in ranges of jet \pt\ for (left) signal $W$-jets in \ttbar{} events and (right) multijet background. The $\epsilon_{W}^{\mathrm{G}\&\mathrm{T}} \sim$ 50\% working points obtained with the combined mass window and \TauTwoOneWTA requirements are shown in (a) and (b), and the $\sim$ 25\% working points are shown in (c), (d). The deviations from 50\% and 25\% in (a) and (c) respectively are due to the optimisations being based on $W$-jets in a different $W^{\prime} \rightarrow WZ$ topology, as discussed in the text. The lower panels show ratios of the efficiency measured in data to the efficiency in two different MC simulations.}
\label{fig:Eff_t21}
\end{figure}

The $W$-jet tagging efficiency in \ttbar{} events using the R2-trimmed jet mass window and the medium and tight \CTwoBetaOne selections is measured in top-enriched data and in MC provided by \PowhegBox~+~\Pythia and \Mcatnlo+~\Herwig. The background efficiency with the same selection is measured in multijet-enriched data and in \Pythia and \Herwigpp simulations. The results of these measurements are shown in \figref{Eff_c2}. In both the signal and background efficiency distributions, the ratio of data to each of the two MC models is shown in the lower panels. The corresponding signal and background efficiency distributions for \DTwoBetaOne and \TauTwoOneWTA are shown in \figref{Eff_d2} and \figref{Eff_t21} respectively. Systematic errors from background modeling are added for the signal data points, while no background modeling is involved in the derivation of background efficiencies, whose points only show statistical error. Good agreement is observed between data and predictions. 

The signal efficiency at the medium working point is not exactly 50\% because the selection requirements for the $\epsilon_{W}^{\mathrm{G}\&\mathrm{T}} = 50\%$ working point are calculated using $W$-jets from $W^\prime \rightarrow WZ \rightarrow qq\ell\ell$ events, and are applied here to $W$-jets in \ttbar{} events.

The data points are the result of fits using templates extracted from \PowhegBox~+~\Pythia; the difference with respect to the results that would be obtained using templates from \Mcatnlo+~\Herwig is added in quadrature as an additional source of systematic uncertainty. 

The \DTwoBetaOne tagger has the smallest background efficiency for the medium and tight working points in all \pt ranges except for the lowest, $200 < \pt < 250\GeV$. The background efficiencies decrease with increasing \pt{}, with the exception of the \CTwoBetaOne tagger, for which the background efficiency increases for jets in the range $250 < \pt < 350\GeV$. This behaviour can be explained by the stronger \pt\ dependence of the \CTwoBetaOne tagger compared to the \DTwoBetaOne and \TauTwoOneWTA taggers.

For the signal efficiencies, the uncertainty bands of the ratios account for the correlations in the systematic uncertainties between data and MC. In general, data and \PowhegBox+~\Pythia agree better than data and \Mcatnlo+~\Herwig. For the medium working point, there is agreement between the two MC models within 1$\sigma$ except in the range $200 < \pt < 250\GeV$, while for the tight working point ($\epsilon_{W}^{\mathrm{G}\&\mathrm{T}} \sim 25\%$) the efficiency of \Mcatnlo+~\Herwig is 1.5$\sigma$ to 2$\sigma$ higher than both the efficiency predicted by \PowhegBox+~\Pythia and the measurements in data. There is a potential bias towards \PowhegBox+~\Pythia, as this generator provides the signal template used in determining the background subtraction that is necessary to define the signal efficiency in data. However, even when using \Mcatnlo+~\Herwig for the templates in the subtraction, \Powheg+~\Pythia gives a better description of the signal efficiency measured in data. The differences in the MC signal efficiencies stem from the differences in the signal mass distributions between models; the mass peak has a different width, so the fraction of signal in the mass window (which is the same for both Monte Carlo samples) is already significantly different after the requirement on the groomed jet mass is applied (see for example \figref{mass_diff_mcs}).

\begin{figure}
\begin{center}
\begin{subfigure}[]{\label{fig:toporoc_200250}
\includegraphics[width=0.45\textwidth]{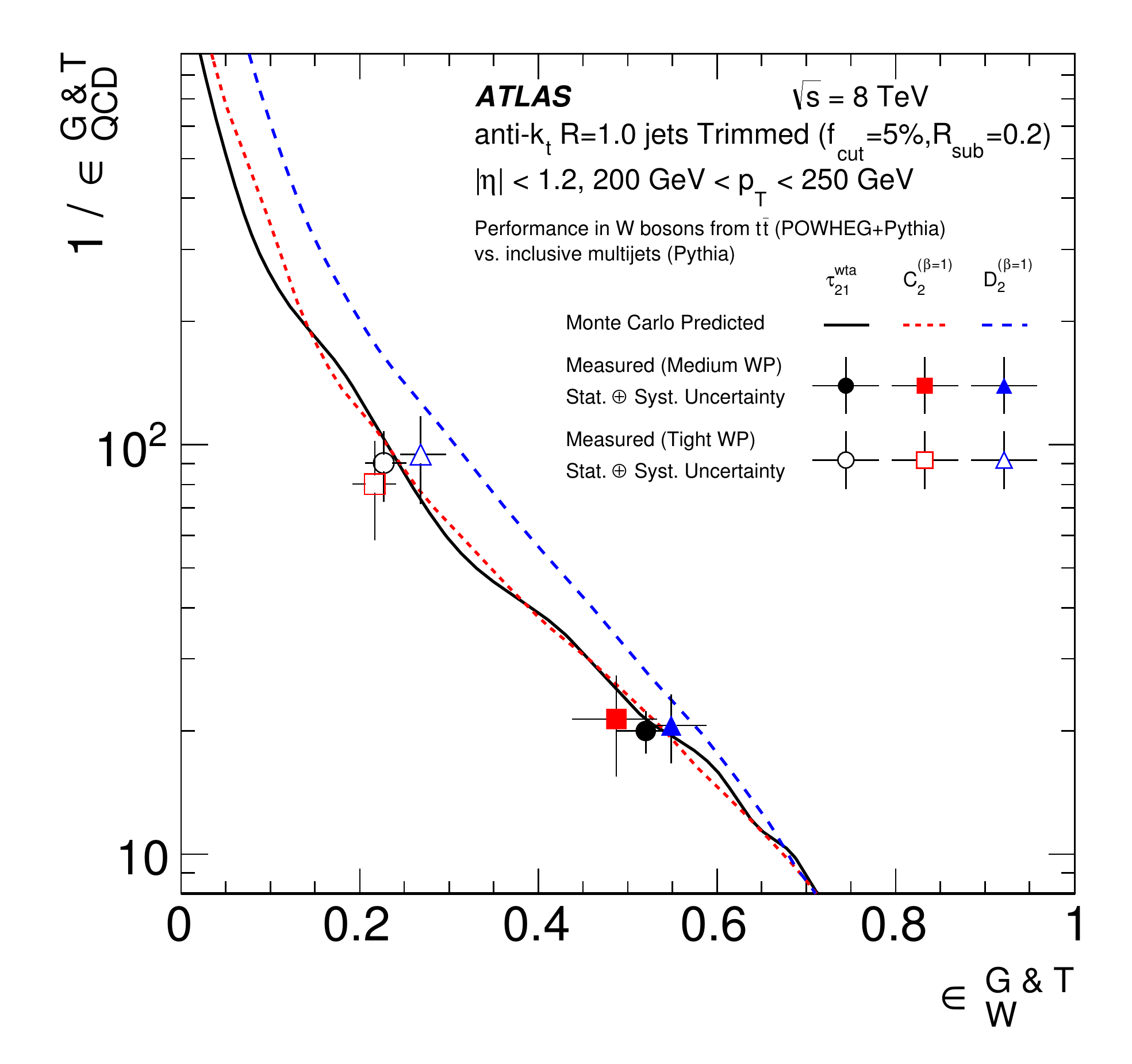}
}
\end{subfigure}
\begin{subfigure}[]{\label{fig:toporoc_250350}
\includegraphics[width=0.45\textwidth]{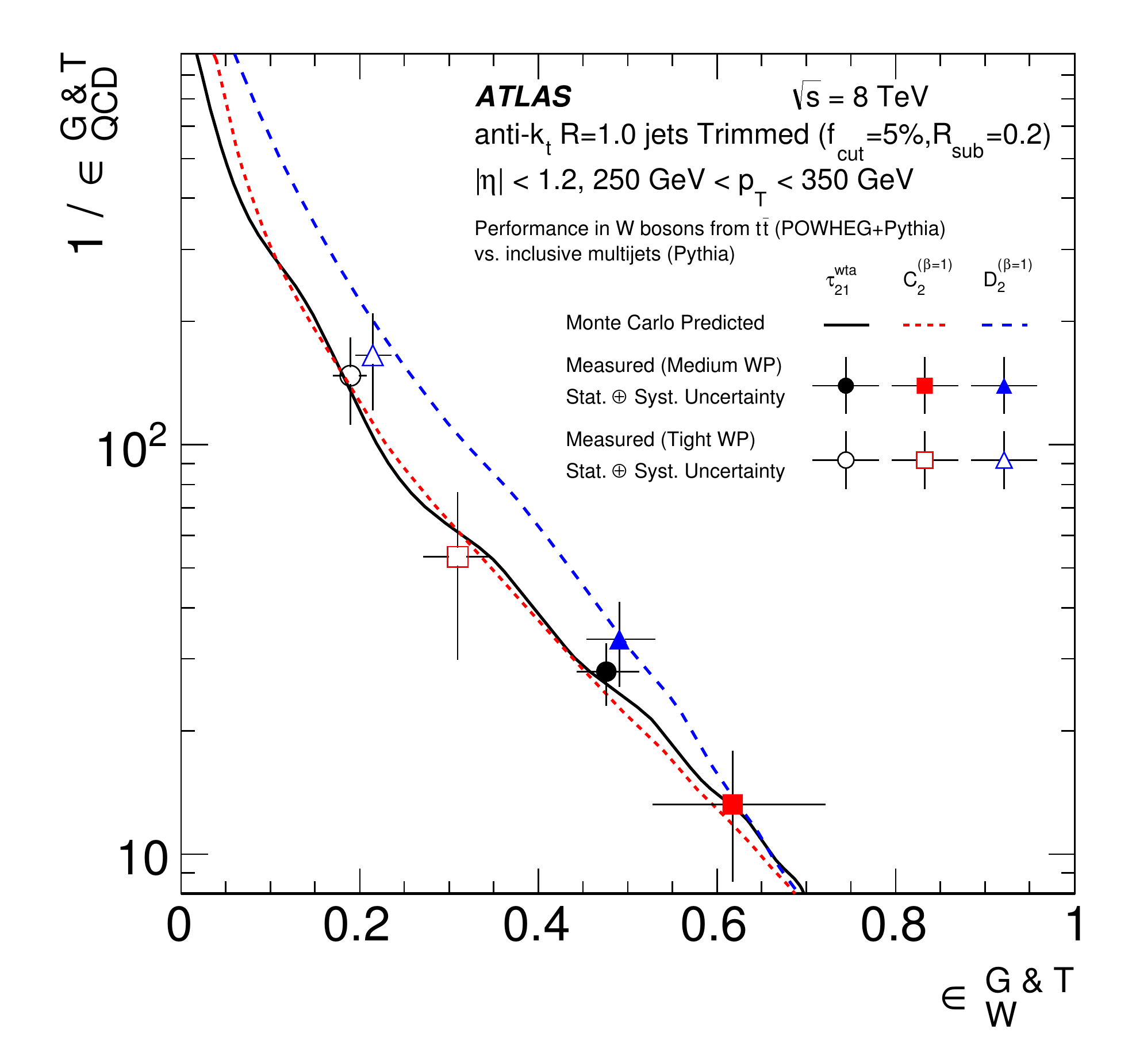}
}
\end{subfigure}
\begin{subfigure}[]{\label{fig:toporoc_350500}
\includegraphics[width=0.45\textwidth]{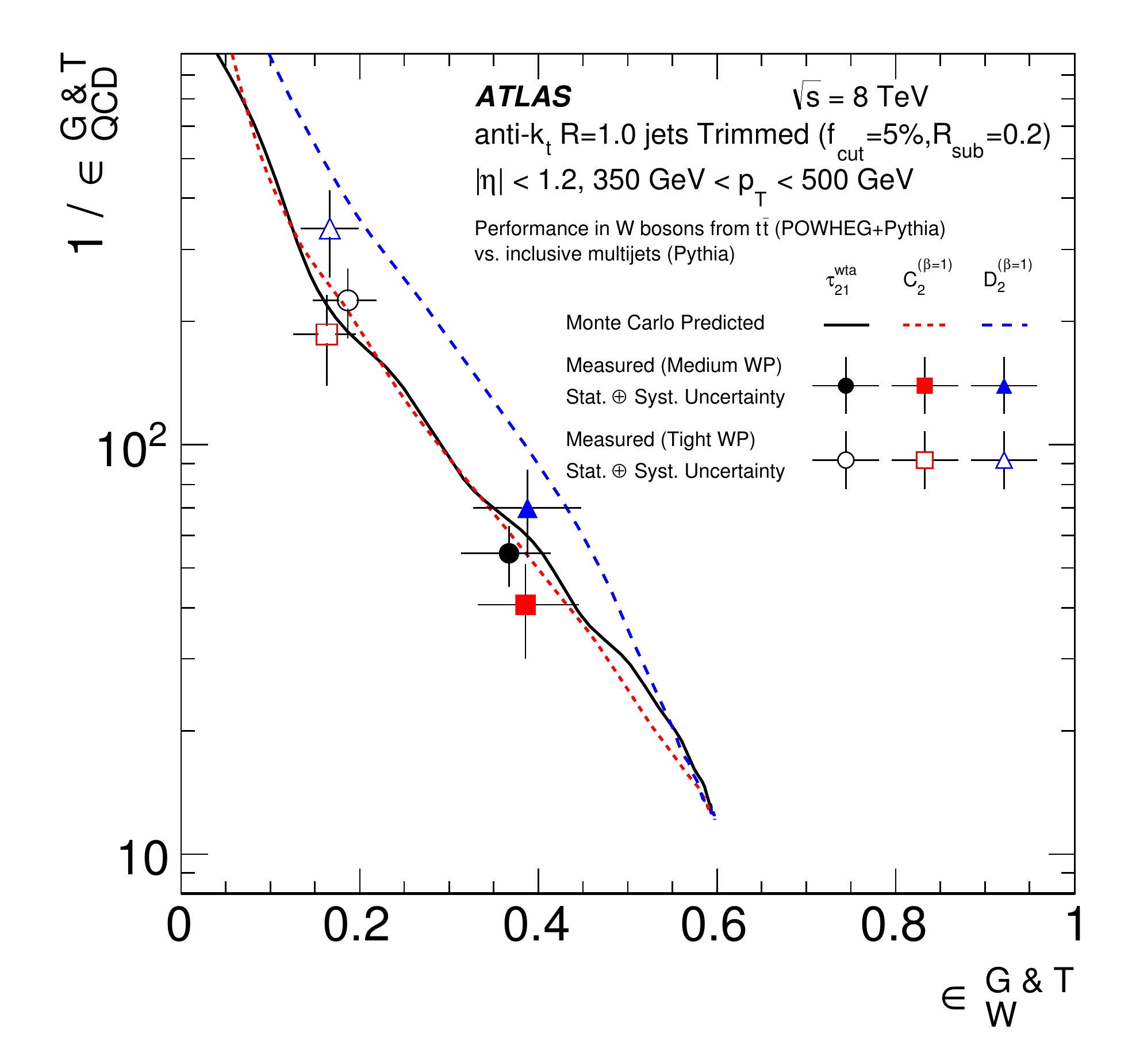}
}
\end{subfigure}
\end{center}
\vspace{-20 pt}
\caption[ROC curves from \PowhegBox+~\Pythia compared with data.]{Signal efficiency versus background rejection power (1 / background efficiency) curves derived using \PowhegBox+~\Pythia signal efficiencies and \Pythia background efficiencies compared with points from data. Three \pt ranges are shown: (a) 200--250~\GeV, (b) 250--350~\GeV, and (c) 350--500~\GeV. The data points include systematic uncertainties on the signal efficiency measurement in \ttbar{} events and the uncertainties on the \Pythia background efficiency predictions.}
\label{fig:topologies_roc}
\end{figure}

Figure~\ref{fig:topologies_roc} shows the \ttbar MC efficiency versus rejection curves with data measurements at the medium and tight working points, including systematic uncertainties on the signal and background efficiencies. Generally good agreement between data and MC simulation is observed in all \pt ranges for these measurements.

\FloatBarrier

\section{Conclusions}
\label{sec:conclusions}

Several combinations of jet grooming algorithms and tagging variables have
been studied to find an optimal $W$-jet tagger in terms of (a) maximising multijet background rejection power for given values of $W$-jet signal efficiency; (b) minimising systematic uncertainties and the effects of pileup; and (c) the modelling of the jet mass and substructure variables in Monte Carlo simulations. 

The signal efficiency working point $\epsilon_{W}^{\mathrm{G}} = 68\%$ is chosen as a suitable baseline for the comparison of grooming algorithms. The performances of the best few configurations of trimming, pruning and split-filtering are similar at this working point, and the \antiktten jet trimmed with $\fcut=5\%$ and $\subjetr=0.2$ (`R2-trimming') does particularly well in terms of removing pileup-dependence. Cambridge-Aachen pruning also provides significant discrimination for $W$-jet tagging, as does split-filtering without the mass-drop requirement. The irrelevance of the mass-drop requirement was shown previously in phenomenological studies~\cite{Dasgupta:2013ihk}, and is verified here in MC samples with a full ATLAS detector simulation. Trimming with $\subjetr=0.1$ shows promise in terms of the jet mass; it is not pursued further in these studies because it is challenging in terms of systematic uncertainties, as one is entering the arena of single-cluster jet, but it may well be considered in future extensions of these studies (for example in tagging $W$ bosons with \pt\ $>$ 1~\TeV).

The energy correlation ratios \DTwoBetaOne, \CTwoBetaOne are found to be particularly good variables for tagging $W$-jets, as shown for the first time here in data. However, there is some evidence of the \CTwoBetaOne variable having a higher background efficiency for low-\pt{} jets. Similarly good is the N-subjettiness ratio \TauTwoOneWTA, which performs better than its predecessor \TauTwoOne. 

The signal and background efficiencies obtained using pairwise combinations of the R2-trimmed mass and three different substructure variables are measured in \ttbar{} and multijet events from \SI{20.3}{\per\fb} of 8~\TeV\ $pp$ collisions recorded by ATLAS at the LHC. These are compared to various MC predictions which show in general good agreement within the uncertainties with the data measurements of signal efficiencies around $50\%$ for background efficiencies around $2\%$.

In some configurations, significant differences are observed in both the signal and background efficiencies from different Monte Carlo predictions. This can provide important information to improve the Monte Carlo simulations for searches for physics beyond the Standard Model. It further highlights the potential for data measurements such as these to be utilised for tuning Monte Carlo simulations. 

These studies are necessarily limited in scope to comparing simple two-variable taggers, made up of a groomed mass window and a substructure variable requirement, both of which are sensitive to \pt\ and therefore optimised for three different \pt\ ranges. Extensions to these studies could include combining three or more variables and using multivariate techniques to further boost the signal efficiency and/or reduce the background; investigating how these conclusions change if dedicated pileup-removal techniques are used alongside grooming; and varying the $\epsilon_{W}^{\mathrm{G}}$ baseline at which the grooming algorithms are compared.

\section*{Acknowledgements}


We thank CERN for the very successful operation of the LHC, as well as the
support staff from our institutions without whom ATLAS could not be
operated efficiently.

We acknowledge the support of ANPCyT, Argentina; YerPhI, Armenia; ARC, Australia; BMWFW and FWF, Austria; ANAS, Azerbaijan; SSTC, Belarus; CNPq and FAPESP, Brazil; NSERC, NRC and CFI, Canada; CERN; CONICYT, Chile; CAS, MOST and NSFC, China; COLCIENCIAS, Colombia; MSMT CR, MPO CR and VSC CR, Czech Republic; DNRF, DNSRC and Lundbeck Foundation, Denmark; IN2P3-CNRS, CEA-DSM/IRFU, France; GNSF, Georgia; BMBF, HGF, and MPG, Germany; GSRT, Greece; RGC, Hong Kong SAR, China; ISF, I-CORE and Benoziyo Center, Israel; INFN, Italy; MEXT and JSPS, Japan; CNRST, Morocco; FOM and NWO, Netherlands; RCN, Norway; MNiSW and NCN, Poland; FCT, Portugal; MNE/IFA, Romania; MES of Russia and NRC KI, Russian Federation; JINR; MESTD, Serbia; MSSR, Slovakia; ARRS and MIZ\v{S}, Slovenia; DST/NRF, South Africa; MINECO, Spain; SRC and Wallenberg Foundation, Sweden; SERI, SNSF and Cantons of Bern and Geneva, Switzerland; MOST, Taiwan; TAEK, Turkey; STFC, United Kingdom; DOE and NSF, United States of America. In addition, individual groups and members have received support from BCKDF, the Canada Council, CANARIE, CRC, Compute Canada, FQRNT, and the Ontario Innovation Trust, Canada; EPLANET, ERC, FP7, Horizon 2020 and Marie Skłodowska-Curie Actions, European Union; Investissements d'Avenir Labex and Idex, ANR, Region Auvergne and Fondation Partager le Savoir, France; DFG and AvH Foundation, Germany; Herakleitos, Thales and Aristeia programmes co-financed by EU-ESF and the Greek NSRF; BSF, GIF and Minerva, Israel; BRF, Norway; the Royal Society and Leverhulme Trust, United Kingdom.

The crucial computing support from all WLCG partners is acknowledged
gratefully, in particular from CERN and the ATLAS Tier-1 facilities at
TRIUMF (Canada), NDGF (Denmark, Norway, Sweden), CC-IN2P3 (France),
KIT/GridKA (Germany), INFN-CNAF (Italy), NL-T1 (Netherlands), PIC (Spain),
ASGC (Taiwan), RAL (UK) and BNL (USA) and in the Tier-2 facilities
worldwide.

\clearpage

\printbibliography

\newpage 

\begin{flushleft}
{\Large The ATLAS Collaboration}

\bigskip

G.~Aad$^{\rm 85}$,
B.~Abbott$^{\rm 113}$,
J.~Abdallah$^{\rm 151}$,
O.~Abdinov$^{\rm 11}$,
R.~Aben$^{\rm 107}$,
M.~Abolins$^{\rm 90}$,
O.S.~AbouZeid$^{\rm 158}$,
H.~Abramowicz$^{\rm 153}$,
H.~Abreu$^{\rm 152}$,
R.~Abreu$^{\rm 116}$
% ...
\bigskip
\\
% ...
$^{11}$ Institute of Physics, Azerbaijan Academy of Sciences, Baku, Azerbaijan\\
% ...
$^{85}$ CPPM, Aix-Marseille Universit{\'e} and CNRS/IN2P3, Marseille, France\\
% ...
$^{90}$ Department of Physics and Astronomy, Michigan State University, East Lansing MI, United States of America\\
% ...
$^{107}$ Nikhef National Institute for Subatomic Physics and University of Amsterdam, Amsterdam, Netherlands\\
% ...
$^{113}$ Homer L. Dodge Department of Physics and Astronomy, University of Oklahoma, Norman OK, United States of America\\
% ...
$^{116}$ Center for High Energy Physics, University of Oregon, Eugene OR, United States of America\\
% ...
$^{151}$ Institute of Physics, Academia Sinica, Taipei, Taiwan\\
$^{152}$ Department of Physics, Technion: Israel Institute of Technology, Haifa, Israel\\
$^{153}$ Raymond and Beverly Sackler School of Physics and Astronomy, Tel Aviv University, Tel Aviv, Israel\\
% ...
$^{158}$ Department of Physics, University of Toronto, Toronto ON, Canada\\
% ...
\end{flushleft}


\end{document}